\documentclass[twocolumn]{aastex631}

\accepted{for publication in the Astronomical Journal on \today}

\shorttitle{Small and Close-In Planets are Uncommon around A-type Stars}
\shortauthors{Giacalone $\&$ Dressing}
\graphicspath{{./}{figures/}}

\usepackage{enumitem}
\usepackage{amsmath}
\usepackage{amssymb}
\usepackage{hyperref}

\begin{document}

\title{Small and Close-In Planets are Uncommon around A-type Stars}

\correspondingauthor{Steven Giacalone}
\email{giacalone@astro.caltech.edu}

\author[0000-0002-8965-3969]{Steven Giacalone}
\altaffiliation{NSF Astronomy and Astrophysics Postdoctoral Fellow}
\affiliation{Department of Astronomy, California Institute of Technology, Pasadena, CA 91125, USA}
\affiliation{Department of Astronomy, University of California Berkeley, Berkeley, CA 94720, USA}

\author[0000-0001-8189-0233]{Courtney D. Dressing}
\affiliation{Department of Astronomy, University of California Berkeley, Berkeley, CA 94720, USA}



\begin{abstract}

The \textit{Kepler} and \textit{K2} missions enabled robust calculations of planet occurrence rates around FGKM-type stars. However, these missions observed too few stars with earlier spectral types to tightly constrain the occurrence rates of planets orbiting hotter stars. Using \textit{TESS}, we calculate the occurrence rate of small ($1 \, R_\oplus < R_{\rm p} < 8 \, R_\oplus$), close-in ($P_{\rm orb} < 10$~days) planets orbiting A-type stars for the first time. We search a sample of { 20,257} bright ($6 < T < 10$) A-type stars for transiting planets using a custom pipeline and vet the detected signals, finding no reliable small planets. We characterize the pipeline completeness using injection/recovery tests and determine the $3\sigma$ upper limits of the occurrence rates of close-in sub-Saturns ($4 \, R_\oplus < R_{\rm p} < 8 \, R_\oplus$), sub-Neptunes ($2 \, R_\oplus < R_{\rm p} < 4 \, R_\oplus$), and super-Earths ($1 \, R_\oplus < R_{\rm p} < 2 \, R_\oplus$). We find upper limits of $2.2 \pm 0.4$ sub-Saturns and $9.1 \pm 1.8$ sub-Neptunes per 1000 A-type stars, which may be more than $3\times$ and $6\times$ lower than \textit{Kepler}-era estimates for Sun-like stars. We calculate an upper limit of $186 \pm 34$ super-Earths per 1000 A-type stars, which may be more than $1.5\times$ lower than that for M dwarfs. Our results hint that small, close-in planets become rarer around early-type stars and that their occurrence rates decrease faster than that of hot Jupiters with increasing host star temperature. We discuss plausible explanations for these trends, including star-disk interactions and enhanced photoevaporation of planet atmospheres.

\end{abstract}



\section{Introduction} \label{sec:intro}

The \textit{Kepler} spacecraft \citep{borucki2010} provided a glimpse into the demographics of planets orbiting close to FGKM-type stars \citep[e.g.,][]{howard2012planet, dressing2013occurrence, dressing2015occurrence, fressin2013false, petigura2013, mulders2015stellar,mulders2015increase, hsu2019occurrence, hardigree2019, kunimoto2020, bryson2021, bergsten2022demographics, bergsten2023evidence, zink2023K2, christiansen2023K2}. Because the primary goal of the \textit{Kepler} mission was to calculate the occurrence rate of Earth-like planets orbiting Sun-like stars, more massive and hotter stars were largely neglected in the search for transiting planets, with only $1.6 \%$ of the 200,000-star \textit{Kepler} mission target list being comprised of stars hotter than 7500~K \citep{mathur2017dr25}. Ground-based transit surveys have found over a dozen hot Jupiters around stars hotter than 7500~K \citep[e.g.,][]{gaudi2017kelt, lund2017kelt}, but their relatively low photometric precisions and lack of continuous monitoring have made it challenging to find planets smaller than Jupiter or with orbital periods longer than a few days. In addition, radial velocity surveys, which are similarly sensitive to close-in planets, typically have not targeted hot stars due to their rapid rotation rates that greatly limit radial velocity precision and inhibit detection of orbiting planets \citep{griffin2000, galland2005}. As a consequence, our understanding of short-period planets has historically been limited to cooler FGKM-type stars.

With the launch of the Transiting Exoplanet Survey Satellite \citep[\textit{TESS};][]{ricker2010transiting}, which observes stars nearly indiscriminately across almost the entire sky, our ability to search for transiting planets around early-type stars improved drastically. Some studies have utilized this ability to explore the occurrence rate of hot Jupiters orbiting main-sequence A-type stars, finding evidence that the occurrence rate of hot Jupiters decreases with increasing stellar mass, with about half as many hot Jupiters around A-type stars compared to G-type stars \citep{zhou2019,beleznay2022hot}. However, this finding is at tension with findings from earlier surveys. For instance, using data from the \textit{K2} mission \citep{howell2014K2}, \citet{zink2023K2} found no significant change in the occurrence rate of hot Jupiters across FGK-type stars. In addition, radial velocity surveys that targeted slowly rotating ``retired'' A-type stars have reported an increase in giant planet occurrence rate with increasing stellar mass \citep{johnson2010}.\footnote{However, we note that these calculations included giant planets with longer orbital periods than those in \citet{zhou2019} and \citet{beleznay2022hot}, which may have different formation and migration histories.} More detailed calculations, which utilize a larger sample of stars and a longer {\it TESS} baseline, are needed to definitively resolve these discrepancies.

The occurrence rate of small ($R_\mathrm{p} < 8 \, R_\oplus$) close-in ($P_\mathrm{orb} < 10$ days) planets orbiting A-type stars is, to this day, completely unknown. The demographics of these planets can reveal novel information about how planets form and evolve around stars of different masses. For example, it is known that more massive stars have more dust mass in their protoplanetary disks \citep{ansdell2016disk}, providing more solid material with which to build planets. It has been posited that this increased dust mass is responsible for the observation that wide-separation ($10-100$ AU) giant planets are more than twice as common around high-mass ($M_\star > 1.5 \, M_\odot$) stars than low-mass ($M_\star < 1.5 \, M_\odot$) stars \citep{nielsen2019}. More massive stars are also known to become depleted in gas at a higher rate than less massive stars \citep{ribas2015}. Thus, planets orbiting A-type stars likely have relatively little time to migrate towards their host stars via disk migration, a process that relies on the presence of gas in the disk, in comparison to lower mass stars.

After the dispersal of the protoplanetary disk, planets orbiting close to A-type stars are subjected to different environments than those orbiting cooler stars. Unlike low-mass stars, which have convective outer layers, A-type stars are primarily radiative in structure. It has been inferred that these radiative layers are less efficient at dissipating tides excited by short-period planets \citep[e.g.,][]{winn2010}, potentially leading to slower tide-induced orbital evolutions relative to cooler stars. The radiative interiors of A-type stars also lead to lower levels of X-ray and extreme-ultraviolet (XUV) radiation \citep{schroder2007}, which is thought to drive the photoevaporation of planetary atmospheres around young low-mass stars \citep[e.g.,][]{lammer2003, ribas2005evolution, murrayclay2009escape, jackson2012coronal}. However, unlike their low-mass counterparts, A-type stars have high levels of near-ultraviolet continuum emission, which may drive efficient photoevaporation of planet atmospheres over the entire main-sequence lifetime of the star \citep{munoz2019}. The impact of photoevaporation on the landscape of short-period planets around Sun-like stars and cooler has been explored in depth \citep[e.g.,][]{lopez2013role, owen2013kepler, owen2017evaporation, lopez2018formation, owen2018photoevaporation}, but the role this mechanism plays in sculpting the population of planets orbiting hotter stars is less well understood.

The population of short-period planets around A-type stars is also relevant to our understanding of planetary systems orbiting white dwarfs. Because A-type stars have short lifetimes and masses too low for the formation of neutron stars or black holes upon their collapse, many of the white dwarfs we observe today are the remnants of A-type stars \citep[e.g.,][]{heger2003stardeath, cummings2018whitedwarf}. Evidence has been found for the presence of transiting planets, asteroids, and metal-rich disks around these compact objects \citep{becklin2005dustydisk, gansicke2006metaldisk, vanderburg2015disintegrating, vanderburg2020giant, rappaport2016asteroid, vanderbosch2020circumstellar}. { In addition, observations have found that up to $50 \%$ of white dwarfs have heavy element contamination in their photospheres \citep{zuckerman2003whitedwarfs, zuckerman2010whitedwarfs, koester2014whitedwarfs}}. Some studies have proposed planetary, asteroidal, or cometary engulfment to be responsible for this metal enhancement \citep{bonsor2015pollution, petrovich2017engulfment, harrison2021pollution, mcdonald2021engulfment, buchan2022pollutants, stock2022pollution, oconnor2023exocomet}. In general, planetary mass companions and rocky material around white dwarfs are believed to originate at wide separations and migrate inwards via common-envelope evolution \citep{lagos2021commonenvelope,merlov2021commonenvelope}, Kozai-Lidov cycles \citep{munoz2020kozai, oconnor2021kozai, stephan2021kozai}, or planet-planet scattering \citep{veras2015unpacking, maldonado2021instabilities, veras2021hr8799} after the star has evolved off of the main sequence. { Planets that are close-in while their stars are on the main sequence are unlikely to contribute to this contamination, as they would be engulfed by their post-main-sequence host stars as they expand. Any surviving material accreted onto the white dwarf would quickly sink into its interior via gravitational sedimentation \citep[e.g.,][]{bauer2019polluted}. However, the demographics of planets in the inner regions of these systems may inform planet formation and evolution theories at more distant orbits and provide clues for the origin of this white dwarf pollution phenomenon.}

To gain a better understanding of how planets form and evolve around early-type stars, we calculate the occurrence rate of small short-period planets orbiting A-type stars using \textit{TESS} \citep{mast}. The calculation provides a closer look into the population of planets smaller than Jupiter around these hot stars, providing new insight into the demographics of small planets in the galaxy. This work complements other recent studies that use \textit{TESS} to explore the demographics of planetary systems in previously inaccessible regions of stellar parameter space \citep[e.g.,][]{fernandes2022tess, fernandes2023tess, bryant2023, gan2023rate, ment2023occurrence, temmink2023occurrence, vach2024occurrence}. 

This paper is organized as follows. In Section \ref{sec:sample}, we discuss the selection of the stellar sample used in this study. In Section \ref{sec:search}, we describe our planet detection pipeline and planet candidate vetting procedure. In Section \ref{sec:completeness}, we outline the calculation of our pipeline completeness. In Section \ref{sec:occurrence}, we calculate the occurrence rate of small, short-period planets around our sample of stars. In Section \ref{sec:discussion}, we compare our results to previous studies and discuss possible physical explanations for our calculated occurrence rate. Lastly, in Section \ref{sec:conclusions}, we provide a summary of our findings and brief concluding remarks.

\begin{figure*}[hbtp]
  \centering
    \includegraphics[width=1.0\textwidth]{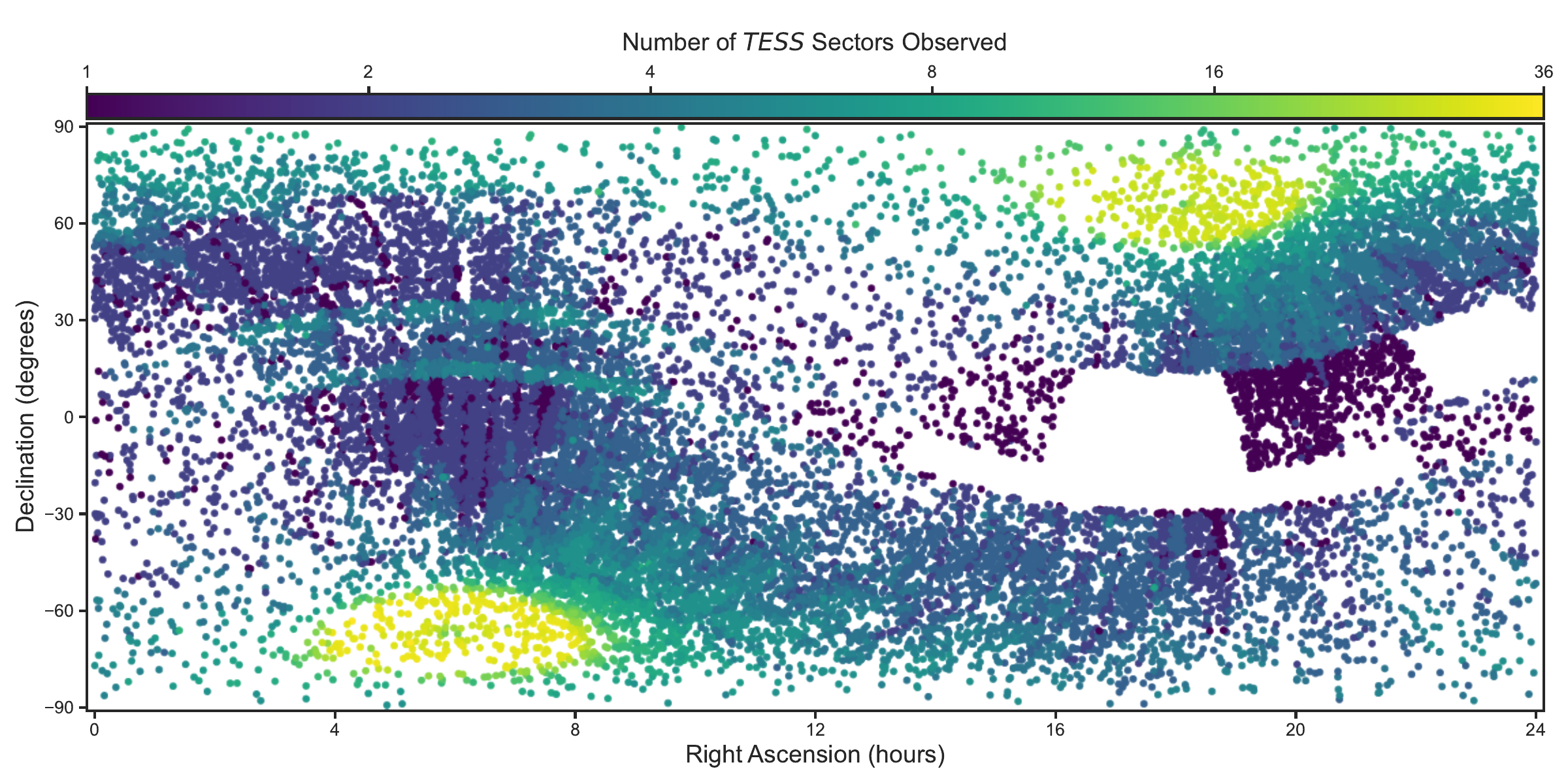}
    \includegraphics[width=0.325\textwidth]{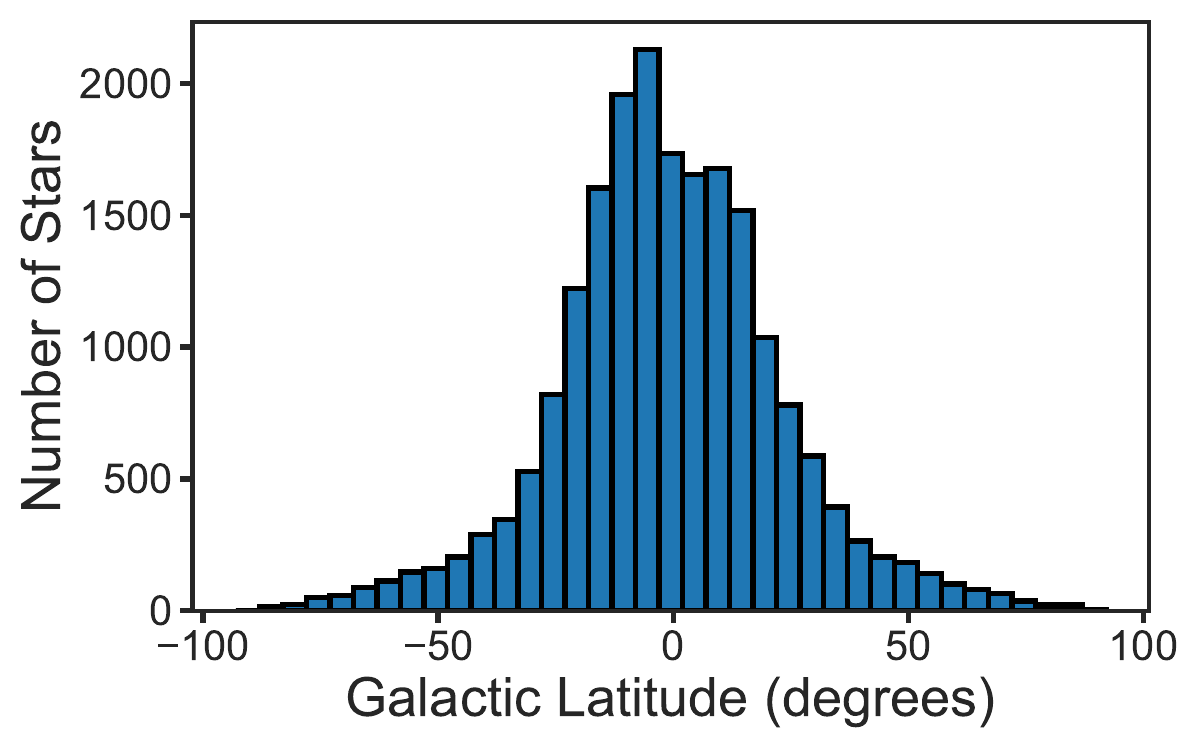}
    \includegraphics[width=0.325\textwidth]{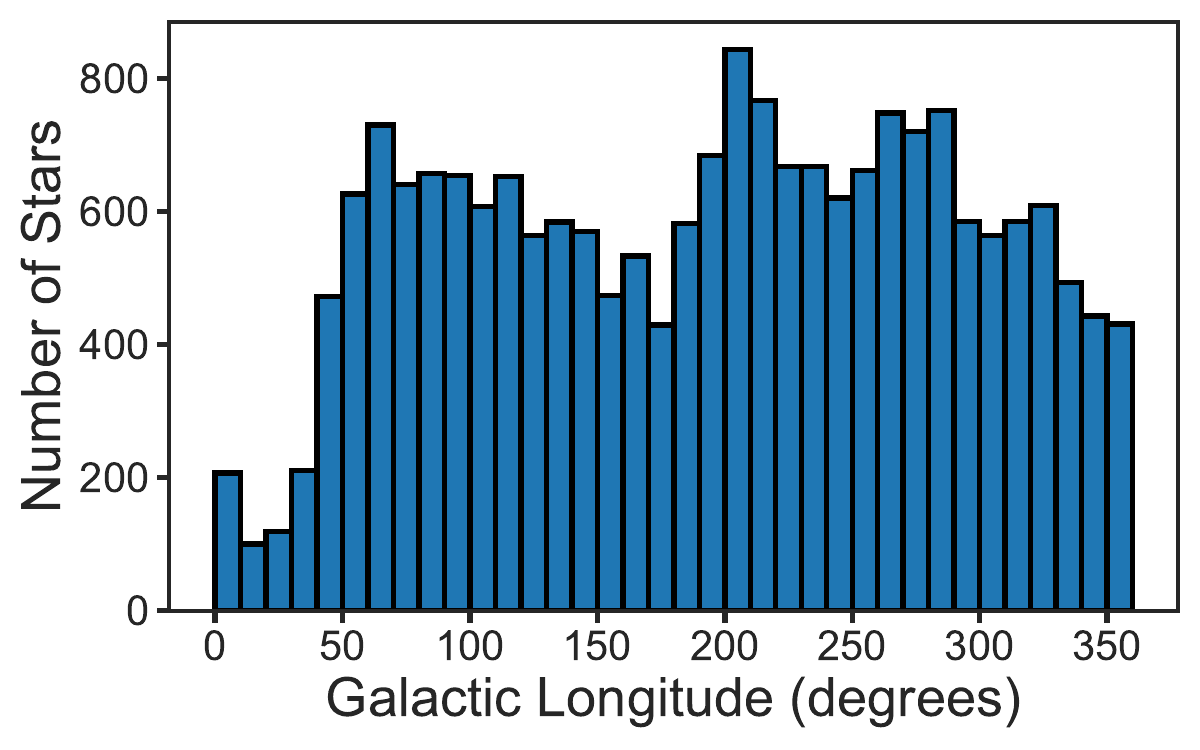}
    \includegraphics[width=0.325\textwidth]{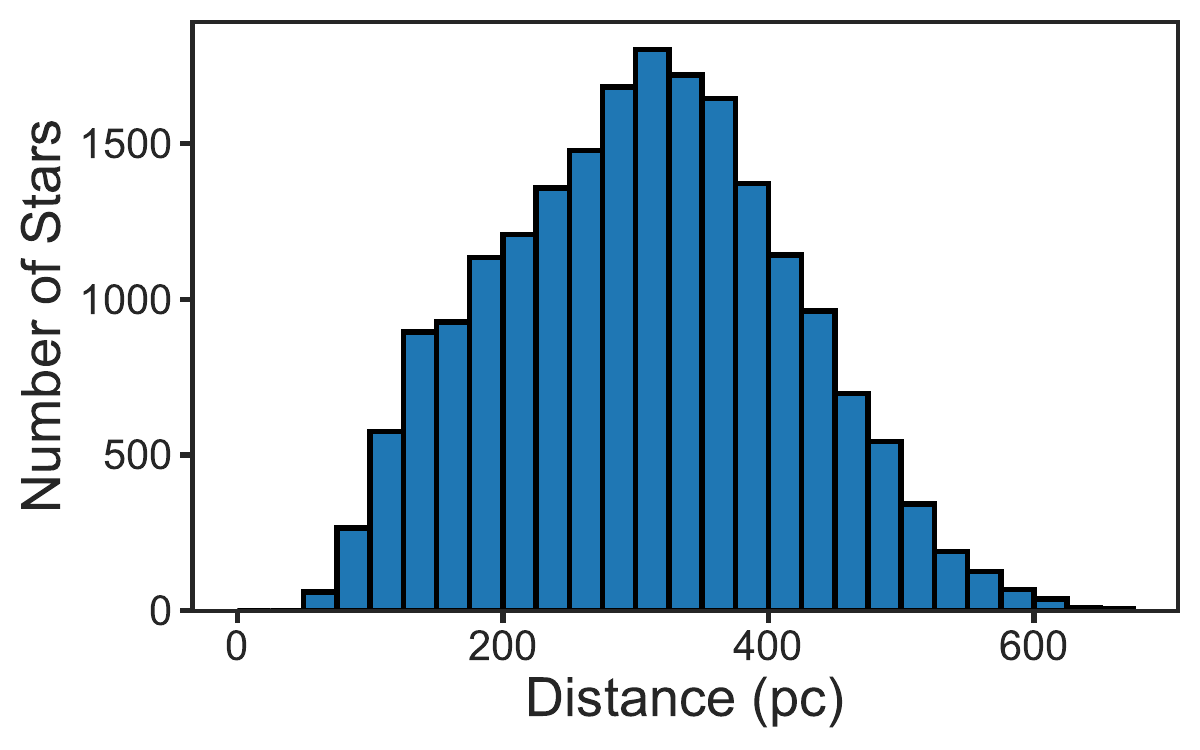}
    \includegraphics[width=0.325\textwidth]{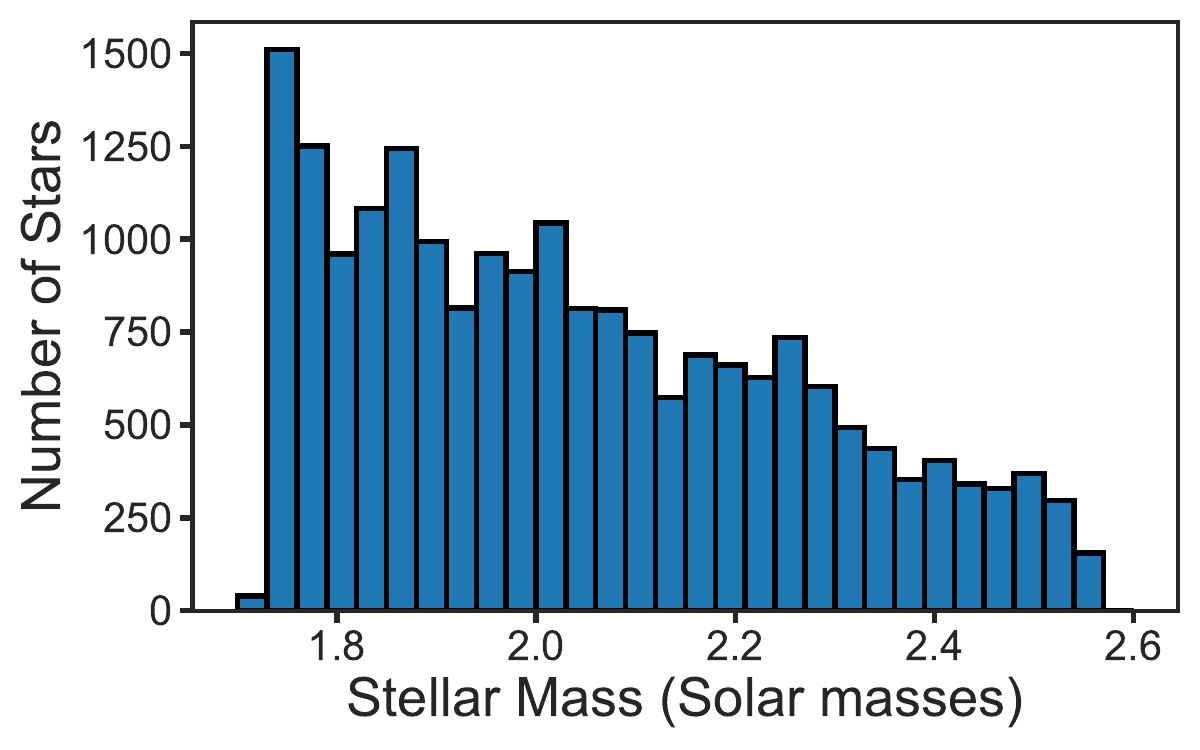}
    \includegraphics[width=0.325\textwidth]{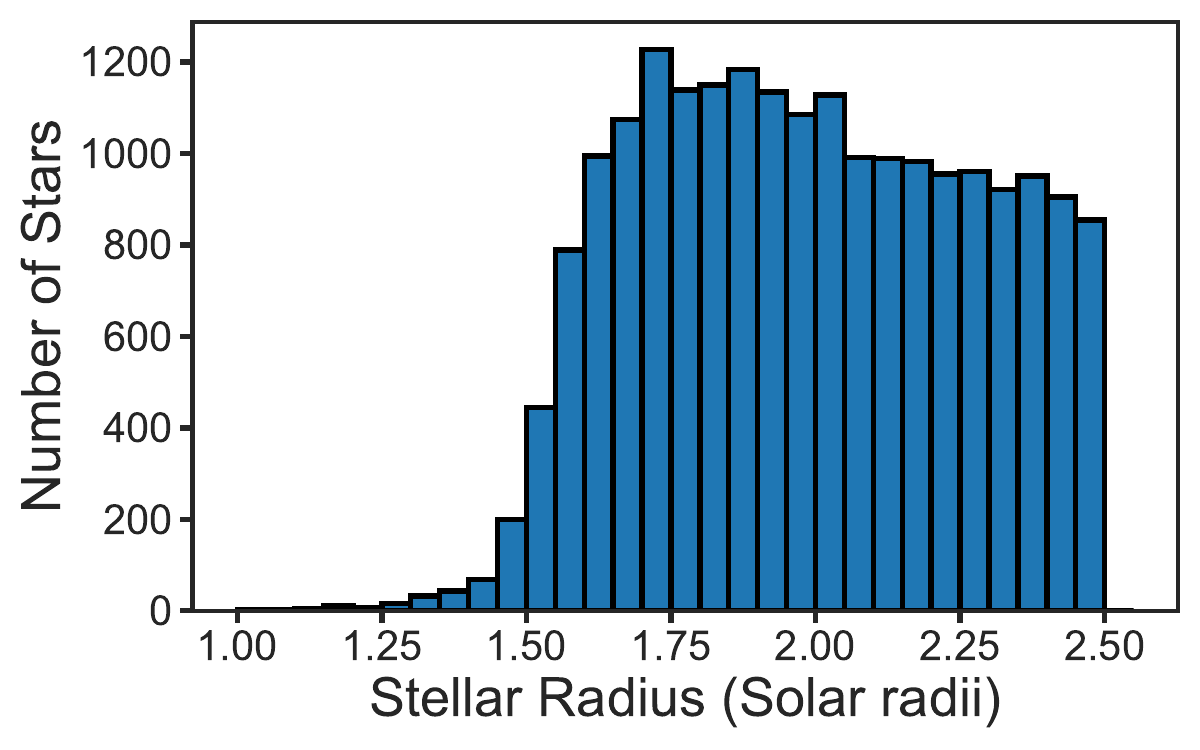}
    \includegraphics[width=0.325\textwidth]{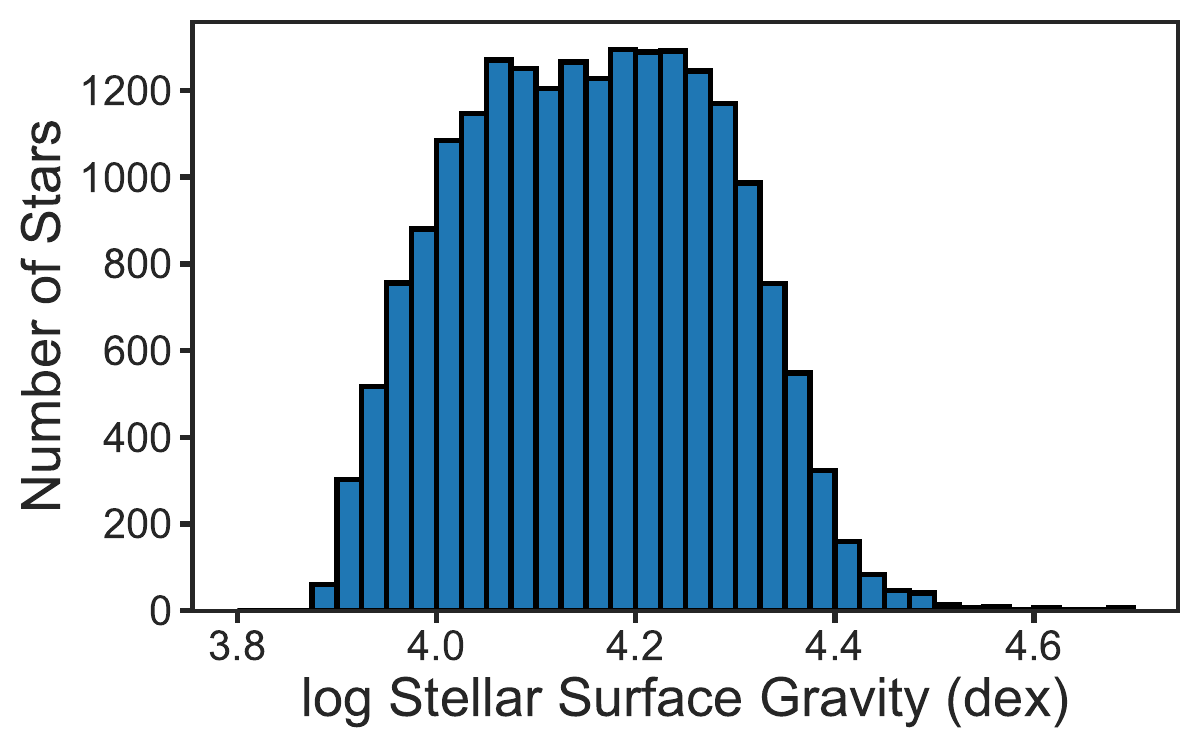}
    \includegraphics[width=0.325\textwidth]{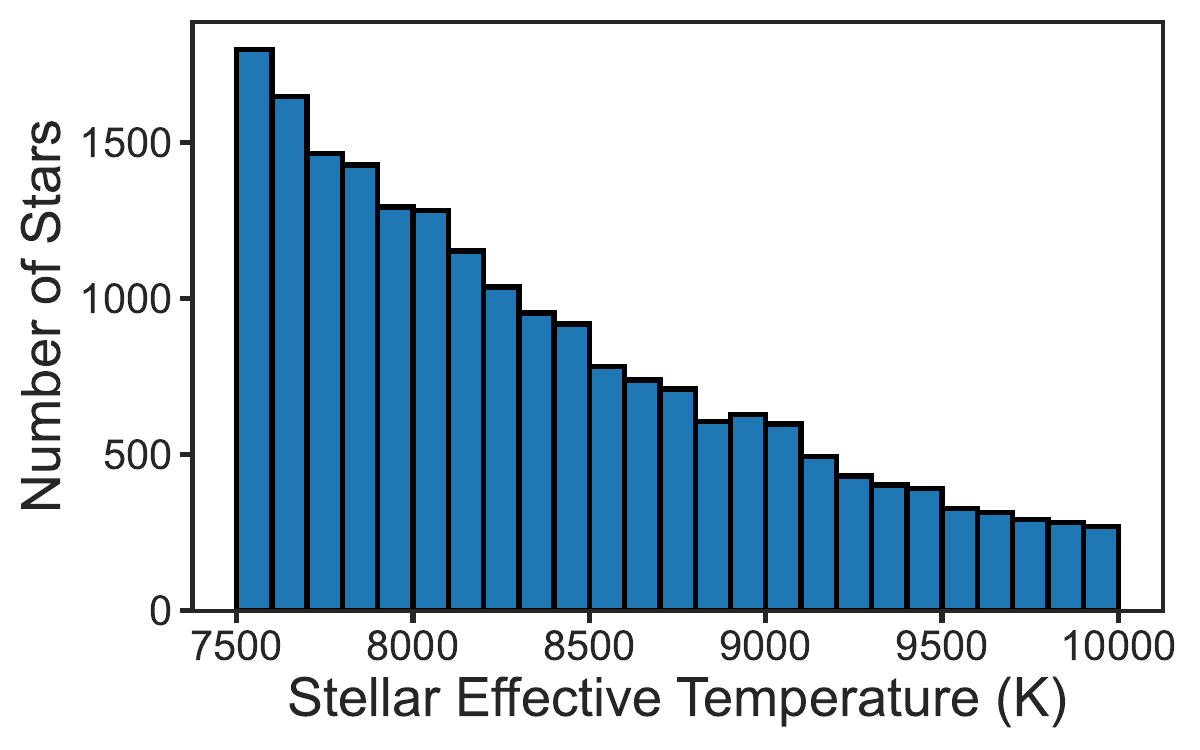}
    \includegraphics[width=0.325\textwidth]{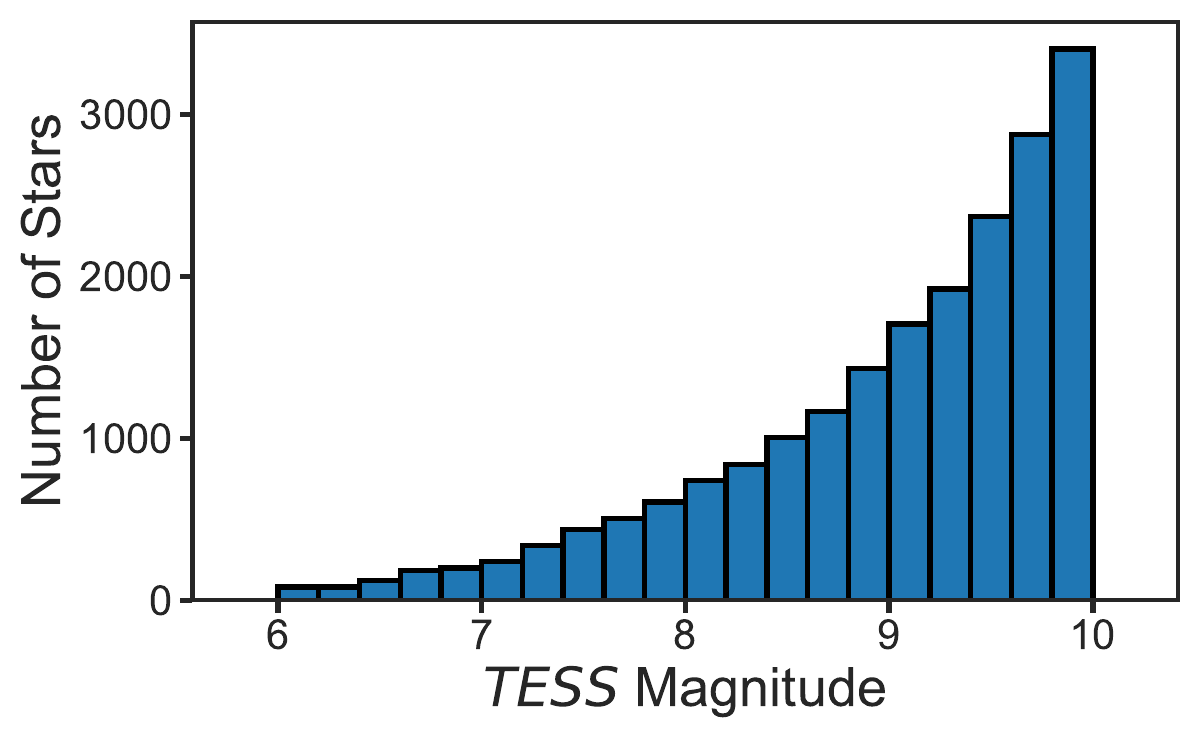}
    \includegraphics[width=0.325\textwidth]{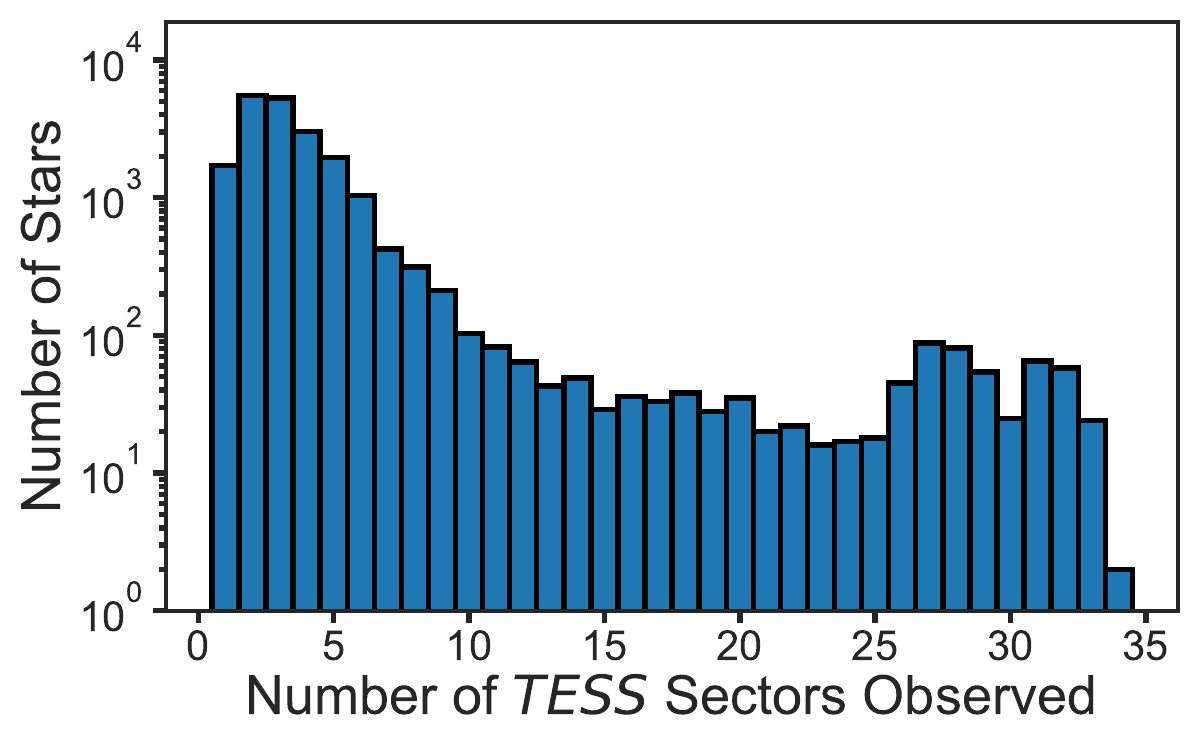}
    \caption{Coordinates and properties of the { 20,257} A-type stars that are used in this study. In the top-most panel, each point is a star, with color indicating the number of sectors it has been observed by \textit{TESS}.}
    \label{fig: Astar_sample}
\end{figure*}

\section{Sample Selection}\label{sec:sample}

We begin by defining a sample of main-sequence A-type stars around which to search for planets. We build this sample by querying version 8.2 of the \textit{TESS} Input Catalog (TIC; \citealt{stassun2018tess,stassun2019}), selecting stars that have all of the following properties:
\begin{itemize}[noitemsep]
    \item An effective temperature ($T_{\rm eff}$) between 7500 and 10000 K,
    \item A radius ($R_\star$) less than or equal to 2.5 $R_\odot$,
    \item A \textit{TESS} magnitude ($T$) { between 6 and 10},
    \item A non-zero mass ($M_\star$) estimate.
\end{itemize}
In addition, we require all stars to have data products from the MIT Quick Look Pipeline (QLP; \citealt{huang2020photometry1,huang2020photometry2}) available on the Mikulski Archive for Space Telescopes (MAST). The $T_\mathrm{eff}$ requirement is meant to ensure that the stars are of the desired spectral type.\footnote{We note that similar $T_{\rm eff}$ bins have been used by previous occurrence rate studies, such as \citet{mulders2015stellar}, with which we compare our results in Section \ref{sec:discussion}.} The $M_\star$ and $R_\star$ requirements ensure that the stars are on the main sequence and also remove particularly large stars for which it is difficult to detect the transits of small planets. { The lower $T$ requirement removes very bright stars that are likely to have extreme systematics due to saturation of the {\it TESS} cameras}. The { upper} $T$ requirement removes particularly faint stars, for which \textit{TESS} has a lower photometric precision and therefore a lower sensitivity to planetary transits \citep[e.g.,][]{barclay2018yield, kunimoto2022predicting}. The QLP requirement is needed because the QLP-processed light curves are used in the occurrence rate calculation, which is described further in the following sections. The final sample is shown in Figure \ref{fig: Astar_sample} and consists of { 20,257} stars observed in \textit{TESS} full frame images between sectors 1 and 69 (i.e., the first five years of the \textit{TESS} mission).

\begin{figure}[t!]
  \centering
    \includegraphics[width=0.48\textwidth]{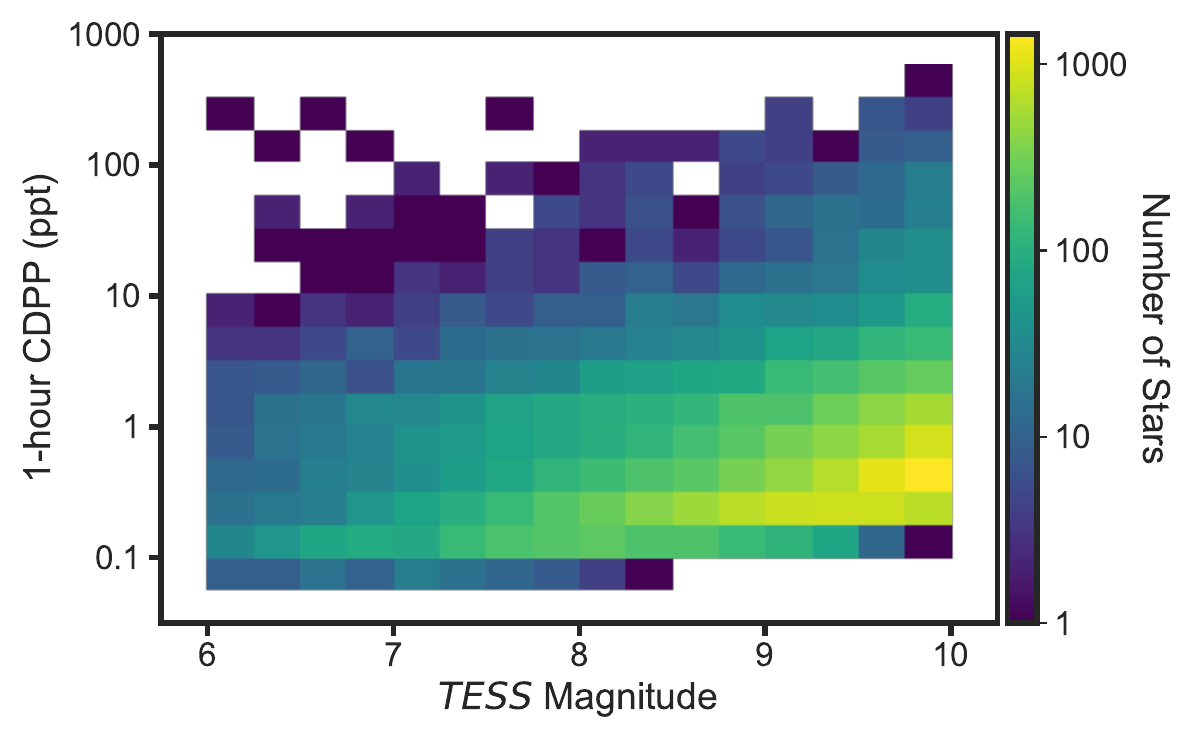}
    \caption{ 2D histogram of the stars in our sample, binned according to their 1-hour combined differential photometric precisions (CDPPs) and their {\it TESS} magnitudes. Most stars in the sample have CDPPs between 0.1 and 1 parts per thousand.}
    \label{fig: CDPP}
\end{figure}

{ To better understand the noise properties of our sample, we performed a preliminary exploration of the combined differential photometric precisions (CDPPs) of the stars \citep{gilliland2011cdpp, vancleve2016cdpp} using the \texttt{estimate$\_$cdpp} function in the \texttt{Lightkurve} Python package \citep{lightkurve2018}. Figure \ref{fig: CDPP} displays the distribution of CDPP as a function of $T$, assuming a 1-hour window length. A majority of the stars in our sample have CDPPs between 0.1 and 1 parts per thousand, in general agreement  with previously reported noise values \citep[e.g.,][]{kunimoto2022qlp, kunimoto2022predicting}. Given these noise levels, we anticipated being able to detect the transits of sub-Jovian planets around our sample of stars. We direct the reader to Section~\ref{sec:completeness} for a more rigorous analysis of the sensitivity.}

{ We note that $\sim 18 \%$ of the stars in our sample have radii that fall below the typical cutoff for main-sequence A-type stars ($R_\star \approx 1.7 \, R_\odot$). To assess the reliability of the TIC properties of these relatively small stars, we compare their values of $T_{\rm eff}$ and surface gravity ($\log g$) to those estimated in \textit{Gaia} DR3 \citep{gaia2023dr3}. Specifically, we compare them to the quantities estimated by the Apsis pipeline using the GSP-Phot module, which determines stellar parameters using photometry and BP/RP spectra \citep{fouesneau2023apsisII}. We find that the $T_{\rm eff}$ values agree well in the two catalogs; the distribution of $T_{\rm eff, DR3}/T_{\rm eff, TIC}$ has a mean and standard deviation of $1.010 \pm 0.096$. We find that the $\log g$ values are systematically lower in \textit{Gaia} DR3 than they are in the TIC; the distribution of $\log g_{\rm \, DR3}/\log g_{\rm \, TIC}$ has a mean and standard deviation of $0.969 \pm 0.026$. While this discrepancy is statistically significant, it only corresponds to a modest $R_\star$ increase of $\sim 20\%$ when going from TIC to \textit{Gaia} DR3 (assuming a fixed mass). We ultimately conclude that TIC parameters for these stars are relatively reliable, although we account for this potential $20 \%$ systematic uncertainty in $R_\star$ in our occurrence rate calculation (see Section \ref{sec:occurrence}).

Assuming the radii of these stars are correct, it is alternatively possible that some of them are subdwarf stars. In general, there are two classes of subdwarfs: hot subdwarfs, which are stars that have evolved off of the main sequence and have lost significant fractions of their envelopes as they expanded (most likely due to stripping from a close-in degenerate companion; \citealt{heber2009hot}), and cool subdwarfs, which are main-sequence stars with very low metallicities \citep{kuiper1939sd}. Hot subdwarfs are typically O-type or B-type with surface gravities of $\log g \gtrsim 5$ and $T_{\rm eff} > 20,000$~K \citep{heber2009hot}. Our stars are much too cool to fall under this category, but bear a slightly closer resemblance to subdwarf A-type stars (sdAs), which have similarly high surface gravities but $T_{\rm eff} < 20,000$~K \citep{kepler2016sdA}. It was originally hypothesized that sdAs are extremely low-mass white dwarfs that formed in a way similar to hot subdwarfs. If this were true and the smallest stars in our sample (which are generally easier to find transiting planets around) are in fact evolved, it would introduce age-related biases into our analysis that would likely compromise the validity of our results. Nonetheless, the general consensus today is that the vast majority of sdAs are indeed metal-poor main-sequence A-type stars \citep{brown2017sdA, hermes2017sdA, pelisoli2018asdA, pelisoli2018bsdA, pelisoli2019sdA, kepler2019sdA, yu2019sdA}. We therefore keep these stars in our sample with high confidence that they are indeed on the main sequence.}

\section{Planet Search}\label{sec:search}

\subsection{Light Curve Retrieval and Flattening}\label{sec:lightcurves}

For a given star, we download the QLP light curve file using the \texttt{Lightkurve} Python package \citep{lightkurve2018}. Downloaded light curves have cadences of 30 minutes (sectors 1--26), 10 minutes (sectors 27--55), and 200 seconds (sectors 56--69). We record the ``raw'' flux provided by the QLP, rather than that which has been flattened using a spline-fitting procedure, and mask out all data flagged as having poor quality (quality flag $>0$). We use the raw flux, rather than the spline-flattened flux provided by the QLP, for the purpose of calculating the pipeline sensitivity. Any light curve flattening procedure has the potential to remove or distort embedded transits. The impact this has on the ability of the pipeline to detect planets can only be quantified by injecting artificial transits into raw light curves before the flattening is performed (this injection/recovery process is described in detail in Section~\ref{sec:completeness}). We therefore apply a custom light curve flattening routine on the raw data.

The custom flattening routine is performed using the \texttt{w{\={o}}tan} Python package \citep{hippke2019wotan}. We first separate the light curve into segments by identifying gaps $> 0.5$ days in length. We then discard segments $< 2$ days long, which cannot be flattened easily without affecting the shapes of planet transits. We flatten each remaining segment using the ``robust penalized pspline'' algorithm, which fits the data to a spline curve through iterative sigma-clipping. In each iteration, data points that are $> 3 \sigma$ outliers are removed, and the iterations continue until no outliers remain or until 10 iterations occur. The algorithm calculates the optimal number of knots per segment using Ridge regression, in which the data is fit using a cost function that penalizes solutions in which a greater number of knots are used in order to combat overfitting. The minimum distance between knots is set to 0.5 days to prevent the algorithm from significantly impacting transits in the data, which will always have durations under 12 hours for orbital periods under 10 days. We determined this minimum knot distance visually using real {\it TESS} light curves with artificial injected transits. We then remove 0.25 days of data from the edges of each segment, due to the flattening procedure struggling to fit the data properly near the start and end of each segment.

Next, we remove data collected near \textit{TESS} momentum dumps. These momentum dumps cause temporary changes in spacecraft pointing, leading the positions of stars to change slightly within the \textit{TESS} pixels. These shifts, which are not accounted for in the simple aperture photometry light curve extraction used by the QLP, typically cause the measured flux of the target star to increase or decrease as more or less light is captured by the predefined aperture. Because momentum dumps are regularly spaced in any given sector, these dips in flux are often erroneously identified as periodic transits by transit-detection algorithms. In order to prevent our pipeline from labeling these false alarms as potential planets, we do the following:
\begin{itemize}[noitemsep]
    \item For data collected in sectors 1--13, we remove data within 0.25 days of each momentum dump. Each of these sectors typically experienced 8 momentum dumps separated by 3--4 days.
    \item For data collected in sectors 13--26, we remove data within 0.5 days of each momentum dump. Each of these sectors experienced 4--6 momentum dumps separated by 2--5 days.
    \item For all remaining sectors, we remove data within 0.75 days of each momentum dump. After sector 26, each sector experienced 2--4 momentum dumps separated by 5--15 days.
\end{itemize}
Sectors with fewer momentum dumps have more data removed per dump because those events tend to induce a larger change in pointing and therefore a larger change in measured flux.

Lastly, we clean our flattened light curves by clipping $10 \sigma$ outliers and removing data points labeled as \texttt{NaN}. The flattened and cleaned light curves are then analyzed for transiting planets using the methods described below.

\begin{figure*}[t!]
  \centering
    \includegraphics[width=1.0\textwidth]{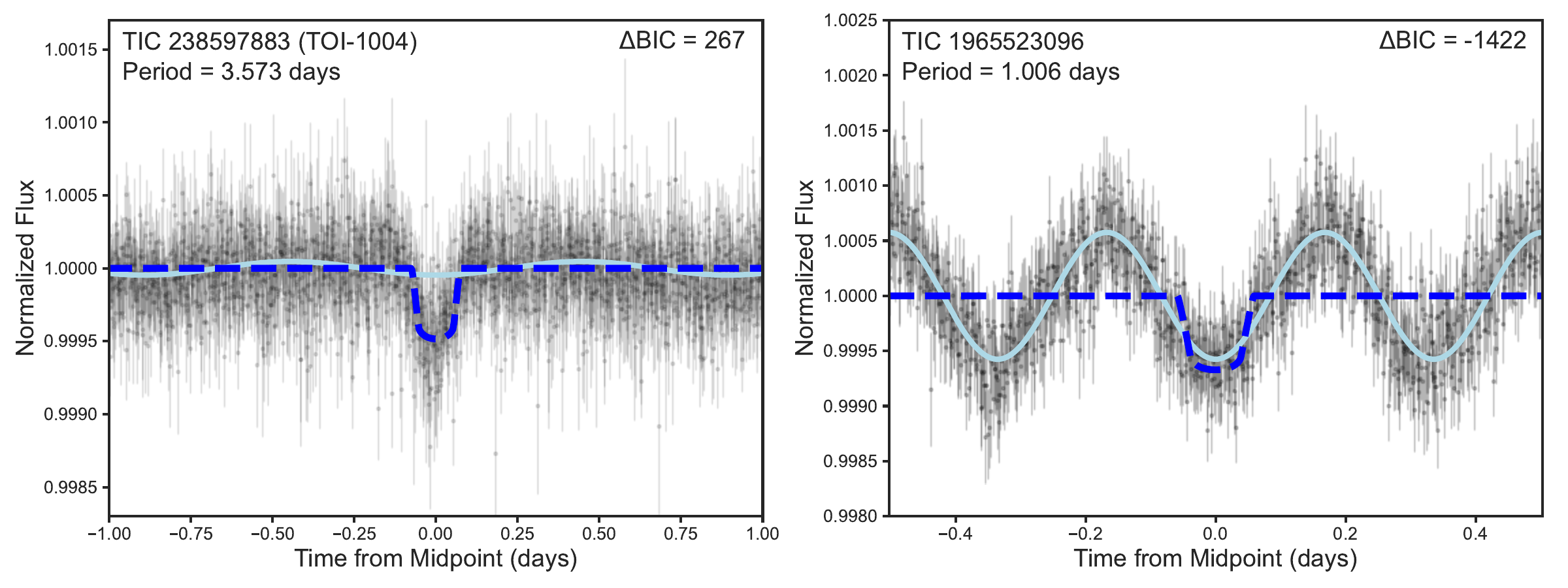}
    \caption{Visualization of sinusoid test described in Section \ref{sec:TCEpipeline}. The black points show the phase-folded data of the event detected by the BLS periodogram. The solid light blue curves show the best-fit sinusoid models and the dashed dark blue curves show the best-fit transit models. The left-side panel shows the results for TIC~238597883 (TOI-1004), which strongly favors the transit model, and the right-side panel shows the results for TIC~1965523096, which strongly favors the sinusoid model. With our thresold of $\Delta {\rm BIC} = 50$, the former passes the sinusoid test and the latter fails.}
    \label{fig: sinusoid_test}
\end{figure*}

\subsection{Automated TCE Detection Pipeline}\label{sec:TCEpipeline}

After flattening and cleaning the raw QLP light curve, we employ an automated pipeline to detect threshold-crossing events (TCEs) and eliminate those that display non-transit-like morphologies. We begin by searching for periodic signals using the box least-squares algorithm \citep[BLS;][]{kovacs2002bls}, as implemented in \texttt{Lightkurve}. The BLS searches for events using a grid of 100,000 orbital periods between 0.5 days and 10 days that are uniformly separated in frequency space.\footnote{ We note that other choices of frequency spacings, such as $f^{2/3}$, have been shown to be more efficient \citep{ofir2014optimizing}. Our decision to use a uniform frequency spacing was purely due to it being the default option in \texttt{Lightkurve}.} { Next, we split the grid of orbital periods into 0.5-day chunks. Within each chunk, the search utilizes several possible box widths (i.e., transit durations, $T_{\rm dur}$), with a maximum possible box width determined by the following equation:
\begin{equation}
    T_{\rm dur, max} = \frac{P_{\rm orb, max}}{\pi} \arcsin{\left(\frac{R_\star}{a_{\rm max}}\right)},
\end{equation}
where $P_{\rm orb, max}$ is the maximum period searched within the chunk, $a_{\rm max}$ is the maximum possible semi-major axis (given $P_{\rm orb, max}$, a circular orbit, and the properties of the star), and an orbital inclination of $90^\circ$ is assumed. At the end of this search, the BLS periodograms generated for the chunks are recombined so as to span the full 0.5--10 day period range. The purpose of this procedure is to decrease the number of false alarm detections caused by signals with durations too long to be caused by a planet transiting the target star at a given orbital period.} Using the periodogram generated by this procedure, we calculate the signal detection efficiency (SDE) for each tested orbital period using the Equation~6 of \citet{kovacs2002bls} and record all periods with $\mathrm{SDE} \geq 10$.

For signals with $\mathrm{SDE} \geq 10$, we calculate the signal-to-noise ratio (S/N) following \citet{christiansen2012cdpp}:
\begin{equation}
    {\rm S/N} = \frac{(R_{\rm p}/R_\star)^2}{\rm CDPP_{\rm 1 hr}} \sqrt{N_{\rm tra}}
\end{equation}
where ${\rm CDPP_{\rm 1 hr}}$ is the 1-hour combined differential photometric precision for the star calculated in Section~\ref{sec:sample} and $N_{\rm tra}$ is the number of transits observed. The purpose of this S/N test is to eliminate signals with high probabilities of being false alarms. For instance, the \textit{Kepler} mission pipeline required signals to have S/N $>$ 7.1 to advance to TCE status, which \citet{jenkins2010tps} determined to ensure no more than one statistical false alarm in the \textit{Kepler} data set \citep{jenkins2002transits}. For our pipeline, we apply a more liberal requirement of S/N $>$ 1 for a signal to advance to the next stage of vetting. While this threshold is expected to result in more false alarms than relatively high thresholds, we apply additional automated and manual tests to eliminate non-planetary signals from our planet candidate list, which we outline below.

\begin{figure*}[t!]
  \centering
    \includegraphics[width=1.0\textwidth]{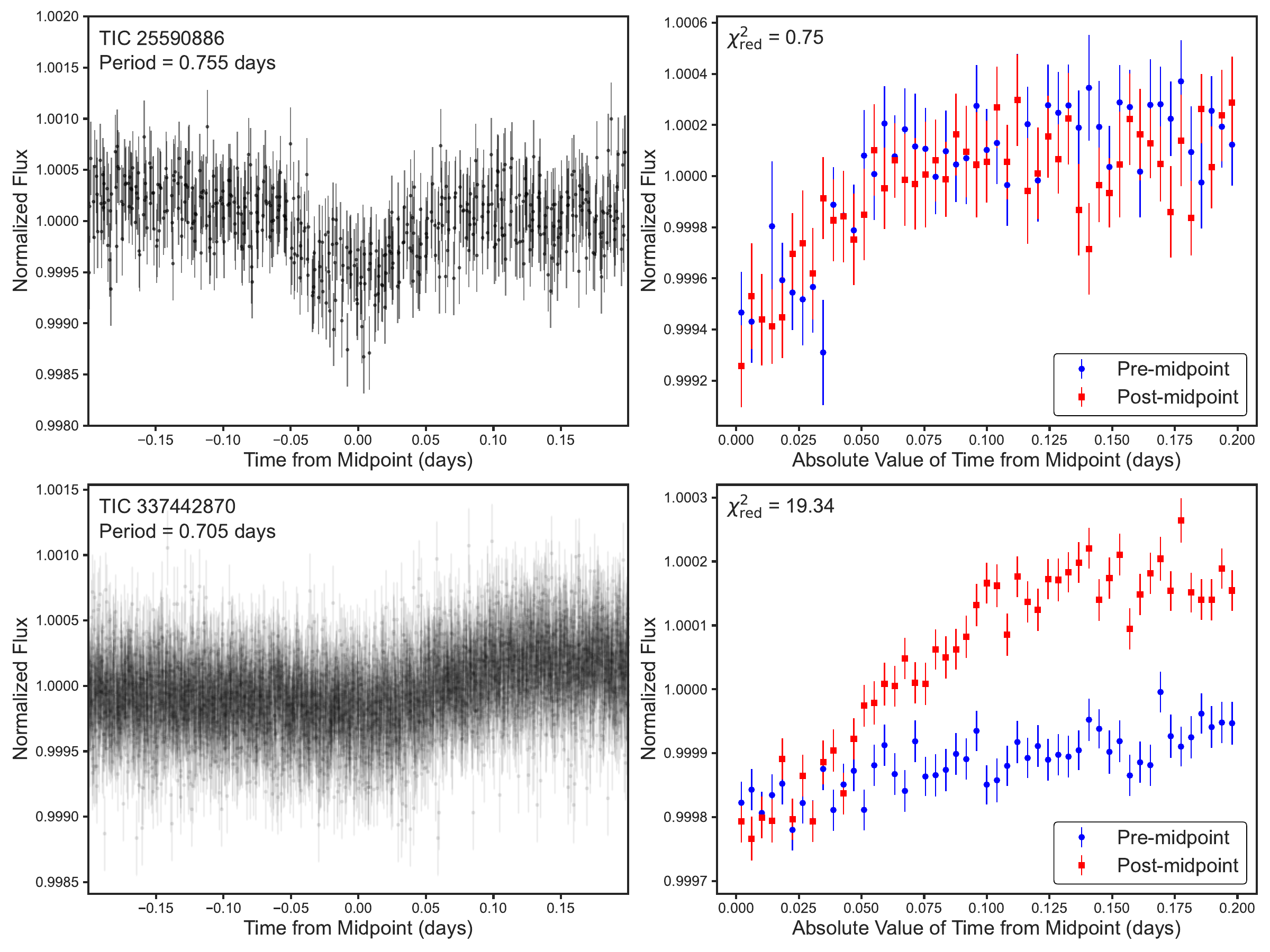}
    \caption{Visualization of event symmetry test described in Section \ref{sec:TCEpipeline}. The left-side panels show the phase-folded events detected by the BLS periodogram. The right-side panels compare the binned pre-midpoint (blue circles) and post-midpoint (red squares) data. The top two panels show the results for a symmetric event detected around TIC~25590886, which returns $\chi^2_{\rm red} = 0.75$. The bottom two panels show the results for an asymmetric event detected around TIC~337442870, which returns $\chi^2_{\rm red} = 19.34$. With our threshold of $\chi^2_{\rm red} = 2$, the former passes the symmetry test and the latter fails.}
    \label{fig: symmetry_test}
\end{figure*}

All signals that surpass the SDE and S/N thresholds are analyzed with a test to determine if they more closely resemble transits or sinusoids. \citet{kunimoto2023falsealarm} noted a high rate of sinusoid-like false alarms in \textit{TESS} data at orbital periods under $\sim 1$ day, likely as a result of scattered light from the rotation of the Earth, stellar variability, and blended light from nearby contact eclipsing binaries. However, it was found that the Sine Wave Evaluation Event Test (SWEET), utilized by \citet{thompson2018planetary} for the \textit{Kepler} data set, is able identify and remove a significant fraction of these false alarms. We employ a test similar to SWEET in our pipeline. First, we phase fold the light curve using the event period and midpoint returned by the BLS. We then fit the phase-folded data to two models using a simple grid-based optimization approach. The first model is a sinusoid model with three free parameters: amplitude, period, and phase. The second model is a transit model, as implemented by the \texttt{batman} Python package \citep{kreidberg2015batman}, with two free parameters: planet radius and orbital inclination. For the transit model, { the orbital ephemeris is fixed to that returned by the BLS search,\footnote{ While the orbital period and transit epoch returned by the BLS are not as precise as those that would be obtained by a proper fit of the data, we determined them to be sufficient for the purpose of this test based on the injection-recovery analysis outlined in Section~\ref{sec:completeness}.} eccentricity is fixed to zero, and the} stellar parameters are fixed to those listed in the TIC and quadratic limb darkening coefficients are selected based on $T_\mathrm{eff}$ and $\log g$ using the values provided by \citet{claret2017limb}. Using the best-fit parameters, we calculate the Bayesian Information Criterion (BIC) for each model using the equation
\begin{equation}
    \mathrm{BIC} = k \log{n} - 2 \log{\mathcal{L}}
\end{equation}
where $k$ is the number of free parameters in the model, $n$ is the number of data points, and $\mathcal{L}$ is the maximized value of the Gaussian likelihood function
\begin{equation}
    \mathcal{L} = - 0.5 \sum_i \frac{(y_i - f(t_i))^2}{\sigma_{y,i}^2}
\end{equation}
where $t_i$ is the time of data point $i$, $y_i$ is the flux of data point $i$, $\sigma_{y,i}$ is the flux uncertainty at data point $i$, and $f(t)$ is the value of the model at time $t$.
We calculate the difference between the two as $\Delta \mathrm{BIC} = \mathrm{BIC_{sinusoid}} - \mathrm{BIC_{transit}}$, with the requirement that $\Delta \mathrm{BIC} \geq 50$ for a signal to pass the test. This test is visualized in Figure \ref{fig: sinusoid_test}.

In addition to the sinusoid test, we perform a test to evaluate the symmetry of the detected signal. Signals originating from astrophysical transits or eclipses should be symmetrical, to first order. However, signals of instrumental origin, such as those arising due to flux variations near \textit{TESS} momentum dumps, often have non-symmetric morphologies. To evaluate the symmetry of the signal, we phase fold the light curve using the event period and midpoint returned by the BLS, masking data more than two event durations from the midpoint. We then divide the data in half at the midpoint and flip one of the sequences in time space, such that the pre-midpoint and post-midpoint data can be directly compared. The two sequences are binned onto a common time grid and we calculate the reduced $\chi^2$ between the two data sets as
\begin{equation}
    \chi^2_{\rm red} = \frac{1}{n} \sum_i \frac{(y_{\mathrm{pre},i} - y_{\mathrm{post},i})^2}{\sigma_{y, \mathrm{pre},i}^2 + \sigma_{y, \mathrm{post},i}^2}
\end{equation}
where $n$ is the number of data points in each binned sequence, $y_{\mathrm{pre},i}$ and $y_{\mathrm{post},i}$ are the pre-midpoint and post-midpoint flux values at data point $i$, and $\sigma_{y, \mathrm{pre},i}$ and $\sigma_{y, \mathrm{post},i}$ are their uncertainties. We find (through the injection/recovery tests described in Section \ref{sec:completeness}) that planetary signals generally return $\chi^2_{\rm red} < 2$. We therefore require signals to have $\chi^2_{\rm red} < 2$ to pass the test. This test is visualized in Figure \ref{fig: symmetry_test}.

Signals that pass the two previously described tests have their approximate planet radii calculated using the signal depth estimated by the BLS and the radius of the star. Because this study is focused on smaller planets ($R_{\rm p} < 8 \, R_\oplus$), we eliminate all signals with approximate planet radii larger than $8 \, R_\oplus$.

\begin{figure*}[t!]
  \centering
    \includegraphics[width=1.0\textwidth]{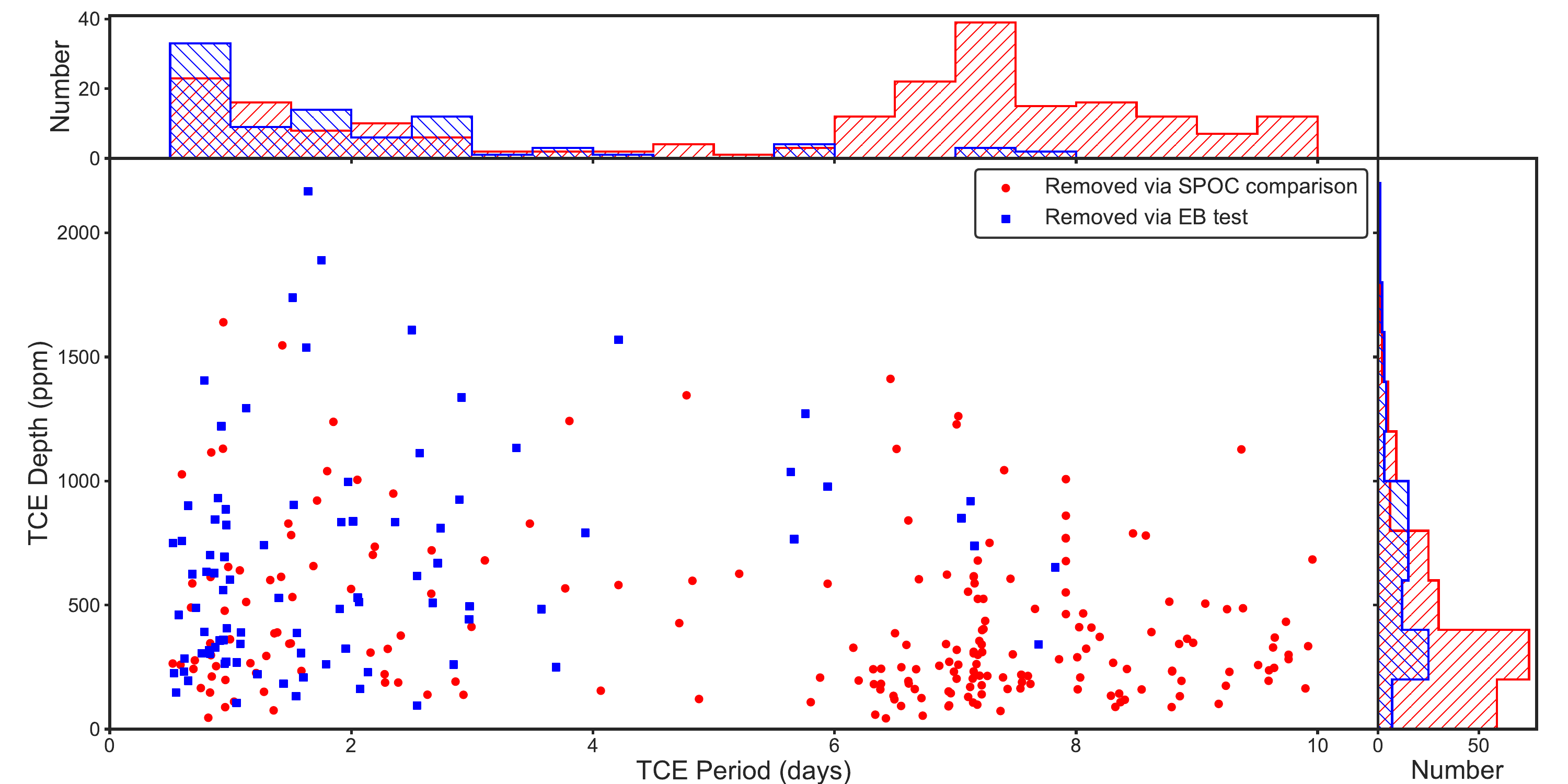}
    \caption{Periods and depths of TCEs detected by the automated pipeline described in Section \ref{sec:TCEpipeline}. The points are divided into two categories: red circles, which were determined to be false alarms or false positives in the SPOC comparison vetting test (Section \ref{sec:SPOC_comparison}), and blue squares, which were determined to be false positives in the secondary eclipse test (Section \ref{sec:secondary_eclipse}), centroid offset test (Section \ref{sec:centroid_offset}), ExoFOP cross-match (Section \ref{sec:exofop}), or transit shape analysis (Section \ref{sec:triceratops}). The period distribution of TCEs is somewhat bimodal, with peaks between 0.5 and 1 day and between  6 and 9 days. Nearly all of the TCEs with longer periods were ruled out as false alarms in the SPOC comparison test, which suggests that they are artifacts associated with the orbit of the \textit{TESS} spacecraft or its momentum dumps \citep[e.g.,][]{kunimoto2023falsealarm}. Most of the shorter period TCEs are consistent with being false positives caused by eclipsing binary stars or stellar variability.}
    \label{fig: TCE_scatterplot}
\end{figure*}

After running the full sample of stars through this automated pipeline, we recover 299 TCEs, which we list in Table \ref{tab:1}. In addition, we visualize the detected TCEs in Figure \ref{fig: TCE_scatterplot}. 14 of these TCEs correspond to previously reported \textit{TESS} Objects of Interest (TOIs; \citealt{guerrero2021tois}), which appear on the Exoplanet Follow-up Observing Program (ExoFOP) website (\dataset[10.26134/ExoFOP3]{https://exofop.ipac.caltech.edu/}).\footnote{\url{https://exofop.ipac.caltech.edu/tess/}} In order to identify reliable planet candidates, we vet these 299 TCEs manually using the steps outlined in the following subsection.

{ We acknowledge that five stars in our sample have TOIs with reported $P_{\rm orb} < 10$~days and $R_{\rm p} < 8 \, R_\oplus$ that were not detected by our pipeline: TOI-1037.01, TOI-1497.01, TOI-1570.01, TOI-4180.01, and TOI-5387.01. TOI-1037.01 was not detected by the BLS search due to light curve systematics. TOI-1497.01, TOI-1570.01, and TOI-5387.01 were detected by the BLS search but failed the $\Delta$BIC test. TOI-4180.01 was not detected by the BLS search, but this is unsurprising given that the \textit{TESS} pipelines only detected this TOI in short-cadence \textit{TESS} data and not in the relatively long-cadence full frame images that the QLP pipeline utilizes. Lastly, we note that TOI-1037.01, TOI-1497.01, and TOI-1570.01 are all listed as astrophysical false positives in ExoFOP. TOI-4180.01 and TOI-5387.01 are currently listed as planet candidates, although we note that spectroscopic observations have revealed evidence that both of these systems are false positives.\footnote{A high-resolution spectrum of TOI-4180 has revealed a binary companion (\url{https://exofop.ipac.caltech.edu/tess/edit_obsnotes.php?id=126737992}). Radial velocities of TOI-5387 have revealed a signal that is in-phase with the orbital ephemeris of the planet candidate, but is far too large in amplitude to correspond to a planet smaller than $8 \, R_\oplus$ (\url{https://exofop.ipac.caltech.edu/tess/edit_obsnotes.php?id=166086403}).} No other reliable TOIs were missed by our search.}

\subsection{Manual Vetting}

\subsubsection{Comparison to SPOC Light Curves}\label{sec:SPOC_comparison}

We begin our manual vetting of TCEs by searching for the signals detected in our flattened QLP light curves in the analogous light curves generated by the \textit{TESS} Science Processing Operations Center (SPOC) \citep{jenkins2016spoc}. The SPOC pipeline originally delivered processed light curves only for a select set of stars that had been observed with a short-cadence ``postage stamps,'' but the pipeline has since been utilized to generate light curves for a large fraction of stars with $T < 10$ that have been observed only in long-cadence full frame images. The SPOC pipeline is able to identify and remove systematics from light curves more robustly than the QLP pipeline, which was optimized for efficiency in order to search for transiting planets around a larger number of stars. We can therefore use the SPOC light curves to determine if the signals detected by our pipeline are false alarms. We eliminate TCEs as false alarms if any of the following criteria are true.
\begin{enumerate}[noitemsep]
    \item The phase-folded signal is clearly not transit-like in shape in the QLP light curve, the SPOC light curve, or both. Signals are visually assessed as non-transit-like based on features such as degree of in-transit flux variability, out-of-transit flux variability immediately preceding or following the event, and event symmetry.\footnote{We note that while our event symmetry test removes a majority of clearly asymmetric signals, a small fraction pass the test and are classified as TCEs.}
    \item The phase-folded signal that appears in the QLP light curve is absent in the SPOC light curve, indicating that the signal is a systematic that our pipeline failed to remove. A signal is determined to be absent in the SPOC data if there is no visible decrease in flux at the reported period and epoch with respect to the baseline.
    \item The phase-folded signal that appears in the QLP light curve is inverted in the SPOC light curve, indicating that the signal originates in the background or from a different nearby star. Inverted signals are visually identified as events that are transit-like in shape but increase in flux over the event window, rather than decrease in flux.
    \item The star has been observed in more than one sector but the signal is only apparent in one of them, indicating that the signal is likely a sector-dependent systematic. We assess whether a signal is apparent in a given sector by phase-folding the data from each sector separately and performing the three previously listed checks. We acknowledge that this test has the potential to remove true planets with orbital periods close to 10 days, especially if some of the transits occur close to intra-sector data gaps. We therefore assess longer period TCEs more conservatively. 
\end{enumerate}
Using this comparison technique, we rule out 211 TCEs as false alarms, leaving 88 for further scrutiny. { TOI-1354.01 and TOI-6260.01 are among the TCEs that are removed by this step, the former of which is designated a false positive on ExoFOP and the latter of which is currently designated a planet candidate, pending further follow-up observations. However, we note that TOI-6260.01 has a particularly V-shaped transit that is common for astrophysical false positives.} In Figure \ref{fig: TCE_scatterplot}, we display the TCEs ruled out by this vetting step in period-depth space. In general, these TCEs have relatively shallow depths and long periods, with a peak in the distribution near 7 days. This is consistent with the findings of \citet{kunimoto2023falsealarm}, who found an excess of false alarms near half of the 13.7-day \textit{TESS} orbital period. These longer periods are also consistent with the spacings between \textit{TESS} momentum dumps.

\subsubsection{Secondary Eclipse Search}\label{sec:secondary_eclipse}

\begin{figure*}[t!]
  \centering
    \includegraphics[width=1.0\textwidth]{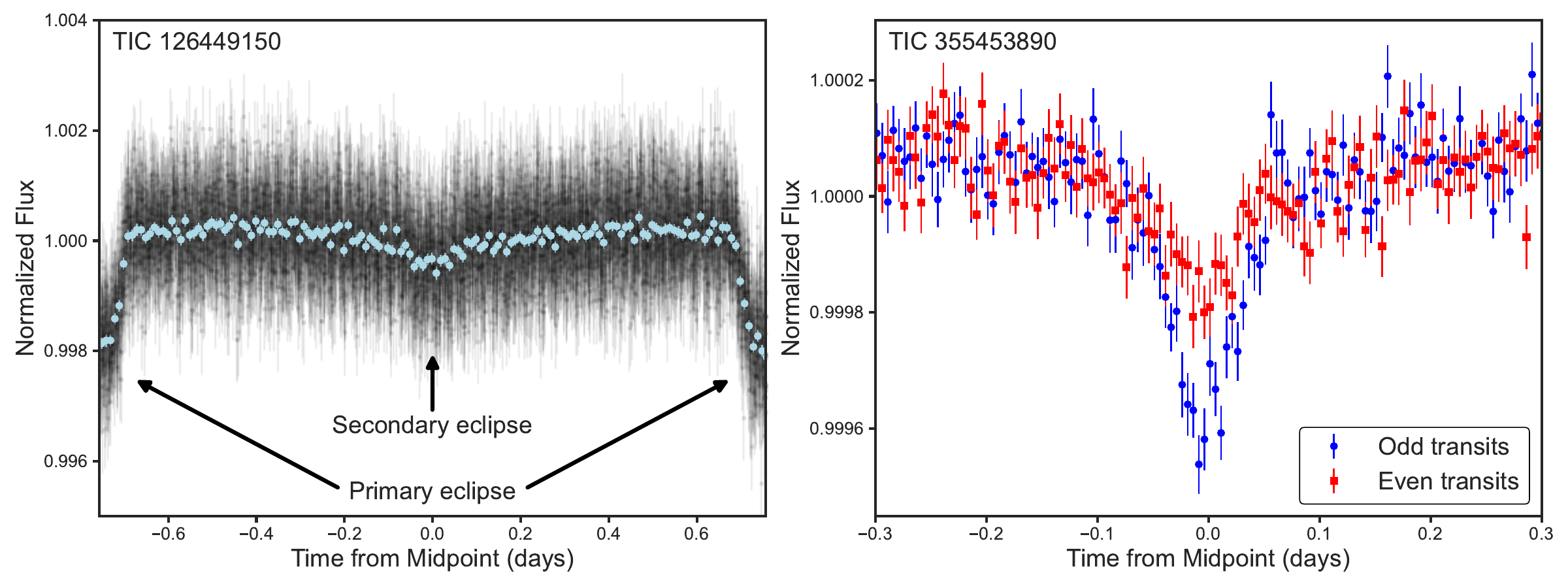}
    \caption{Visualization of signals that fail the manual secondary eclipse search. The left-side panel shows the signal detected in the light curve of TIC~126449150, which has a convincing secondary eclipse. The phase-folded light curve is shifted by half of the detected period so that the secondary eclipse appears at the center of the plot. The right-side panel shows the phase-folded odd-numbered events (blue circles) and even-numbered events (red squares) of the signal detected in the light curve of TIC~355453890, which exhibit significantly different depths.}
    \label{fig: secondary_eclipse_test}
\end{figure*}

Next, we visually inspect the light curves of the surviving TCEs for evidence of secondary eclipses and variations in transit depth between odd-numbered and even-numbered transits, both of which are strong evidence that the TCE is actually an eclipsing binary. We assess whether a secondary eclipse is present by masking out the events detected by the automated pipeline and rerunning the BLS periodogram on the light curve. If the periodogram detects a signal with ${\rm SDE} \geq 10$ and with a period within $0.1 \%$ of the originally detected period, we rule a secondary eclipse to be present. Odd-even transit depth variation is assessed visually, where a variation is ruled to be present if the depths of the phase-folded transits are inconsistent given the $1 \sigma$ error bars. Examples of TCEs that are eliminated by this inspection are shown in Figure \ref{fig: secondary_eclipse_test}. In total, 31 TCEs are found to have secondary eclipses and 13 TCEs are found to have odd-even transit depth variations, all of which we discard as false positives. { We note that TOI-1387.01, which is designated a false positive on ExoFOP, is among the discarded TCEs.} After this inspection, 44 TCEs remain.

\subsubsection{Centroid Offset Test}\label{sec:centroid_offset}

\begin{figure*}[t!]
  \centering
    \includegraphics[width=0.335\textwidth]{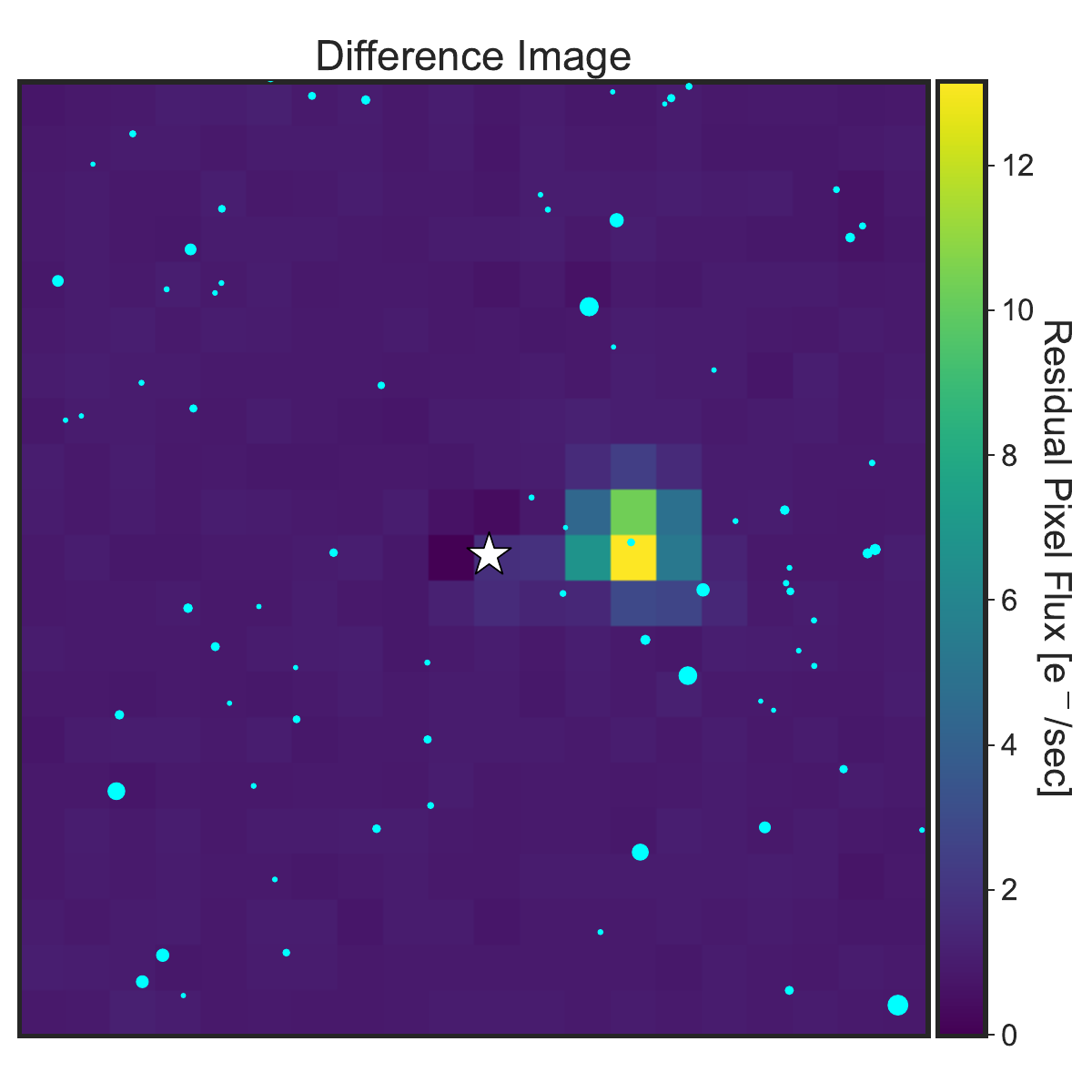}
    \includegraphics[width=0.335\textwidth]{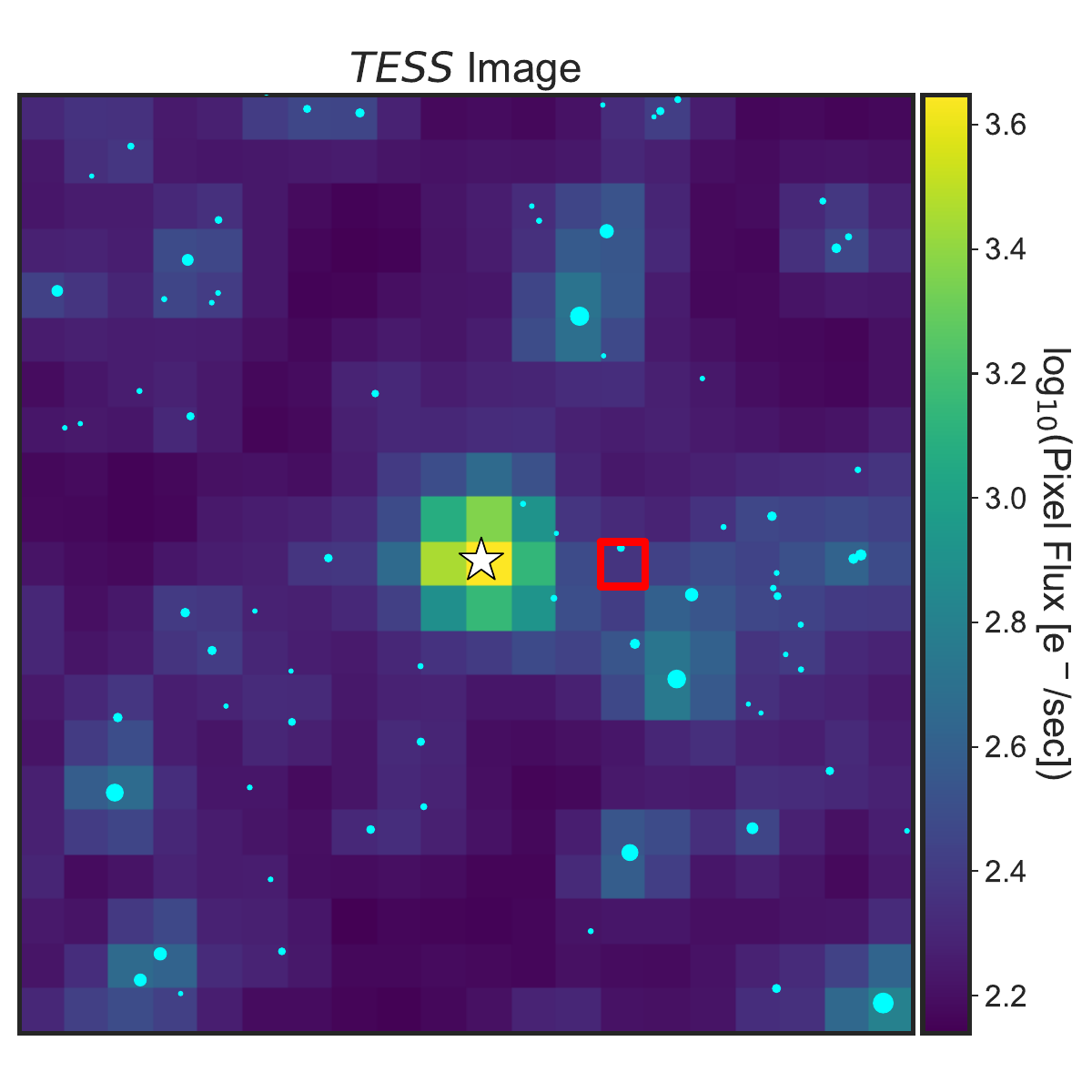}
    \includegraphics[width=0.31\textwidth]{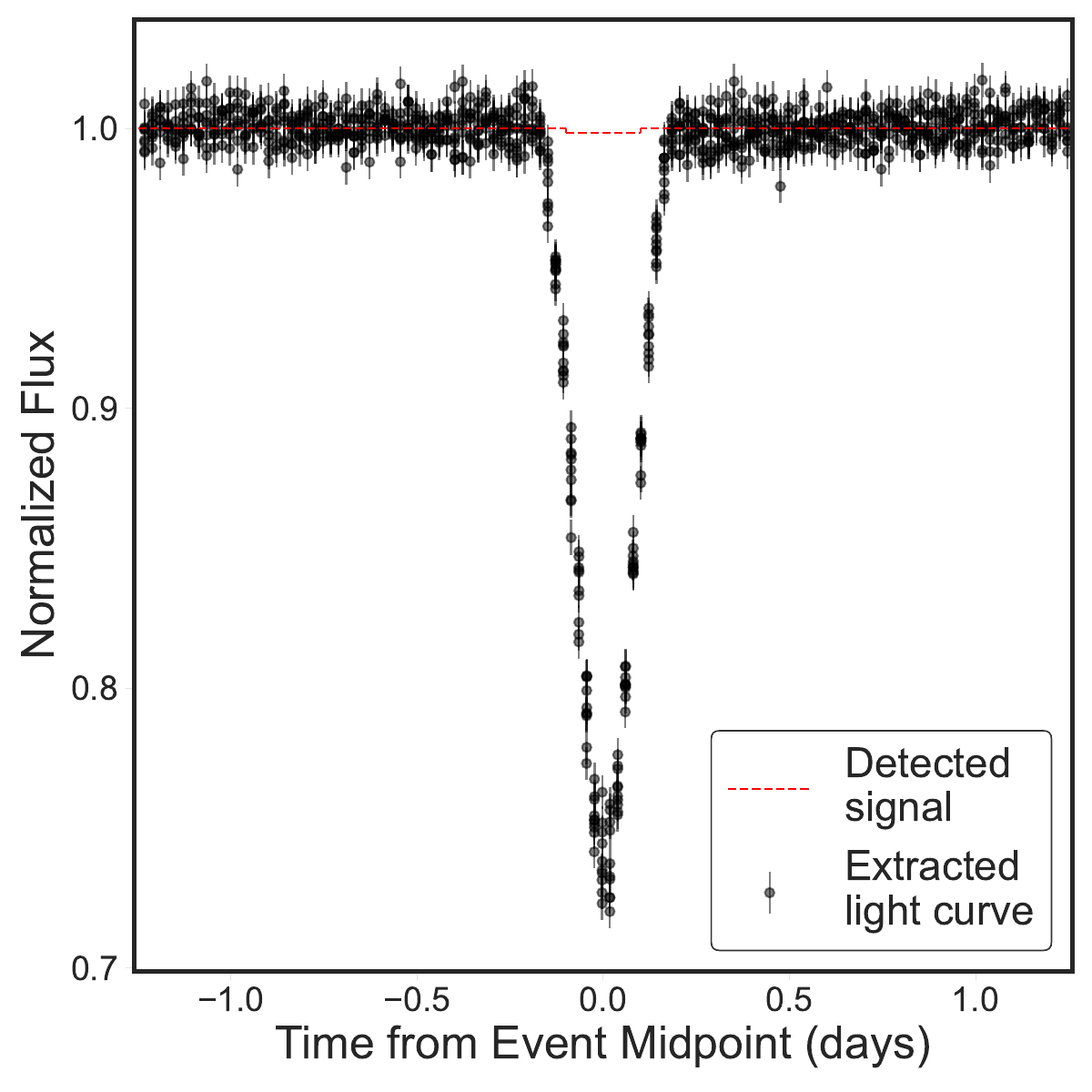}
    \caption{Visualization of the centroid offset test for TIC 177120452, which was found to have 2.5-day signal by the TCE detection pipeline. \emph{Left:} The difference image of the \textit{TESS} sector 7 data of the star, which shows a clear offset 3 pixels (approximately $1\arcmin$) to the east. TIC 177120452 is indicated by a white star and bright nearby stars are indicated by cyan circles, where size corresponds to relative brightness. \emph{Center:} The time-averaged \textit{TESS} sector 7 image of the target. The red square corresponds to the pixel with the peak residual flux in the difference image. \emph{Right:} The light curve extracted from the highlighted pixel, phase-folded to the ephemeris detected by the TCE pipeline. The red dashed line shows the signal predicted by the TCE pipeline for photometry extracted from TIC 177120452. The true source if the detected signal appears to be a pair of eclipsing binary stars originating from TIC 177014166 ($\Delta T = 4.71$, sep = $62"$).}
    \label{fig: diff_image_test}
\end{figure*}

It is possible to identify false positives in the form of nearby eclipsing binary stars at the pixel level using difference imaging \citep{bryson2013}. With this method, one subtracts the in-transit images from the out-of-transit images and measures the location of the difference image centroid, which should be close to the true source of the detected signal. For the TCEs that pass the vetting steps described previously, we calculate difference images using the publicly available \texttt{TESS-plots} tool \citep{kunimoto2022faint}. For each \textit{TESS} sector in which a TCE is observed, we download $21 \times 21$ pixel cutouts of the \textit{TESS} full frame images, separate them into in-transit frames and out-of-transit frames using the periods and transit times reported by the TCE detection pipeline (discarding data within one transit duration before transit ingress and following transit egress), and subtract the average of the two to generate a difference image. Next, we search for an offset between the pixel containing the maximum difference image signal and the pixel containing the target star. If an offset is present and is consistent in location across sectors, we extract a light curve from the pixel with the maximum difference image using \texttt{Lightkurve}. If a signal is recovered that matches the ephemeris of the TCE, but has a deeper transit than that detected by the QLP, we rule out the TCE as a nearby eclipsing binary. We demonstrate this process in Figure \ref{fig: diff_image_test}. We identify 34 false positives using this method, leaving 10 TCEs to be further analyzed. { TOI-957.01, TOI-1004.01, and TOI-2115.01, which are all designated false positives on ExoFOP, are among the TCEs eliminated by this analysis.}

\subsubsection{Cross Matching with ExoFOP}\label{sec:exofop}

For the remaining 10 TCEs, we consult ExoFOP to determine if any have been previously identified as TOIs and, if so, any have been determine to be astrophysical false positives or bona fide planets via follow-up observations. Eight of the TCEs have corresponding TOIs on ExoFOP: TIC 373424049 (TOI-742), TIC 54390047 (TOI-998), TIC 136274063 (TOI-1094), TIC 267489265 (TOI-1132), TIC 367102581 (TOI-1522), TIC 315350812 (TOI-4373), TIC 259230140 (TOI-4384), and TIC 350575997 (TOI-4386). According to publicly available information on ExoFOP, the first five of these TOIs have been determined to be false positives based on follow-up observations. The signal around TIC 259230140 has not yet been classified as a false positive or bona fide planet, but has a true orbital period (14.32~days) that is twice that detected by our pipeline (7.16~days). Because this planet candidate falls outside of the orbital period range considered by this study, we consider the signal to be a false alarm. The signals around TIC 315350812 and TIC 350575997 are, as of the writing of this paper, still classified as planet candidates on ExoFOP. We discuss these two candidates and the remaining two non-TOI TCEs (TIC 100588438 and TIC 120155231) in the following subsection.

\begin{figure*}[t!]
  \centering
    \includegraphics[width=1.0\textwidth]{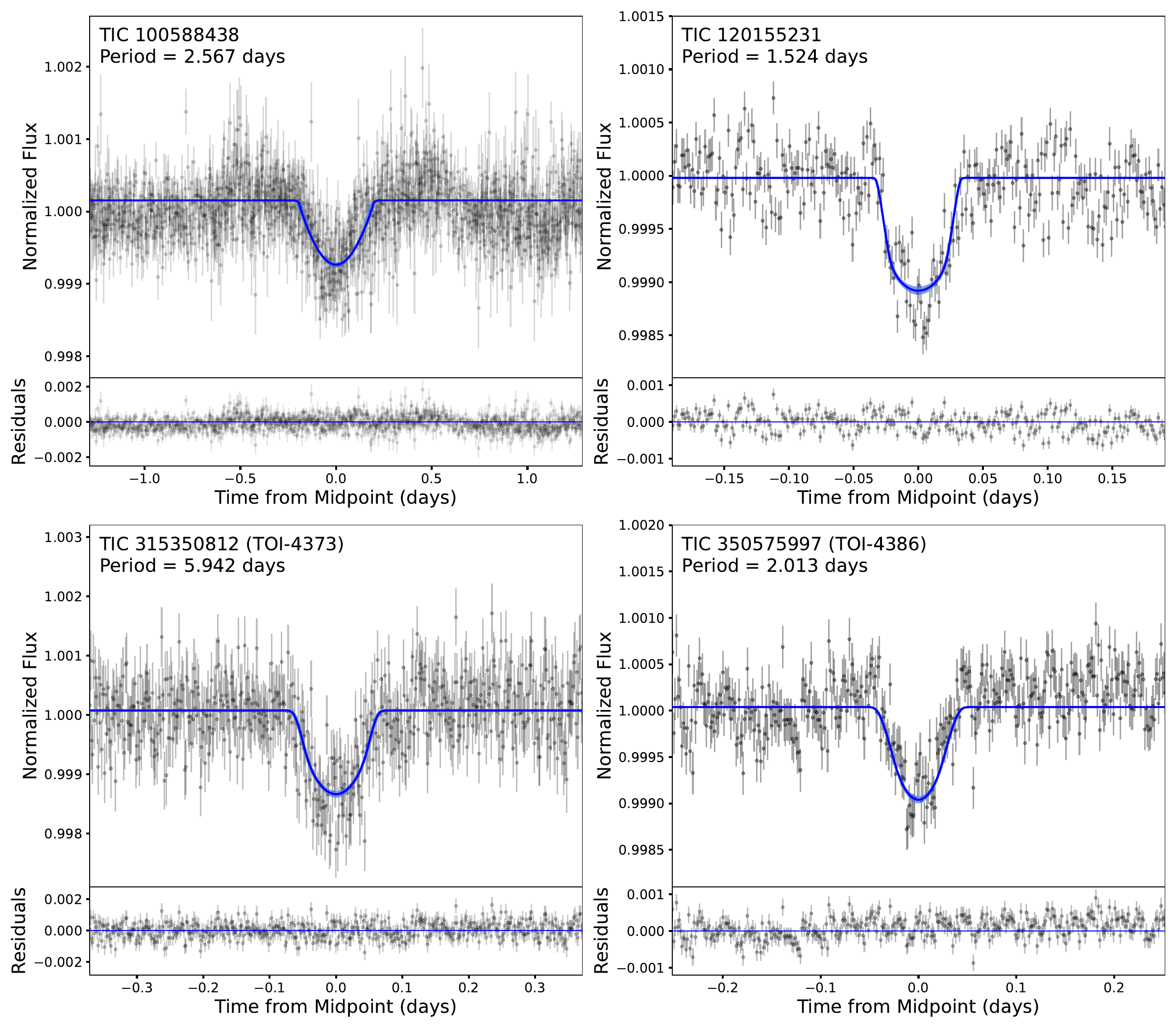}
    \caption{Phase-folded light curves of the four signals discussed in Section \ref{sec:triceratops}, binned to 2-minute intervals for clarity. Best-fit transit models with $1 \sigma$ uncertainties are shown in blue. After fitting the signal around TIC 120155231 with a transit model, we determined the planet candidate to be larger than $8 \, R_\oplus$. We determined the remaining signals to be false positives based on analyses with \texttt{TRICERATOPS}.}
    \label{fig: transits}
\end{figure*}

\subsubsection{Transit Shape Analysis}\label{sec:triceratops}

Lastly, we analyze the shapes of the transits of the remaining four TCEs using \texttt{TRICERATOPS}, which assesses the probability that each is an astrophysical false positive \citep{giacalone2020ascl, giacalone2021triceratops}. \texttt{TRICERATOPS} uses a Bayesian model to calculate the likelihood of a transit-like signal being caused by a number of astrophysical scenarios, including: (1) a planet transiting the target star, (2) a stellar companion eclipsing the target star, (3) a planet transiting an unresolved star that is gravitationally bound to or chance aligned with the target star, (4) an unresolved pair of eclipsing binary stars that are gravitationally bound to or chance aligned with the target star, (5) a planet transiting a known nearby star that is resolved from the target star, or (6) a pair of nearby eclipsing binary stars that are resolved from the target star. This calculation returns both a False Positive Probability (FPP; the overall probability that the signal is caused by an astrophysical false positive) and a Nearby False Positive Probability (NFPP; the probability that the signal is caused by an astrophysical false positive originating from a known nearby star). \citet{giacalone2021triceratops} showed that, in general, TCEs with ${\rm FPP} < 50\%$ and ${\rm NFPP} < 0.1\%$ are likely bona fide transiting planets, whereas those with higher values of FPP or NFPP are most often false positives.

Prior to running \texttt{TRICERATOPS} on each TCE, we refine our estimates of $T_0$ (the transit midpoint time) and $P_{\rm orb}$ by fitting the \textit{TESS} data to a light curve model using \texttt{exoplanet} \citep{foremanmackey2021}, which implements \texttt{pymc3} to estimate planet properties via Markov Chain Monte Carlo \citep{salvatier2016}. We initialize each model with the following priors: Gaussian priors on $M_\star$ and $R_\star$ based on the values in the TIC; Gaussian priors on $T_0$, the log of $P_{\rm orb}$, and the log of the planet-star radius ratio ($R_{\rm p}/R_\star$), all centered on the values estimated by our TCE detection pipeline; and a uniform prior on the transit impact parameter ($b$) that ranges from 0 to $1 + R_{\rm p}/R_\star$. We apply uniform priors to the quadratic limb darkening coefficients and assumed circular orbits for all fits. For each TCE, we run a 10 walker ensemble with 20,000 steps and discarded the first 10,000 steps as burn-in, ensuring convergence using the Gelman-Rubin statistic \citep{gelmanrubin1992}. We then use the best-fit $T_0$ and $P_{\rm orb}$ to create phase-folded light curves for the TCEs, which are used as input in \texttt{TRICERATOPS}. We display these transits in Figure \ref{fig: transits}.

\textit{TIC 100588438}: TIC 100588438 was observed in sectors 12, 39, and 66 with cadences of 30 minutes, 10 minutes, and 200 seconds, respectively. We use the SPOC-processed light curves from sectors 12 and 39 to calculate updated planet properties and analyze the signal with \texttt{TRICERATOPS}, binning the data to the longest cadence at which the star was observed (i.e., 30 minutes). Based on our light curve model fit of the data, we estimate the planet candidate to have radius of $R_{\rm p} = 5.29_{-0.31}^{+0.28} \, R_\oplus$. After running \texttt{TRICERATOPS} 20 times, we consistently found ${\rm FPP} > 99\%$ and ${\rm NFPP} > 99\%$, indicating that this TCE is most likely an astrophysical false positive. This star is located close to the galactic plane (galactic latitude = $0.19^\circ$) in a dense stellar field. \texttt{TRICERATOPS} estimates 27 nearby stars that could plausibly cause the observed signal. We also note that the transit of this planet candidate exhibits ``shoulder''-like features before ingress and after egress, which may be ellipsoidal variations from a nearby pair of eclipsing binary stars. Based on this analysis, we classify this TCE as a false positive.

\textit{TIC 120155231}: TIC 120155231 was observed in sectors 10, 36, 37, 63 and 64 with cadences of 30 minutes, 10 minutes, 10 minutes, 200 seconds, and 200 seconds, respectively. We found the SPOC-processed data in sectors 63 and 64 to contain the best-quality data, so we used only these two sectors to calculate updated planet properties and analyze the detected signal with \texttt{TRICERATOPS}. Although, we note that even this data suffers from high photometric scatter (especially out of transit, which can be seen in Figure \ref{fig: transits}). Based on our light curve model fit of the data, we estimate the planet candidate to have radius of $R_{\rm p} = 8.08_{-0.65}^{+0.64} \, R_\oplus$, with a high impact parameter of $b = 0.91 \pm 0.01$. Because this planet candidate has a best-fit size outside of the range we test in this study, we classify it a false positive. In addition, we note that the best-fit transit model for the planet candidate does not provide an excellent fit to the data, by visual inspection. In Figure \ref{fig: transits}, we see that the data is much more V-shaped than the transiting planet model is able to achieve, hinting that the signal may actually be an unresolved pair of background eclipsing binary stars.


\textit{TIC 315350812}: TIC 315350812 (TOI-4373) was observed in \textit{TESS} sectors 11, 38, and 65 with cadences of 30 minutes, 10 minutes, and 20 seconds, respectively. We use the SPOC-processed light curves to calculate updated planet properties and analyze the signal with \texttt{TRICERATOPS}, binning the data to the longest cadence at which the star was observed (i.e., 30 minutes). Based on our light curve model fit of the data, we estimate the planet candidate to have radius of $R_{\rm p} = 7.09_{-0.55}^{+0.56} \, R_\oplus$. After running \texttt{TRICERATOPS} 20 times, we consistently found ${\rm FPP} > 92\%$ and ${\rm NFPP} > 27\%$, indicating that this TCE is most likely an astrophysical false positive. The high NFPP is partially a result of the star residing in a region of the galactic plane (galactic latitude = $-2.72^\circ$) that has a very high stellar density. \texttt{TRICERATOPS} estimates 149 nearby stars that could plausibly cause the observed signal. Based on this analysis, we classify this TCE as a false positive.

\textit{TIC 350575997}: TIC 350575997 (TOI-4386) was observed in \textit{TESS} sectors 12, 39, and 66 with cadences of 30 minutes, 120 seconds, and 20 seconds, respectively. We use the SPOC-processed light curves to calculate updated planet properties and analyze the signal with \texttt{TRICERATOPS}, binning the data to the longest cadence at which the star was observed (i.e., 30 minutes). Based on our light curve model fit of the data, we estimate the planet candidate to have radius of $R_{\rm p} = 7.58_{-0.81}^{+0.80} \, R_\oplus$. After running \texttt{TRICERATOPS} 20 times, we consistently found ${\rm FPP} > 99\%$ and ${\rm NFPP} > 22\%$, indicating that this TCE is most likely an astrophysical false positive. Like the previously discussed TCE hosts, this star resides close to the galactic plane (galactic latitude = $-9.95^\circ$) in a dense stellar field. \texttt{TRICERATOPS} estimates 28 nearby stars that could plausibly cause the observed signal. Based on this analysis, we classify this TCE as a false positive.

Based on the analyses described here and in the previous subsections, we determine that all of the TCEs detected by our pipeline for our stellar sample are likely to be false alarms or false positives. We therefore report no bona fide planets with $P_{\rm orb} < 10$ days and $1 \, R_\oplus < R_{\rm p} < 8 \, R_\oplus$ orbiting the { 20,257} A-type stars searched in this study. We quantify this non-detection into an upper limit on the occurrence rate of such planets in the following sections.

\subsection{ Caveats}

{ We acknowledge that our analysis neglects some astrophysical phenomena that impact the detectability of planetary transits. Here, we call attention to these caveats and encourage others to include them in future studies.

A-type stars often display variability in the form of $\delta$ Scuti pulsations. Because these pulsations have similar durations and flux variations as planetary transits, they can inhibit the detection of transiting planets. While it is possible to remove these pulsations and improve the sensitivity of a planet-detection pipeline \citep[e.g.,][]{ahlers2019deltascuti, hey2021deltascuti}, the task is non-trivial. Rather than removing this variability from our light curves, we simply incorporate the loss of sensitivity into our occurrence rate calculations (see Section~\ref{sec:completeness}).

A-type stars also often rotate so rapidly that their shapes deviate significantly from spherical. This distortion results in a lower temperature near the equator and a higher temperature near the poles, with a corresponding change in brightness across the stellar surface. This gravity darkening can affect the shapes, depths, and durations of planetary transits depending on the orientation of the orbit. For instance, \citet{barnes2009gravdark} predicted that a planet transiting Altair could have a depth that varies by $\sim 50 \%$ for different orbital orientations. Such variations would certainly impact the ability of a transit-detection algorithm to find planets, and would likely impact the estimation of the physical properties of transiting planets. However, we note that the Altair scenario is an extreme example. Altair rotates extremely rapidly, with a $v \sin{i} > 200$~km/s, a rotation period of less than 9 hours, and an oblateness (defined as $1 - R_{\rm pole}/R_{\rm eq}$, where $R_{\rm pole}$ is the polar radius and $R_{\rm eq}$ is the equatorial radius) of $\sim 0.2$ \citep{monnier2007altair}. However, the average A-type star in our sample likely has a much lower rotational speed and oblateness, resulting in less prominent gravity darkening effects. For instance, \citet{zorec2012Astar} found the average A-type star with $M_\star < 2.6 \, M_\odot$ to have a $v \sin{i}$ between 100 and 150~km/s. Gravity darkened transit observations of A-type stars with these more moderate rotation rates have been found to exhibit relatively minor variations to transit shape and have yielded oblateness measurements below 0.1 \citep[e.g.,][]{zhou2019, ahlers2020gravdarka, ahlers2020gravdarkb, hooton2022gravdark}. Consequently, we expect the impacts of gravity darkening on our occurrence rate calculation to be relatively minor. Nonetheless, we encourage future studies to take this effect into account for more robust calculations.}

\section{Pipeline Completeness}\label{sec:completeness}

\begin{figure*}[hbtp]
  \centering
    \includegraphics[width=1.0\textwidth]{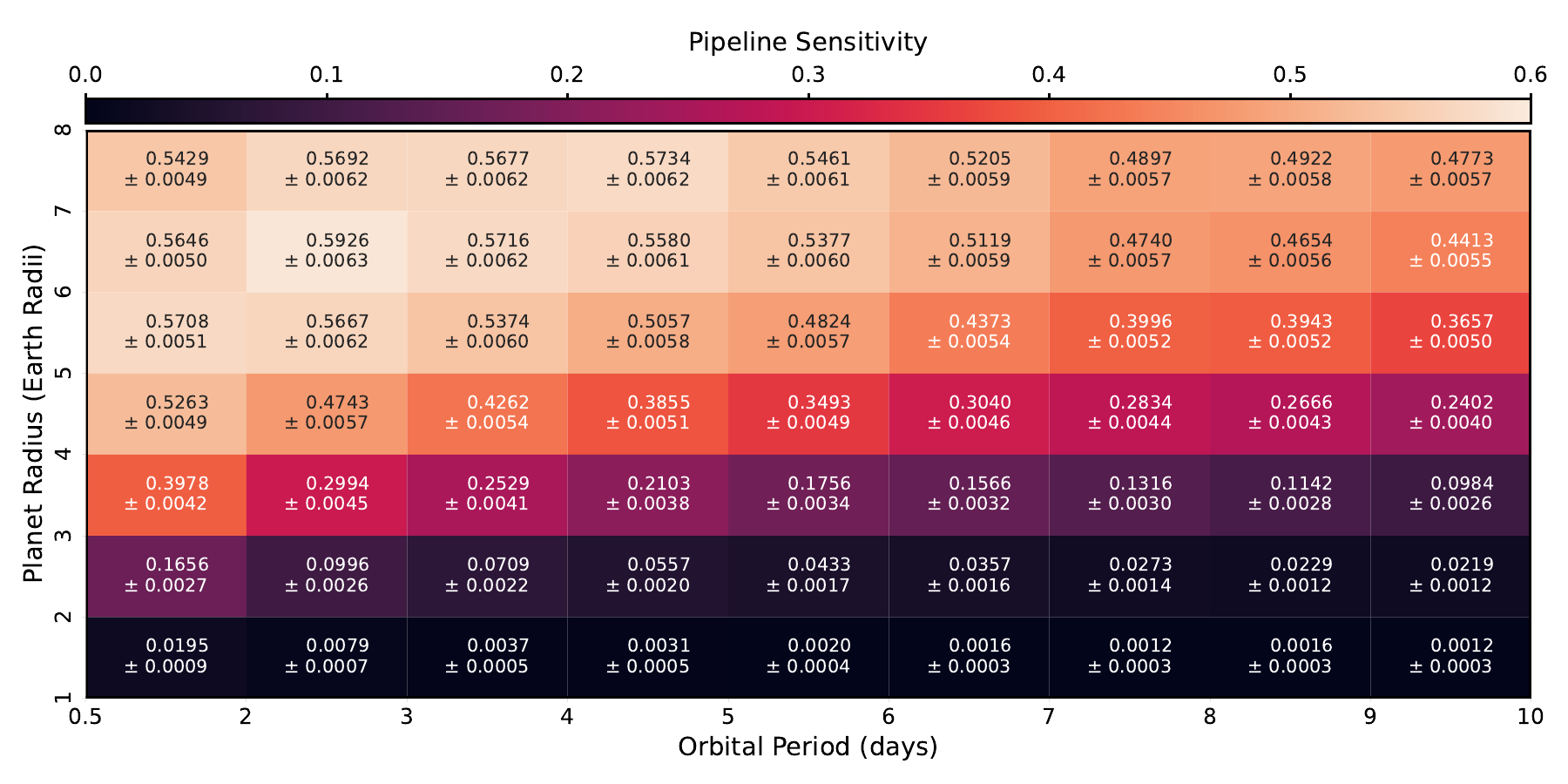}
    \includegraphics[width=1.0\textwidth]{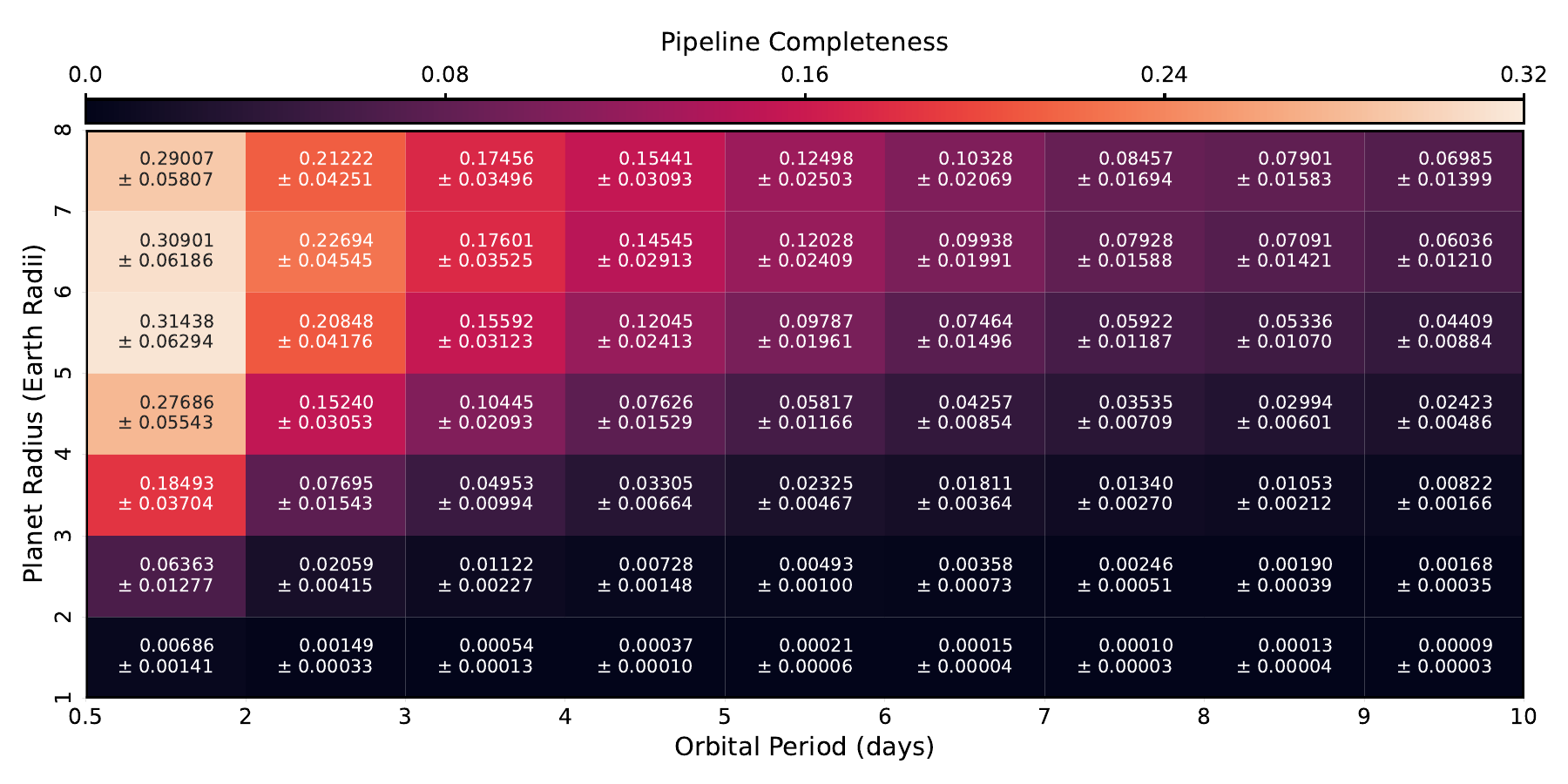}
    \caption{Pipeline sensitivity (top) and completeness (bottom) for different planet radii and orbital periods for our sample of { 20,257} stars. The color of each cell corresponds to the value displayed within. { Uncertainties in sensitivity are Poisson errors associated with injected planets that were recovered in each cell. Uncertainties in completeness include both Poisson errors and an additional $20 \%$ systematic uncertainty to account for the precision of stellar mass and radius estimates.} Non-monotonicity near large $R_{\rm p}$ and low $P_{\rm orb}$ are a product of the automated vetting tests described in Section \ref{sec:TCEpipeline}. The overall completeness of the pipeline (i.e., integrated over the full $P_{\rm orb}$ and $R_{\rm p}$ range) is $7.2\%$. The completenesses for planets with radii of $4-8 \, R_\oplus$, $2-4 \, R_\oplus$, and $1-2 \, R_\oplus$ are $13.1 \%$, $3.2 \%$, and $0.2 \%$, respectively.}
    \label{fig: sensitivity_completeness}
\end{figure*}

Calculating the occurrence rate of planets around our sample of stars requires knowledge of the completeness of the transit detection pipeline. Completeness accounts for two factors: the probability of the pipeline detecting a planet that is transiting (also known as the pipeline sensitivity) and the geometric probability of a planet being in an orientation such that it transits along our line of sight. We outline the procedure for calculating these probabilities here.

The pipeline sensitivity is determined using injection/recovery tests. These tests involve injecting artificial transit signals into real data and quantifying the ability of the pipeline to detect them. Following the precedent established by previous occurrence rate studies \citep[e.g.,][]{dressing2013occurrence, dressing2015occurrence, petigura2013, bryant2023, gan2023rate}, we calculate pipeline sensitivity with the following steps. The results of this pipeline sensitivity calculation are shown in Figure \ref{fig: sensitivity_completeness}.
\begin{enumerate}[noitemsep]
    \item We download the raw QLP light curves for all { 20,257} stars defined in Section \ref{sec:sample}, discarding the 300 stars with signals detected by our automated TCE detection pipeline. These stars are the basis of the injection-recovery tests. For each star, we simulate 50 instances of transiting planets, producing approximately 1,000,000 instances in total.
    
    \item  For each instance $k$, we draw $R_\mathrm{p}$ and $P_\mathrm{orb}$ from uniform distributions linearly spaced over the intervals ($1 \, R_\oplus$, $8 \, R_\oplus$) and (0.5 days, 10 days), respectively. The midpoint time of the first transit ($T_0$) is randomly selected between the starting time of the light curve ($t_\mathrm{min}$) and $t_\mathrm{min} + P_\mathrm{orb}$. The transit impact parameter ($b$) of the simulated planet is drawn from from a uniform distribution over the interval [0, 0.9] and the eccentricity is set to zero.\footnote{We note that this upper limit on $b$ leads to a slight overestimate in completeness, but we expect this difference to have a negligible impact on our final occurrence rate estimates.} The quadratic limb-darkening coefficients of the star are determined based on the $T_\mathrm{eff}$ and $\log g$ of the star using the values reported by \citep{claret2017limb}, assuming Solar metallicity. Transits are then injected into the raw light curves using \texttt{batman} \citep{kreidberg2015batman}, with the data supersampled to account for the different observation cadences used in different \textit{TESS} sectors.

    
    \item For each injected planet, we flatten the light curve using the procedure outlined in Section \ref{sec:lightcurves} and run the automated TCE detection pipeline described in Section \ref{sec:TCEpipeline}. If an injected planet is detected with a $\mathrm{SDE} \geq 10$, { a S/N $>$ 1}, a $\Delta \mathrm{BIC} \geq 50$, a $\chi_\mathrm{red}^2 < 2$, and a BLS $R_\mathrm{p} < 8 \, R_\oplus$, { we more closely examine if the parameters predicted by the BLS periodogram match those of the injected planet}. If the BLS $P_\mathrm{orb}$ is within one transit duration of the injected $P_\mathrm{orb}$ or its four nearest harmonics (i.e., $1/2\times$, $2/3\times$, $3/2\times$, or $2\times$ the injected $P_\mathrm{orb}$), the BLS $T_0$ is within one transit duration of any injected $T_0$, { and the BLS $R_{\rm p}$ is within a factor of two of the injected $R_{\rm p}$, the injected planet} is considered recovered and we set $r_{k} = 1$. If the injected planet fails any of the automated TCE detection pipeline tests or is detected at the incorrect $P_\mathrm{orb}$ or $T_0$, we set $r_{k} = 0$.
    
    \item After all instances are run, we generate a sensitivity map by defining a grid in $P_\mathrm{orb}$--$R_\mathrm{p}$ space, where each cell is indexed as ($i,j$). We calculate the fraction of recovered planets in a given grid cell using the equation 
    \begin{equation}\label{eq: sensitivity}
        \mathcal{R}_{i,j} = \sum_{k \in K_{i,j}} r_{k} \bigg/ \sum_{k \in K_{i,j}} 1
    \end{equation}
    where $K_{i,j}$ is the set of instances that fall within the cell.
\end{enumerate}

The full pipeline completeness is determined by also taking into account the geometric transit probability, which, for a given injection instance $k$, can be approximated as
\begin{equation}
    p_{\mathrm{geo}, k} \approx \frac{R_{\star, k}}{a_{k}} = R_{\star, k} \left( \frac{G M_{\star, k} P_{\mathrm{orb}, k}^2}{4 \pi^2} \right)^{-1/3} ,
\end{equation}
which assumes circular orbits and that the radii and masses of the planets are negligible compared to those of their stars. We take this factor into account by treating it as a weight in Equation \ref{eq: sensitivity}, such that the completeness within a given grid cell ($\mathcal{C}_{i,j}$) is given by the equation
\begin{equation}
    \mathcal{C}_{i,j} = \sum_{k \in K_{i,j}} r_{k} p_{\mathrm{geo}, k} \bigg/ \sum_{k \in K_{i,j}} 1.
\end{equation}
The results of these sensitivity and completeness calculations are shown in Figure \ref{fig: sensitivity_completeness}. { Pipeline sensitivity uncertainties are calculated as Poisson errors associated with the number of injected planets that are recovered in each cell. Pipeline completeness uncertainties are a combination of Poisson errors and a $20\%$ systematic uncertainty on the geometric transit probability, which stems from the precision with which $M_\star$ and $R_\star$ are known. We note that $>99\%$ of the stars in our sample have fractional $M_\star$ and $R_\star$ uncertainties less than $20\%$ in the TIC; our choice of systematic uncertainty is therefore conservative.} The overall completeness of our pipeline, when integrated across all planet radius and orbital period bins, is $7.2 \pm 1.4 \%$. When split into different planet radius ranges, we achieve a completeness of $13.2 \pm 2.6\%$ for planets $4 - 8 \, R_\oplus$ in size, $3.1 \pm 0.6 \%$ for planets $2 - 4 \, R_\oplus$ in size, and $0.14 \pm 0.03 \%$ for planets $1 - 2 \, R_\oplus$ in size. This demonstrates that we are primarily sensitive to planets larger than $2 \, R_\oplus$, a testament to how difficult the transits of Earth-size planets are to detect around these large stars.

\subsection{Completeness for Early vs. Late A-type Stars}

Because our sample of A-type stars spans a wide range of properties, there is a possibility that our completeness is high for only a small subset of systems. To test this, we split the sample into early ($T_{\rm eff} \geq 8750$ K) and late ($T_{\rm eff} < 8750$ K) samples and repeat our completeness calculation. For the early stars, which are generally slightly larger, we find completenesses of $16.1 \pm 3.2 \%$, $4.2 \pm 0.8 \%$, and $0.21 \pm 0.04 \%$ for planets with radii $4 - 8 \, R_\oplus$, $2 - 4 \, R_\oplus$, and $1 - 2 \, R_\oplus$, respectively. For the late stars, we find completenesses of $12.3 \pm 2.5 \%$, $2.8 \pm 0.6 \%$, and $0.11 \pm 0.02 \%$ for planets with radii $4 - 8 \, R_\oplus$, $2 - 4 \, R_\oplus$, and $1 - 2 \, R_\oplus$, respectively. The higher completeness for hotter stars is likely a consequence of they being brighter, which allows for better photometric precisions. The completenesses for the early and late samples agree within a factor of two, with the largest discrepancy existing for the smallest planets to which we already had a low sensitivity with the unified sample. Thus, our results are not strongly sensitive to the properties of the underlying stars.

\section{Occurrence Rate}\label{sec:occurrence}

We calculate the occurrence rate following the procedure outlined in studies like \citet{howard2012planet}, \citet{petigura2018}, \citet{zhou2019}, and \citet {gan2023rate}. The effective number of stars searched for planets in a given grid cell, after correcting for search completeness, is calculated with the equation (dropping the index subscripts so as to consider any arbitrarily defined cell)
\begin{equation}
    n_\mathrm{trial} = n_\star \mathcal{C},
\end{equation}
where $n_\star$ is the total number of stars in the sample. We also define the number of observed planets in a given grid cell as
\begin{equation}
    n_\mathrm{obs} = \sum_{l=1}^{n_\mathrm{p}} (1 - f_{\mathrm{FP}, l}),
\end{equation}
where $n_\mathrm{p}$ is the total number of planet candidates in the cell and $f_{\mathrm{FP}, l}$ is the false positive rate, which is set to 1 for known false positives and to 0 for confirmed planets. Finally, the occurrence rate in a given grid cell is given by
\begin{equation}
    f_\mathrm{cell} = n_\mathrm{obs} / n_\mathrm{trial}.
\end{equation}

Because all TCEs detected in our sample are labeled false positives, we calculate the upper limit on the occurrence rate by assuming that the probability of detecting $X$ planets in a given grid cell follows a binomial distribution:
\begin{equation}
    P(n_\mathrm{trial}, X, f_\mathrm{cell}) = N f_\mathrm{cell}^X (1 - f_\mathrm{cell})^{n_\mathrm{trial} - X},
\end{equation}
where
\begin{equation}
    N = \frac{\Gamma (n_\mathrm{trial} + 1)}{\Gamma (X + 1) \Gamma (n_\mathrm{trial} - X + 1)}.
\end{equation}
Thus, for a null detection in a given grid cell, the upper limit on planet occurrence rate to a confidence interval $\mathrm{CI}$ is calculated with
\begin{equation}
    \int_{0}^{f_\mathrm{cell, upper}} (n_\mathrm{trial} + 1) P(n_\mathrm{trial}, 0, f_\mathrm{cell}) df_\mathrm{cell} = \mathrm{CI},
\end{equation}
which is solved to find
\begin{equation}
    f_\mathrm{cell, upper} = 1 - (1 - \mathrm{CI})^{1/(n_\mathrm{trial} + 1)}.
\end{equation}

For ease of comparison with previous occurrence rate studies and to distinguish the effects of various formation and evolution mechanisms, we divide planets into three size regimes: sub-Saturns ($4 \, R_\oplus < R_\mathrm{p} < 8 \, R_\oplus$), sub-Neptunes ($2 \, R_\oplus < R_\mathrm{p} < 4 \, R_\oplus$), and super-Earths ($1 \, R_\oplus < R_\mathrm{p} < 2 \, R_\oplus$).\footnote{In the literature, ``super-Earths'' are generally defined as being smaller than $1.6 \, R_\oplus$, which corresponds roughly to the radius at which planets transition from having thin atmospheres to volatile-rich atmospheres \citep[e.g.,][]{rogers2015most}. However, most previous occurrence rate calculations did not draw a bins according to this definition. Our use of the term is therefore purely for the sake of convenience.} Integrated over the full range of orbital periods tested, we obtain $3 \sigma$ ($2\sigma$) upper limits of $2.2 \pm 0.4$ ($1.1 \pm 0.2$) sub-Saturns per 1000 A-type stars, $9.1 \pm 1.8$ ($4.7 \pm 0.9$) sub-Neptunes per 1000 A-type stars, and $186 \pm 34$ ($101 \pm 19$) super-Earths per 1000 A-type stars. In Figure \ref{fig: occurrence_rates}, we compare these occurrence rates to those calculated for FGKM-type stars by \citet{dressing2013occurrence, dressing2015occurrence}, \citet{mulders2015stellar}, and \citet{kunimoto2020} using {\it Kepler} data. We discuss these results further in the following section. 

\section{Results}

\subsection{The Occurrence Rates of Small Planets as a Function of Stellar Effective Temperature}

\begin{figure*}[t!]
  \centering
    \includegraphics[width=1.0\textwidth]{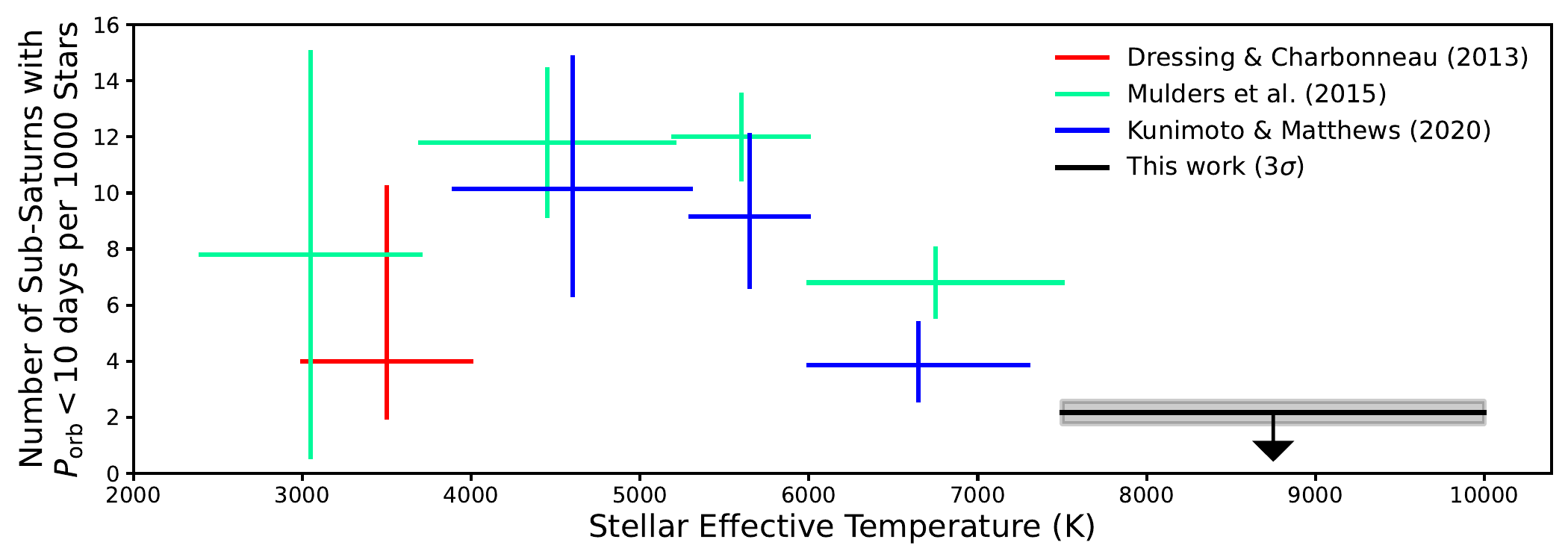}
    \includegraphics[width=1.0\textwidth]{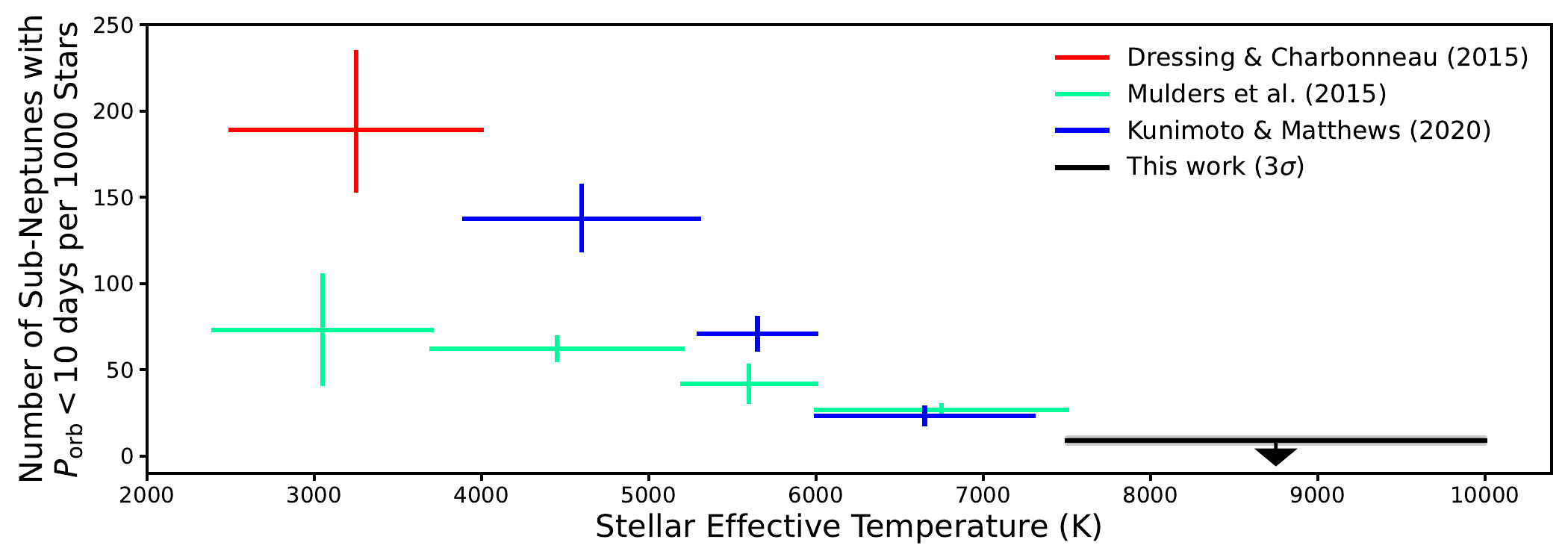}
    \includegraphics[width=1.0\textwidth]{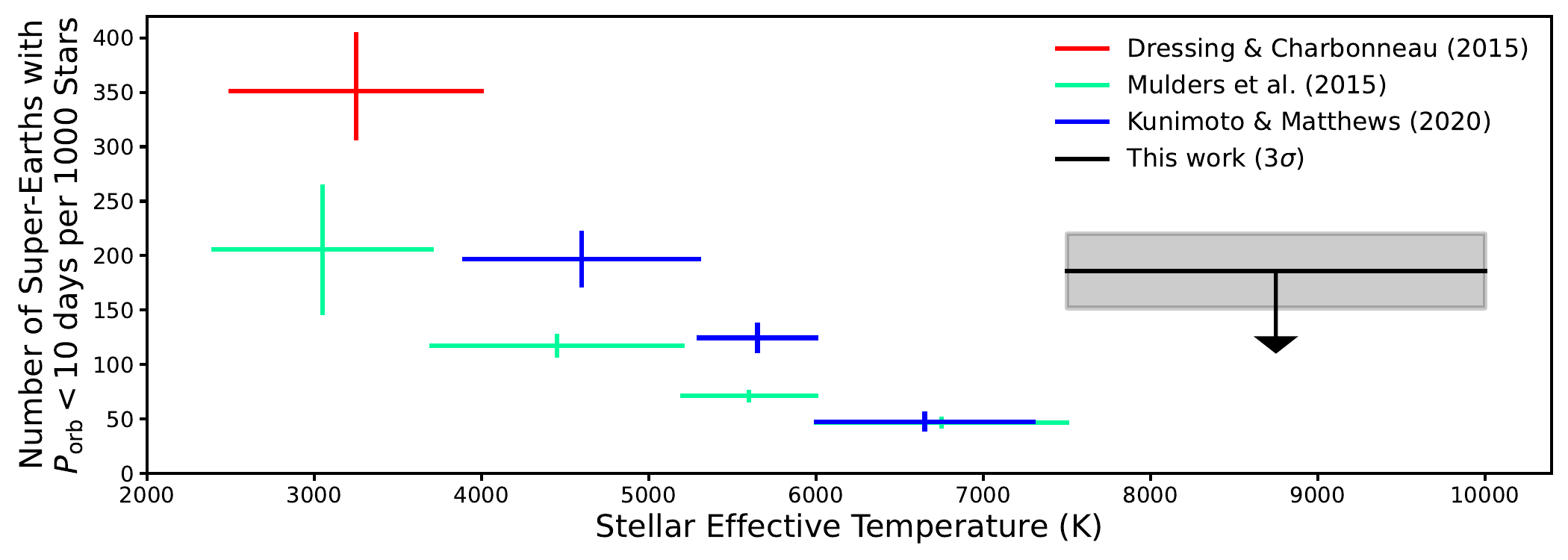}
    \caption{Calculated $3 \sigma$ upper limits on the occurrence rates of sub-Saturns ($4 \, R_\oplus < R_\mathrm{p} < 8 \, R_\oplus$; top), sub-Neptunes ($2 \, R_\oplus < R_\mathrm{p} < 4 \, R_\oplus$; middle), and super-Earths ($1 \, R_\oplus < R_\mathrm{p} < 2 \, R_\oplus$; bottom) with $P_{\rm orb} < 10$ days around A-type stars from {\it TESS} data (black). { Error bars on the  $3 \sigma$ upper limits are shown as gray shading.} Shown for comparison are the corresponding occurrence rates from \citet{dressing2013occurrence} ($P_{\rm orb} = 0.68 - 10$ days; red), \citet{dressing2015occurrence} ($P_{\rm orb} = 0.5 - 10$ days; red), \citet{mulders2015stellar} ($P_{\rm orb} = 0.4 - 10$ days; light green), and \citet{kunimoto2020} ($P_{\rm orb} = 0.78 - 12.5$ days; blue) for FGKM-type stars, which are calculated from {\it Kepler} data. Our results indicate a continual decrease in the occurrence rate of sub-Saturns and sub-Neptunes between G-type and A-type stars, and rules out the possibility of an M-dwarf-like super-Earth occurrence rate around A-type stars at a high level of confidence. These findings suggest that small planets either cannot survive at, form at, or migrate to short separations around hot stars.}
    \label{fig: occurrence_rates}
\end{figure*}

We place our upper limits into context by comparing them to previously calculated occurrence rates for planets of the same size around FGKM-type stars. Here, we specifically focus on the results from \citet{dressing2013occurrence, dressing2015occurrence} and \citet{kunimoto2020}. We judge the results from these studies to be among the most reliable due to their robust methodologies of occurrence rate calculation, which include injection/recovery tests similar to those described above to characterize the search and vetting completeness of their pipelines.\footnote{\citet{mulders2015stellar} likely overestimates the completeness of the \textit{Kepler} pipeline, resulting in lower occurrence rates than \citet{dressing2015occurrence} and \citet{kunimoto2020} in Figure \ref{fig: occurrence_rates}.} In addition, these studies calculated occurrence rates in similar grids as those we use. \citet{dressing2013occurrence} reported occurrence rates for planets with orbital periods between 0.68 and 10 days, \citet{dressing2015occurrence} reported occurrence rates for planets with orbital periods between 0.5 and 10 days, and \citet{kunimoto2020} reported occurrence rates for planets with orbital periods between 0.78 and 12.5 days. All three of these studies reported occurrence rates in the same planet radius bins we use in our calculation. These grids allow for a nearly one to one comparison of the upper limits we report.

For sub-Saturns, \citet{kunimoto2020} found occurrence rates of $3.86^{+1.49}_{-1.24}$, $9.17^{+2.88}_{-2.50}$, and $10.14^{+4.68}_{-3.80}$ planets per 1000 F-type, G-type, and K-type stars, respectively. \citet{dressing2013occurrence} found an occurrence rate of $4^{+6}_{-2}$ planets per 1000 M dwarfs. With our upper limit of 2.17 planets per 1000 A-type stars, this suggests that sub-Saturns may be rarer around A-type stars than FM-type stars and more than $3\times$ as rare around A-type stars than GK-type stars. Put in another way: based on the completeness of our pipeline to sub-Saturns with $P_{\rm orb} < 10$ days ($13.2 \%$), we estimate having detected $10^{+4}_{-3}$, $25^{+8}_{-7}$, $27^{+13}_{-10}$, and $11^{+16}_{-5}$ sub-Saturns around our sample of A-type stars if they had occurrence rates equivalent to those around FGKM-type stars.

\begin{figure*}[t!]
  \centering
    \includegraphics[width=0.339\textwidth]{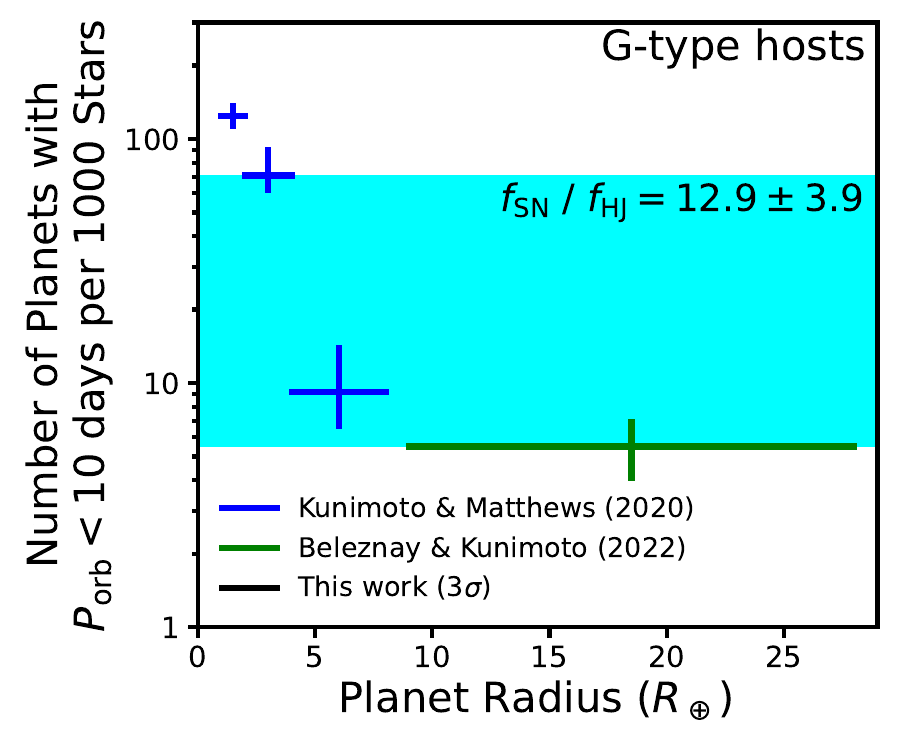}
    \includegraphics[width=0.31\textwidth]{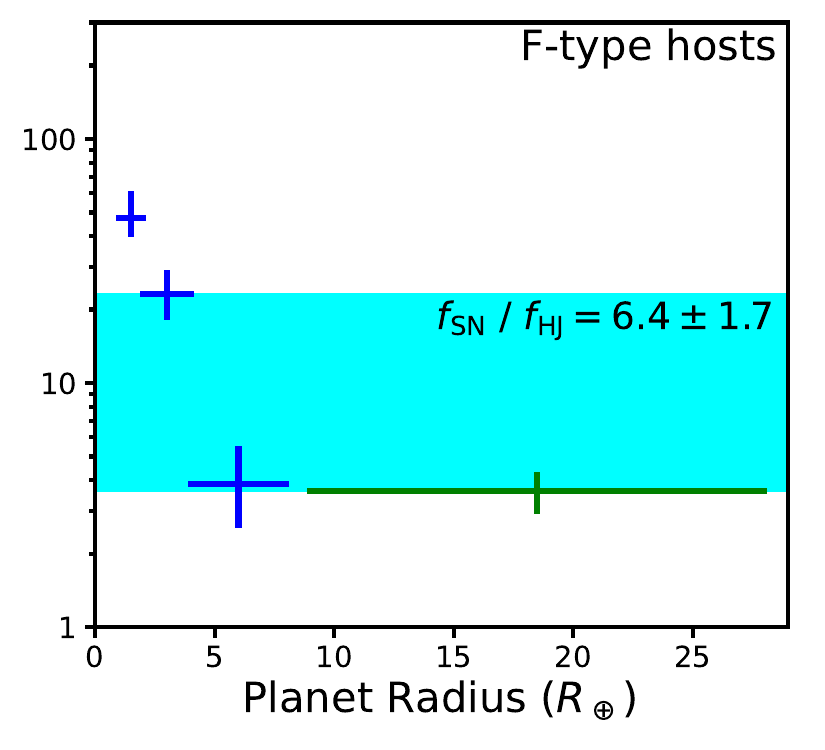}
    \includegraphics[width=0.31\textwidth]{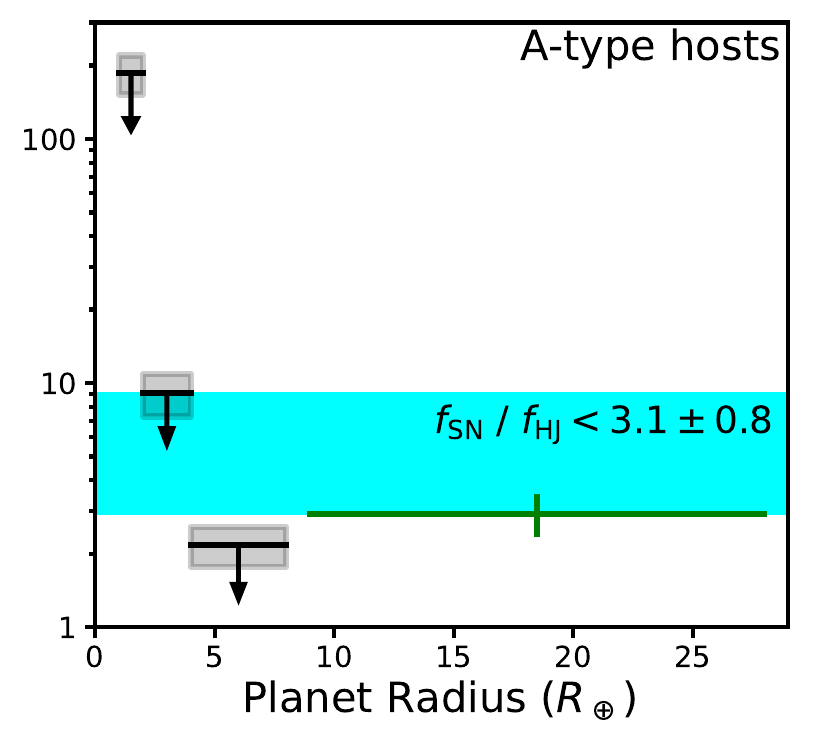}
    \caption{Occurrence rates for super-Earths, sub-Neptunes, sub-Saturns, and hot Jupiters for G-type (left), F-type (center), and A-type (left) stars. Blue data are from \citet{kunimoto2020}, green data are from \citet{beleznay2022hot}, and black data are the $3\sigma$ upper limits calculated in this paper. The limits calculated from each of these papers suggest that the radius cliff may become less steep with increasing host star $T_{\rm eff}$, based on the ratio of sub-Neptune occurrence rate to hot Jupiter occurrence rate ($f_{\rm SN}/f_{\rm HJ}$). In other words, the frequency of close-in sub-Neptunes may decrease more rapidly than that of hot Jupiters around progressively hotter stars.}
    \label{fig: radius_cliff}
\end{figure*}

For sub-Neptunes, \citet{kunimoto2020} found occurrence rates of  $23.21^{+5.03}_{-4.55}$, $70.82^{+9.15}_{-8.95}$, and $137.43^{+19.06}_{-17.83}$ planets per 1000 F-type, G-type, and K-type stars, respectively. \citet{dressing2015occurrence} found an occurrence rate of $189.18^{+44.73}_{-35.25}$ planets per 1000 M dwarfs. With our upper limit of 9.10 planets per 1000 A-type stars, this suggests that sub-Neptunes may be more than $2\times$ as rare around A-type stars than F-type stars, more than $6\times$ as rare around A-type stars than G-type stars, more than $13\times$ as rare around A-type stars than K-type stars, and more than $16\times$ as rare around A-type stars than M dwarfs. Based on the completeness of our pipeline to sub-Neptunes with $P_{\rm orb} < 10$ days ($3.1 \%$), we estimate having detected $15 \pm 3$, $45 \pm 6$, $87^{+12}_{-11}$, and $120^{+28}_{-22}$ sub-Neptunes around our sample of A-type stars if they had occurrence rates equivalent to those around FGKM-type stars.

For super-Earths, \citet{kunimoto2020} found occurrence rates of $47.44^{+7.11}_{-6.77}$, $124.54^{+11.85}_{-11.71}$, and $196.68^{+24.37}_{-23.53}$ planets per 1000 F-type, G-type, and K-type stars, respectively. Comparing with our upper limit of 186 planets per 1000 A-type stars, super-Earths may be slightly less common around A-type stars than K-type stars, but we cannot say confidently that they are also rarer around A-type stars than FG-type stars. This ambiguity is mostly due to the fact that A-type stars are larger than FGK-type stars, on average, which make the transits of super-Earths more difficult to detect. Super-Earths are known to be much more common around M dwarfs than FGK-type stars, with an occurrence rate of $351.37^{+51.63}_{-43.63}$ planets per 1000 M dwarfs as reported by \citet{dressing2015occurrence}. This suggests that super-Earths may be more than $1.5\times$ as rare around A-type stars than M dwarfs. Based on the completeness of our pipeline to super-Earths with $P_{\rm orb} < 10$ days ($0.14 \%$), we estimate having detected $6 \pm 1$ and $10 \pm 1$ super-Earths around our sample of A-type stars if they had an occurrence rates equivalent to those around KM-type stars.

These results resemble the trend for the occurrence rate of small close-in planets to decrease with increasing stellar $T_{\rm eff}$, which has been reported in a number of other studies \citep[e.g.,][]{howard2012planet, hardigree2019, hsu2019occurrence, yang2020occurrence, zink2023K2}. However, all previous calculations have relied on {\it Kepler} and {\it K2} data and have not extended to stars hotter than $7300$ K. In addition, this trend has only been reported conclusively for planets with $R_{\rm p} < 4 \, R_\oplus$. Our finding that sub-Saturns may also decrease in frequency with increasing $T_{\rm eff}$ has not previously been reported.

\subsection{The Slope of the Radius Cliff as a Function of Stellar Effective Temperature}

We also compare our occurrence rate upper limits to those calculated for hot Jupiters ($ 9 \, R_\oplus < R_{\rm p} < 28 \, R_\oplus$, $P_{\rm orb} < 10$ days) around A-type stars. \citet{beleznay2022hot} reported occurrence rates of $5.5 \pm 1.4$, $3.6 \pm 0.6$, and $2.9 \pm 0.5$ hot Jupiters per 1000 G-type, F-type, and A-type stars, respectively. By comparing these statistics with the occurrence rates reported here and in other works, we can examine how the radius distribution of close-in planets for A-type stars differs from those of cooler stars. 

One feature of the radius distribution that we can examine is the ``radius cliff,'' the steep drop-off in the occurrence rate of planets as a function of increasing planet size at $R_{\rm p} \approx 3 \, R_\oplus$ \citep[e.g.,][]{fulton2017gap, dattilo2023cliff, dattilo2024cliff}. The ``steepness'' of the cliff can be described simply as the ratio of the occurrence rates of close-in sub-Neptunes and hot Jupiters ($f_{\rm SN}/f_{\rm HJ}$). Using the close-in sub-Neptune occurrence rates from \citet{kunimoto2020} and the hot Jupiter occurrence rates from \citet{beleznay2022hot}, we calculate $f_{\rm SN}/f_{\rm HJ} = 12.9 \pm 3.9$ for G-type stars and $f_{\rm SN}/f_{\rm HJ} = 6.4 \pm 1.7$ for F-type stars. Using our close-in sub-Neptune occurrence rate upper limits and the hot Jupiter occurrence rates from \citet{beleznay2022hot}, we find that the sub-Neptune-to-hot-Jupiter occurrence ratio for A-type stars is $f_{\rm SN}/f_{\rm HJ} < 3.1 \pm 0.8$, leaving open the possibility that close-in sub-Neptunes are as common or less common than hot Jupiters around these early-type hosts (Fig. \ref{fig: radius_cliff}). Assuming these ratios, $f_{\rm SN}/f_{\rm HJ}$ is greater for G-type stars than A-type stars at $>99\%$ confidence and is greater for F-type stars than A-type stars at $>95\%$ confidence.

This possible decrease in $f_{\rm SN}/f_{\rm HJ}$ hints that the radius cliff may become less steep with increasing stellar $T_{\rm eff}$. In other words, the occurrence rate of close-in sub-Neptunes may decrease faster than the occurrence rate of hot Jupiters around progressively hotter host stars. However, we stress that because this trend relies on combining the results of different occurrence rate calculations with different methodologies, it should not necessarily be taken at face value. A more rigorous calculation involving a uniform sample of stars and a self-consistent analysis is required to judge if this trend really exists.

\section{Discussion}\label{sec:discussion}

Through our analysis of the {\it TESS} data, we find hints that (1) the occurrence rate of small close-in planets may decrease with increasing stellar $T_{\rm eff}$, and that (2) the occurrence rate of close-in sub-Neptunes decreases faster than that of hot Jupiters with increasing $T_{\rm eff}$. Here, we discuss a number of possible explanations for these tentative findings.

\subsection{Disk Truncation and Dust Sublimation}

The landscape of short-period planets is likely influenced by interactions between the young star and its inner protoplanetary disk. For instance, \citet{lee2017magnetospheric} shows that the drop off in occurrence rate of sub-Neptunes and super-Earths at $P_{\rm orb} = 10$ days can be attributed to the magnetospheric truncation of the disk at the corotation radius, inside of which material is channeled by magnetic field lines and accreted onto the star \citep[also see][]{batygin2023inner}. They predict that because A-type stars rotate more rapidly than stars of lower mass, this drop-off should occur at a $P_{\rm orb}$ closer to 1 day. In other words, the small planet occurrence rate around A-type stars should remain roughly constant with increasing $P_{\rm orb}$ exterior to $P_{\rm orb} = 1$ day. Unfortunately, because we detect no small planets interior to 10 days around A-type stars, we are unable to test this prediction. 

Another possibility is that A-type stars are so weakly magnetized (as a consequence of their largely non-convective interiors) that no magnetospheric truncation occurs at all. In this situation, material would accrete directly onto the star via the so-called hot ``boundary layer'' \citep[e.g.,][]{lynden-bell1974evolution, hillenbrand1992accretion, mendigutia2020accretion}. With no disk truncation, planets migrating through the disk could spiral directly into the star, potentially contributing to the death of small planets searched for in this paper. However, we note that most observations of Herbig Ae stars (young A-type stars actively undergoing accretion) favor magnetospheric accretion similar to that inferred for relatively low-mass classical T Tauri stars \citep[e.g.,][]{koenigl1991accretion, vink2002circumstellar, vink2003resolved, eisner2004resolved, mendigutia2011accretion}, whereas there is tentative evidence that accretion via the boundary layer plays a more dominant role for more massive Herbig Be stars \citep[e.g.,][]{mottram2007herbig, ababakr2017herbig}.

Dust sublimation may also play a significant role in sculpting the population of close-in planets. The minimum distance from a star at which a planet can form is set by the dust sublimation radius ($a_\mathrm{sub}$), within which solid planet-building material is depleted. In the simplest terms, $a_\mathrm{sub} \propto \sqrt{L_\mathrm{PMS}}$, where $L_\mathrm{PMS}$ is the pre-main-sequence stellar luminosity. For A-type stars, $L_\mathrm{PMS}$ is approximately an order of magnitude higher than for G-type stars \citep{hayashi1961evolution}, meaning that $a_\mathrm{sub}$ is approximately $3\times$ greater. Assuming young Solar-mass stars have $a_\mathrm{sub} \approx 0.04$ AU \citep{pinte2008inner}, the average A-type star would have $a_\mathrm{sub} \approx 0.12$ AU, which corresponds roughly to an orbital period of 10 days for a star with $M_\star = 2 \, M_\odot$. Thus, the in-situ formation of small close-in planets is likely mitigated around A-type stars. In addition, dust sublimation may hinder the inward migration of planets that form at orbital periods beyond 10 days. \citet{flock2019sublimation} finds that the sublimation of dust induces a pressure bump beyond $a_{\rm sub}$ that halts the inward migration of super-Earths between 10 and 20 days for FGKM-type stars, after which they may migrate inwards due to tidal interactions with the host star. For A-type stars, migration would be halted at even longer orbital periods where star-planet tidal interactions are weak, leaving the planets stranded at wide separations. This may explain why no small planets have been discovered with orbital periods interior to 20 days around A-type stars \citep[e.g.,][]{morton2016validation, masuda2020revisiting, giacalone2022hd56414b}. 

This mechanism also self-consistently explains the tentative decrease in hot Jupiter occurrence with increasing stellar mass reported by \citet{zhou2019} and \citet{beleznay2022hot}, as well as the potential ``flattening'' of the radius cliff around A-type stars reported here. Hot Jupiters are thought to reach their close-in orbits via either disk migration or high-eccentricity migration \citep[e.g.,][]{lin1996migration, rasio1996higheccentricity, fabrycky2007kozai, chatterjee2008scattering, wu2011secular, dawson2018hotjupiters, rice2022origins}, whereas most sub-Neptunes are thought to reach close-in orbits via the former \citep[e.g.,][]{terquem2007migration, ogihara2009nbody, ida2010formation, mcneil2010formation, boley2013formation, chatterjee2014formation, cossou2014migration, izidoro2017breaking, raymond2018migration}. If disk migration is effectively ``turned off'' within $0.1-0.2$ AU for A-type stars, the only planets that can obtain orbital periods under 10 days are hot Jupiters migrating inwards via high-eccentricity migration, leading to a slight decrease in the hot Jupiter occurrence rate and a more drastic decrease in the sub-Neptune occurrence rate.

\subsection{Atmospheric Mass Loss}

For FGKM-type stars, photoevaporation is thought to strip close-in planets of their atmospheres due to high levels of stellar XUV emission \citep{lammer2003, ribas2005evolution, murrayclay2009escape}. This period of high XUV emission is typically thought to persist for the first 100 Myrs of the system lifetime, although it may last longer than 1 Gyr for low-mass M dwarfs \citep{jackson2012coronal}. Unlike their cool counterparts, A-type stars emit very little in the XUV during their youths, largely due to their lack of convective interiors \citep{schroder2007}. As a result, we should not expect XUV photoevaporation to strongly influence their population of close-in planets. However, it has been shown that A-type stars can efficiently strip planets of their atmospheres thanks to high levels of near-ultraviolet continuum emission \citep{munoz2019}. Invoking this mechanism, \citet{giacalone2022hd56414b} showed that the warm Neptune HD 56414 b, which orbits an A-type star with $T_{\rm eff} \sim 8500$ K, would likely completely lose its atmosphere within 1 Gyr timescale if it were located closer than 0.1 AU from its A-type host star. This effect would be most significant for sub-Saturns and sub-Neptunes, which have low bulk densities relative to super-Earths, leaving them more vulnerable to hydrodynamic atmospheric escape \citep[e.g.,][]{hallatt2022subsaturn}.

Core-powered mass loss, the process by which the cooling planetary core gradually strips the planet of its atmosphere, is also thought to influence the occurrence rate of close-in sub-Neptunes and super-Earths \citep{ginzburg2016, ginzburg2018}. The rate of this process is chiefly dependent on the bolometric stellar flux incident on the planet, meaning that it may occur more efficiently for planets around A-type stars than those around cooler stars. However, because models predict this mass loss to occur on timescales comparable to the $1-2$ Gyr main-sequence lifetimes of A-type stars \citep{gupta2019, gupta2020}, its effect may be secondary to those of the relatively rapid photoevaporation. More studies exploring the effects of core-powered mass loss as a function of stellar mass are needed to truly understand its effects on close-in planets around hot stars.

Atmospheric mass loss offers a compelling explanation for the small $f_{\rm SN}/f_{\rm HJ}$ around A-type stars relative to cooler stars. Observations of escaping exoplanet atmospheres have proven these mechanisms to be effective at stripping the envelopes of sub-Neptunes \citep[e.g.,][]{zhang2022escaping, zhang2022massloss, zhang2023outflowing, zhang2023escape} and relatively ineffective at eroding those of hot Jupiters \citep{vissapragada2022stable}. If atmospheric stripping is enhanced around A-type stars, it would naturally lead to a suppression in the occurrence rates of sub-Neptunes. This process would leave behind Earth-sized or super-Earth-sized cores that may exist around our sample of stars, but that we lack the sensitivity to detect with the current \textit{TESS} data set.

\subsection{The Influence of Massive Outer Companions}

More massive stars are known to more frequently have stellar companions and host long-period Jupiter-like planets \citep{johnson2010, moe2017mind, nielsen2019}. It is reasonable to hypothesize that the presence of these massive companions could have an impact on the ability of small planets to form at, migrate to, or survive in close-in orbits. 

N-body simulations suggest that the scattering of long-period giant planets can efficiently destroy small planets in the inner regions of the system \citep[e.g.,][]{raymond2010scattering, mustill2015migration, pu2021scattering}. In addition, it has been shown that stars with stellar-mass companions at separations $< 10$ AU less frequently have small close-in planets relative to single stars \citep{hirsch2021companions, moe2021impact}, although it is unclear if this is a consequence of suppressed planet formation or dynamical star-planet interactions.

Interestingly, there is also evidence that stars with giant planets at wide-separations ($a > 1$ AU) frequently harbor single inner sub-Neptunes or super-Earths \citep{barbato2018exploring, zhu2018super, bryan2019excess, rosenthal2022relationship}, indicating that outer giant planets may sometimes aid the formation of small close-in planets \citep[e.g.,][]{bitsch2023bullies}. Some may therefore view the dearth of small close-in planets around A-type stars as surprising, given the positive correlation between stellar mass and wide-separation giant planet occurrence \citep{nielsen2019}. Perhaps the physical mechanism behind this phenomenon is sensitive to the number of wide-separation giant planets in the system, where systems with more outer giants less frequently harbor small inner planets. This interpretation would be consistent with direct imaging surveys, which have found that planetary systems around A-type stars frequently have multiple giant planets (e.g., $\beta$ Pictoris and HR 8799; \citealt{marois2008hr8799, marois2010hr8799, lagrange2009betapic, lagrange2010betapic, lagrange2019betapic}).

Others have invoked pebble accretion to explain the inverse correlation between long-period giant planets and short-period small planets as a function of host star mass. The core accretion model of giant planet formation posits that giant Jupiter-like planets form in a two-step process: the formation of a massive solid core followed by the runaway accretion of a gaseous envelope from the disk \citep{pollack1996core}. This model therefore requires giant planet cores to form before the gas in the protoplanetary disk is full dissipated. It has been argued that cores can only form on these rapid timescales through the accretion of small millimeter-to-centimeter-sized ``pebbles'' that drift inward from the outer disk \citep{lambrechts2014pebble}. However, if the inward drift of pebbles were to be cut off, the formation of inner planets may be hindered. \citet{mulders2021mdwarfs} predict that pebble accretion forms giant outer planets more rapidly than small inner planets. After the formation of a giant outer planet, a large gap is thought to be carved out of the outer protoplanetary disk, inhibiting the inward flow of pebbles and preventing small inner planets from forming efficiently. This prediction has been supported by observations of protoplanetary disks, which show more massive protostars to have disks with more gaps \citep{vandermarel2021structured}.

\subsection{Flux Dilution from Unseen Stars}

Another possible explanation for the non-detection of small planets around A-type stars is the presence of unresolved stellar companions. Over half of A-type stars have at least one physically associated companion star \citep{moe2017mind}. An unseen star increases the overall brightness of the system, causing the depth of a planetary transit to appear more shallow in the data than it would be for a single star, thereby making the planet more difficult to detect \citep[e.g.,][]{ciardi2015multiplicity, furlan2017kepler, hirsch2017effects}. In our case, this dilution would most significantly impact super-Earths, which are already challenging to detect around A-type stars due to their large size differences. Because the presence of unknown stellar companions has been shown to influence planet occurrence rate calculations in the past \citep[e.g.,][]{bouma2018biases, teske2018effects, savel2020closer, moe2021impact}, it is plausible that the same is true in this paper. We do not attempt to correct for flux dilution due to stellar multiplicity in this analysis, but we note that it is an area that should be afforded more careful attention in future occurrence rate studies.

\subsection{Dependence on Stellar Age}

{ Because main-sequence lifetime decreases with increasing stellar mass, any study of planet occurrence as a function of stellar mass is also a study of planet occurrence as a function of stellar age. While Sun-like stars can remain on the main sequence for upwards of 10~Gyr, A-type stars have an average main sequence lifetime between 1 and 2 Gyr. The average A-type star targeted in this study is likely younger than 1 Gyr, significantly younger than the FGK-type stars observed by \textit{Kepler}. A number of processes are thought to influence planet demographics within the first Gyr, including tidal interactions with the host star \citep[e.g.,][]{matsumura2010tidal, valsecchi2014tidal}, dynamical interactions with other planets \citep[e.g.,][]{hamer2022evidence, hamer2024resonances, dai2024prevalence}, and atmospheric mass loss \citep[e.g.,][]{lammer2003, ribas2005evolution, murrayclay2009escape, lopez2013role, ginzburg2016, ginzburg2018, munoz2019, gupta2019, gupta2020}. Recently, the impact of these processes on the occurrence rates of small, close-in planets has come into view. Using data from the \textit{K2} mission, \citet{christiansen2023K2} found that the occurrence rate of sub-Neptunes around FGK-type stars is $\sim 10 \times$ higher in the 0.6--0.8~Gyr-old Praesepe cluster than in the $>1$~Gyr-old field. \citet{fernandes2022tess, fernandes2023tess}, Fernandes et al. (submitted), and \citet{vach2024occurrence} found a similar result for sub-Netpunes and sub-Saturns in clusters with ages between 0.01 and 1 Gyr. This enhancement points to efficient formation and early inward migration of small planets around FGK-type stars, followed by dynamical or atmospheric evolution that drives down occurrence rate. If all conditions were held constant across stellar mass, one might reasonable extrapolate this trend to young A-type stars and predict a similarly high occurrence rate around our sample. The fact that we do not see this enhancement suggests that A-type stars are particularly hostile to the formation of small planets near the stellar vicinity or their early migration to close-in orbits. Alternatively, it is possible that these planets do arrive at close-in orbits early but experience extremely rapid photoevaporative atmospheric mass loss \citep[e.g.,][]{munoz2019}, shrinking them to sizes below our detection threshold. The youth of the A-type stars in our sample slightly disfavors interpretations in which the observed dearth of planets is caused by longer-term processes such as core-powered atmospheric mass loss or dynamical instabilities with nearby planets \citep[e.g.,][]{dai2024prevalence}.}

\section{Conclusions}\label{sec:conclusions}

Using a sample of { 20,257} bright A-type stars, we search for transiting planets using a custom transit detection pipeline. We calculate the occurrence rate of small close-in planets and find the following:
\begin{itemize}[noitemsep]
    \item Sub-Saturns ($4 \, R_\oplus < R_\mathrm{p} < 8 \, R_\oplus$) with $P_\mathrm{orb} < 10$ days have an occurrence rate of $< 2.2 \pm 0.4$ planets per 1000 A-type stars, at $3 \sigma$ confidence. By comparison with the results of \citet{dressing2013occurrence} and \citet{kunimoto2020}, we find that sub-Saturns may be rarer around A-type stars than FM-type stars and may be more than $3\times$ as rare around A-type stars as GK-type stars. 
    \item Sub-Neptunes ($2 \, R_\oplus < R_\mathrm{p} < 4 \, R_\oplus$) with $P_\mathrm{orb} < 10$ days have an occurrence rate of $< 9.1 \pm 1.8$ planets per 1000 A-type stars, at $3 \sigma$ confidence. By comparison with \citet{dressing2015occurrence} and \citet{kunimoto2020}, we find that sub-Neptunes may be more than $2\times$ as rare around A-type stars as F-type stars, more than $6\times$ as rare around A-type stars as G-type stars, more than $13\times$ as rare around A-type stars as K-type stars, and more than $16\times$ as rare around A-type stars as M dwarfs.
    \item Super-Earths ($1 \, R_\oplus < R_\mathrm{p} < 2 \, R_\oplus$) with $P_\mathrm{orb} < 10$ days have an occurrence rate of $< 186 \pm 34$ planets per 1000 A-type stars, at $3 \sigma$ confidence. By comparison with \citet{dressing2015occurrence} and \citet{kunimoto2020}, we find that super-Earths may be rarer around A-type stars than K-type stars and may be more than $1.5\times$ as rare around A-type stars as M dwarfs. We cannot, however, confidently claim that they are rarer around A-type stars than FG-type stars, primarily due to our low detection sensitivity to small planets.
    \item The occurrence rate of close-in sub-Neptunes may decrease faster than that of hot Jupiters, suggesting that the radius cliff \citep[e.g.,][]{fulton2017gap, dattilo2023cliff, dattilo2024cliff} may become less steep with increasing stellar $T_{\rm eff}$.
\end{itemize} 

These findings are consistent with earlier results that found planets with $R_\mathrm{p} < 4 \, R_\oplus$ to become increasingly rare around progressively hotter main-sequence stars \citep[e.g.,][]{mulders2015stellar, kunimoto2020, zink2023K2}. There are a number of possible explanations for this trend, including mitigated planet formation and migration interior to the dust sublimation front, atmospheric mass loss, dynamical scattering from wide-separation giant planets, depletion of planet-forming pebbles in the inner protoplanetary disk, and flux dilution from unresolved companions stars. If real, the potential decrease in sub-Saturn occurrence rate and ``flattening'' of the radius cliff around hotter stars would be novel results, and would indicate that these planets are either unable to form in or migrate to close-in orbits or that they are rapidly stripped of their atmospheres by the intense and high-energy continuum emission of their hot host stars. Assuming these planets are able to arrive at close-in orbits via a similar mechanism to known hot Jupiters orbiting A-type stars, the cores left over by this rapid photoevaporation may become detectable as \textit{TESS} continues to collect data, or with upcoming exoplanet-hunting missions like \textit{PLATO} \citep{rauer2014}.

These results also have interesting implications for the population of planets around white dwarfs. The dearth of short-period { sub-Jovian planets around our sample of A-type stars supports the notion that white dwarf contaminants migrate inward from wide separations after the star migrates off of the main sequence}, which is consistent with earlier predictions \citep{veras2015unpacking, munoz2020kozai, lagos2021commonenvelope, maldonado2021instabilities, stephan2021kozai, merlov2021commonenvelope, oconnor2021kozai, veras2021hr8799}. Our understanding of planetary systems around white dwarfs and their relation to planets around A-type stars will likely improve in the near future, thanks to data from the Rubin Observatory and \textit{Gaia} \citep{agol2011ztfWD, sanderson2022gaiaWD}.

We make the code for our automated TCE detection pipeline and injection/recovery tests open to the public.\footnote{\url{https://github.com/stevengiacalone/TESS-detection-pipeline}} We hope this code will facilitate future studies of planet statistics using \textit{TESS}.

\bigskip

The authors thank Eugene Chiang, Marshall C. Johnson, Andrew Vanderburg, David R. Ciardi, Rachel B. Fernandes, Jessie Christiansen, Michelle Kunimoto, Andrew W. Mayo, Emma V. Turtelboom, and Caleb K. Harada for their constructive feedback on this work.

SG is supported by an NSF Astronomy and Astrophysics Postdoctoral Fellowship under award AST-2303922. SG and CDD acknowledge support from NASA FINESST award 80NSSC20K1549. CDD also acknowledge support provided by the David and Lucile Packard Foundation via grant 2019-69648, the NASA TESS Guest Investigator Program via grant 80NSSC18K1583, and the NASA Exoplanets Research Program (XRP) via grant 80NSSC20K0250.

This research has made use of the Exoplanet Follow-up Observation Program (ExoFOP; DOI: 10.26134/ExoFOP5) website, which is operated by the California Institute of Technology, under contract with the National Aeronautics and Space Administration under the Exoplanet Exploration Program.


\software{\texttt{exoplanet} \citep{foremanmackey2021}, \texttt{Lightkurve} \citep{lightkurve2018}, \texttt{pymc3} \citep{salvatier2016}, \texttt{w{\={o}}tan} \citep{hippke2019wotan}, \texttt{TRICERATOPS} \citep{giacalone2020ascl}, \texttt{TESS-plots} \citep{kunimoto2022faint}}

\facilities{TESS, MAST, ExoFOP}

\startlongtable
\begin{deluxetable*}{cccccccccc}
\tabletypesize{\footnotesize}
\tablewidth{\textwidth}
\tablecaption{ Threshold-Crossing Events detected by the automated pipeline \label{tab:1}}
\tablehead{
\colhead{TIC ID} & \colhead{TOI} & \colhead{$T$ mag} & \colhead{$T_{\rm eff}$ (K)} & \colhead{$R_\star$ ($R_\odot$)} & \colhead{$M_\star$ ($M_\odot$)} & 
\colhead{$T_0$ (TBJD)} & \colhead{$P_{\rm orb}$ (days)} & \colhead{$\delta$ (ppm)} & \colhead{Vetting Outcome} 
}
\startdata 
144193330 & - & 8.85 & 8450 & 2.10 & 2.05 & 1330.1818 & 9.2691 & 230 & Failed SPOC comparison \\
61851084 & - & 7.05 & 7583 & 1.76 & 2.09 & 1658.5744 & 6.9497 & 271 & Failed SPOC comparison \\
254158364 & - & 9.14 & 7731 & 1.82 & 2.45 & 1662.0146 & 8.7717 & 513 & Failed SPOC comparison \\
265663360 & - & 7.86 & 8680 & 2.17 & 1.67 & 1658.5824 & 0.5857 & 259 & Failed SPOC comparison \\
176109599 & - & 9.13 & 8265 & 2.03 & 1.56 & 1630.3262 & 0.7824 & 1404 & Nearby eclipsing binary detected \\
100588438 & - & 9.85 & 7849 & 1.87 & 2.12 & 1632.0363 & 2.5671 & 1112 & Failed TRICERATOPS analysis \\
376131915 & - & 9.90 & 7771 & 1.84 & 2.17 & 1627.3321 & 0.9514 & 477 & Failed SPOC comparison \\
301152169 & - & 9.00 & 7797 & 1.85 & 1.59 & 1630.7981 & 3.1052 & 679 & Failed SPOC comparison \\
420774265 & - & 9.77 & 8703 & 2.18 & 2.32 & 1627.1023 & 2.6617 & 545 & Failed SPOC comparison \\
246123223 & - & 8.88 & 8023 & 1.94 & 2.47 & 1631.3260 & 2.0644 & 511 & Nearby eclipsing binary detected \\
447299730 & - & 8.49 & 8710 & 2.18 & 1.79 & 1629.4070 & 7.0108 & 1228 & Failed SPOC comparison \\
221200664 & - & 9.01 & 7655 & 1.79 & 1.62 & 1627.6122 & 0.9596 & 885 & Secondary eclipse detected \\
350575997 & 4386 & 9.77 & 8127 & 1.98 & 1.75 & 1628.3568 & 2.0134 & 837 & Failed TRICERATOPS analysis \\
315222615 & - & 8.63 & 7853 & 1.87 & 2.20 & 1630.5659 & 7.1596 & 587 & Failed SPOC comparison \\
215218279 & - & 6.50 & 8253 & 2.03 & 1.93 & 1631.7210 & 7.0108 & 203 & Failed SPOC comparison \\
422823594 & - & 8.81 & 8516 & 2.12 & 2.10 & 1626.4363 & 0.9957 & 361 & Failed SPOC comparison \\
254788512 & - & 9.85 & 8244 & 2.02 & 1.93 & 1630.2811 & 3.4779 & 828 & Failed SPOC comparison \\
255375615 & - & 7.54 & 8864 & 2.23 & 2.32 & 1630.1961 & 0.5489 & 146 & Secondary eclipse detected \\
136274063 & 1094 & 9.98 & 7710 & 1.81 & 2.28 & 1625.6866 & 0.5971 & 758 & False Positive on ExoFOP \\
262922645 & - & 6.84 & 8253 & 2.03 & 2.44 & 1369.1334 & 0.8159 & 45 & Failed SPOC comparison \\
149639802 & - & 9.59 & 8947 & 2.26 & 2.15 & 1630.1970 & 0.8804 & 253 & Failed SPOC comparison \\
143705205 & - & 9.12 & 8382 & 2.07 & 2.15 & 1600.6825 & 1.3870 & 389 & Failed SPOC comparison \\
259230140 & 4384 & 9.66 & 7596 & 1.77 & 1.61 & 1631.3680 & 7.1586 & 737 & True period twice as long \\
261693614 & - & 9.89 & 8701 & 2.18 & 2.07 & 1625.3281 & 0.6929 & 242 & Failed SPOC comparison \\
308830797 & - & 9.81 & 8091 & 1.97 & 1.93 & 1604.1838 & 7.2835 & 750 & Failed SPOC comparison \\
415933085 & - & 8.66 & 7736 & 1.82 & 2.18 & 1599.6684 & 0.8403 & 1114 & Failed SPOC comparison \\
160086759 & - & 9.28 & 7757 & 1.83 & 2.47 & 1600.6677 & 0.7992 & 633 & Secondary eclipse detected \\
454655012 & - & 8.50 & 7974 & 1.92 & 2.41 & 1625.4283 & 7.1441 & 203 & Failed SPOC comparison \\
35905913 & - & 8.50 & 7507 & 1.73 & 1.75 & 1600.1535 & 1.4185 & 613 & Failed SPOC comparison \\
126909342 & - & 8.70 & 7807 & 1.85 & 2.14 & 1599.2440 & 7.6889 & 341 & Secondary eclipse detected \\
259226908 & - & 9.10 & 8646 & 2.16 & 1.79 & 1627.2530 & 2.5432 & 617 & Secondary eclipse detected \\
411906862 & - & 9.31 & 8238 & 2.02 & 1.99 & 1625.9419 & 1.4784 & 828 & Failed SPOC comparison \\
418792366 & - & 8.60 & 7822 & 1.86 & 1.56 & 1631.8611 & 7.2185 & 176 & Failed SPOC comparison \\
448580024 & - & 9.32 & 8459 & 2.10 & 1.72 & 1604.7699 & 8.4710 & 788 & Failed SPOC comparison \\
208570502 & - & 8.26 & 7524 & 1.74 & 1.76 & 1602.0235 & 6.1998 & 195 & Failed SPOC comparison \\
111948529 & - & 8.38 & 7682 & 1.80 & 2.41 & 1606.1763 & 7.6220 & 182 & Failed SPOC comparison \\
326170150 & - & 9.87 & 8414 & 2.08 & 1.59 & 1600.8856 & 1.6265 & 1537 & Nearby eclipsing binary detected \\
83336505 & - & 8.92 & 8172 & 2.00 & 2.06 & 1605.6874 & 7.5456 & 219 & Failed SPOC comparison \\
315350812 & 4373 & 9.85 & 7824 & 1.86 & 1.63 & 1603.5453 & 5.9422 & 976 & Failed TRICERATOPS analysis \\
341865564 & - & 8.96 & 7701 & 1.81 & 2.26 & 1601.4550 & 2.2753 & 221 & Failed SPOC comparison \\
288598068 & - & 9.12 & 7846 & 1.87 & 2.19 & 1601.2357 & 3.9368 & 790 & Nearby eclipsing binary detected \\
418294275 & - & 7.28 & 8500 & 2.11 & 2.43 & 1632.2509 & 8.7907 & 88 & Failed SPOC comparison \\
449608185 & - & 8.98 & 8739 & 2.19 & 2.16 & 1601.4904 & 4.7147 & 427 & Failed SPOC comparison \\
335470158 & - & 9.93 & 8168 & 2.00 & 2.11 & 1602.9014 & 6.9305 & 622 & Failed SPOC comparison \\
228696887 & - & 8.88 & 8800 & 2.21 & 2.46 & 1579.8995 & 9.6445 & 369 & Failed SPOC comparison \\
311588504 & - & 7.87 & 7748 & 1.83 & 1.72 & 1600.4302 & 2.3473 & 949 & Failed SPOC comparison \\
369433644 & - & 9.39 & 8067 & 1.96 & 1.90 & 1602.6055 & 2.9126 & 1336 & Secondary eclipse detected \\
451134562 & - & 9.13 & 8039 & 1.95 & 2.29 & 1602.2054 & 4.8236 & 597 & Failed SPOC comparison \\
327273899 & - & 8.89 & 8236 & 2.02 & 2.41 & 1605.3765 & 7.0503 & 850 & Secondary eclipse detected \\
134907045 & - & 9.50 & 8016 & 1.94 & 2.02 & 1572.5038 & 1.2757 & 741 & Secondary eclipse detected \\
135179024 & - & 7.69 & 8637 & 2.16 & 2.44 & 1572.7138 & 1.0490 & 105 & Nearby eclipsing binary detected \\
179473571 & - & 9.61 & 8029 & 1.94 & 1.96 & 1600.3903 & 2.8942 & 924 & Secondary eclipse detected \\
268660469 & - & 7.85 & 7854 & 1.87 & 2.07 & 1576.8377 & 7.5391 & 163 & Failed SPOC comparison \\
308766739 & - & 9.07 & 8615 & 2.15 & 2.26 & 1600.5067 & 5.6659 & 765 & Secondary eclipse detected \\
94013493 & - & 8.12 & 8338 & 2.06 & 2.30 & 1572.7881 & 7.1228 & 169 & Failed SPOC comparison \\
152738973 & - & 8.67 & 8850 & 2.23 & 1.94 & 1547.4270 & 8.7951 & 232 & Failed SPOC comparison \\
1589794 & - & 7.98 & 8228 & 2.02 & 1.80 & 1550.7670 & 9.5069 & 258 & Failed SPOC comparison \\
398764122 & - & 9.90 & 8325 & 2.05 & 1.72 & 1577.6026 & 9.2496 & 483 & Failed SPOC comparison \\
398540624 & - & 9.99 & 7761 & 1.83 & 2.12 & 1579.2078 & 8.5773 & 780 & Failed SPOC comparison \\
448644231 & - & 9.23 & 8068 & 1.96 & 2.11 & 1572.0740 & 6.6143 & 183 & Failed SPOC comparison \\
392557724 & - & 9.18 & 8946 & 2.26 & 2.25 & 1600.8011 & 5.6374 & 1036 & Nearby eclipsing binary detected \\
188119124 & - & 9.08 & 8929 & 2.25 & 1.82 & 1549.5773 & 7.1257 & 918 & Nearby eclipsing binary detected \\
281019037 & - & 9.56 & 8184 & 2.00 & 1.82 & 1572.1108 & 0.9389 & 560 & Secondary eclipse detected \\
23355734 & - & 9.10 & 8712 & 2.18 & 1.76 & 1550.5019 & 9.7590 & 281 & Failed SPOC comparison \\
57688953 & - & 9.32 & 7882 & 1.88 & 1.88 & 1578.5691 & 6.6756 & 241 & Failed SPOC comparison \\
158548149 & - & 8.24 & 8357 & 2.06 & 1.74 & 1547.5564 & 8.0339 & 207 & Failed SPOC comparison \\
333736725 & - & 8.72 & 8069 & 1.96 & 2.36 & 1547.4514 & 7.3950 & 208 & Failed SPOC comparison \\
437231703 & - & 8.69 & 7712 & 1.81 & 1.94 & 1547.4469 & 8.7951 & 235 & Failed SPOC comparison \\
467891495 & - & 8.09 & 7891 & 1.89 & 2.35 & 1573.9663 & 2.5435 & 94 & Nearby eclipsing binary detected \\
147026702 & - & 6.12 & 7900 & 1.89 & 1.87 & 1546.5062 & 2.0719 & 161 & Nearby eclipsing binary detected \\
120155231 & - & 9.68 & 8257 & 2.03 & 2.00 & 1572.8026 & 1.5237 & 903 & Failed TRICERATOPS analysis \\
36722395 & - & 8.68 & 8753 & 2.20 & 1.85 & 2257.7504 & 1.9042 & 483 & Secondary eclipse detected \\
377385711 & - & 9.25 & 8427 & 2.09 & 2.03 & 1551.0330 & 9.3815 & 487 & Failed SPOC comparison \\
73205080 & - & 9.52 & 7511 & 1.73 & 2.49 & 1550.7465 & 9.6304 & 329 & Failed SPOC comparison \\
370440819 & - & 8.51 & 7698 & 1.81 & 1.83 & 1575.8352 & 6.9442 & 151 & Failed SPOC comparison \\
370709148 & - & 9.78 & 8527 & 2.12 & 2.15 & 1544.7624 & 0.6156 & 231 & Secondary eclipse detected \\
439609369 & - & 9.12 & 8982 & 2.27 & 1.48 & 1552.1835 & 8.8722 & 194 & Failed SPOC comparison \\
447523057 & - & 8.45 & 7962 & 1.92 & 2.03 & 1545.5138 & 2.6301 & 138 & Failed SPOC comparison \\
432563559 & - & 9.75 & 8591 & 2.14 & 1.61 & 1544.7686 & 0.7619 & 305 & Nearby eclipsing binary detected \\
192542930 & - & 8.03 & 7705 & 1.81 & 2.12 & 1547.3825 & 9.2398 & 174 & Failed SPOC comparison \\
220708449 & - & 8.41 & 8591 & 2.14 & 1.75 & 1546.8487 & 2.9743 & 442 & Secondary eclipse detected \\
297732015 & - & 9.07 & 8847 & 2.23 & 2.48 & 1547.6288 & 6.6646 & 161 & Failed SPOC comparison \\
373424049 & 742 & 8.63 & 8508 & 2.12 & 1.82 & 1545.4035 & 0.9619 & 271 & False Positive on ExoFOP \\
304029917 & - & 8.50 & 8539 & 2.13 & 2.08 & 1574.8250 & 7.1470 & 106 & Failed SPOC comparison \\
60984804 & 6260 & 7.12 & 7707 & 1.81 & 1.96 & 1519.4095 & 2.3880 & 187 & Failed SPOC comparison \\
44692683 & - & 9.82 & 7650 & 1.79 & 1.56 & 1546.2414 & 6.6110 & 840 & Failed SPOC comparison \\
293670551 & - & 9.55 & 8780 & 2.21 & 2.49 & 2256.8325 & 0.6833 & 586 & Failed SPOC comparison \\
38033320 & - & 8.86 & 8752 & 2.20 & 2.09 & 1523.7470 & 6.1556 & 327 & Failed SPOC comparison \\
386657306 & - & 6.72 & 8244 & 2.02 & 1.90 & 1546.4282 & 6.5510 & 93 & Failed SPOC comparison \\
375043797 & - & 9.44 & 7531 & 1.74 & 2.30 & 1598.9759 & 0.6185 & 284 & Secondary eclipse detected \\
469255408 & - & 9.77 & 8649 & 2.16 & 1.71 & 1545.0033 & 0.7040 & 276 & Failed SPOC comparison \\
391627710 & - & 9.95 & 7610 & 1.77 & 2.08 & 1518.8825 & 0.5212 & 264 & Failed SPOC comparison \\
447885888 & - & 9.76 & 8773 & 2.20 & 1.83 & 1518.4977 & 1.0509 & 268 & Nearby eclipsing binary detected \\
51945029 & - & 8.97 & 8057 & 1.95 & 1.47 & 1521.9120 & 6.3205 & 241 & Failed SPOC comparison \\
296865162 & - & 9.61 & 8593 & 2.15 & 1.83 & 1545.5787 & 6.5008 & 386 & Failed SPOC comparison \\
309485321 & - & 8.67 & 8079 & 1.96 & 1.83 & 1602.9750 & 7.2185 & 140 & Failed SPOC comparison \\
143438876 & - & 8.90 & 7830 & 1.86 & 1.68 & 1494.2959 & 9.5970 & 236 & Failed SPOC comparison \\
342888057 & - & 7.79 & 8460 & 2.10 & 1.53 & 1541.0819 & 6.3794 & 160 & Failed SPOC comparison \\
147152188 & - & 8.69 & 7961 & 1.91 & 1.61 & 1518.6162 & 1.0805 & 343 & Secondary eclipse detected \\
89936838 & - & 10.00 & 8180 & 2.00 & 1.62 & 1518.7675 & 4.2115 & 580 & Failed SPOC comparison \\
93275805 & - & 8.30 & 8130 & 1.98 & 1.51 & 1520.6472 & 6.3849 & 182 & Failed SPOC comparison \\
147656922 & - & 9.33 & 8387 & 2.08 & 2.44 & 1518.5864 & 0.9942 & 601 & Secondary eclipse detected \\
54390047 & 998 & 9.70 & 8322 & 2.05 & 2.35 & 1492.6013 & 0.9412 & 358 & False Positive on ExoFOP \\
308448327 & - & 9.79 & 7716 & 1.82 & 2.34 & 1327.8751 & 2.3624 & 834 & Nearby eclipsing binary detected \\
419318737 & - & 9.16 & 7723 & 1.82 & 2.08 & 1492.5417 & 1.3286 & 600 & Failed SPOC comparison \\
340006157 & - & 8.72 & 8158 & 1.99 & 2.02 & 1469.8532 & 4.8782 & 121 & Failed SPOC comparison \\
143843200 & - & 9.81 & 7946 & 1.91 & 2.02 & 1493.1766 & 2.7159 & 668 & Nearby eclipsing binary detected \\
178575480 & - & 8.83 & 7718 & 1.82 & 1.67 & 1495.3634 & 4.7750 & 1345 & Failed SPOC comparison \\
355453890 & - & 9.24 & 7540 & 1.74 & 2.04 & 1492.3810 & 0.5312 & 225 & Secondary eclipse detected \\
140163913 & - & 8.19 & 8407 & 2.08 & 1.73 & 1499.4125 & 8.0131 & 160 & Failed SPOC comparison \\
177120452 & - & 9.91 & 8501 & 2.11 & 1.50 & 1492.4975 & 2.5000 & 1608 & Nearby eclipsing binary detected \\
173756330 & - & 8.82 & 8650 & 2.16 & 1.52 & 1493.5672 & 1.7535 & 1889 & Secondary eclipse detected \\
99417525 & - & 8.47 & 8301 & 2.05 & 2.45 & 1492.1267 & 0.6493 & 194 & Nearby eclipsing binary detected \\
174420083 & - & 8.97 & 8026 & 1.94 & 1.66 & 1492.4370 & 1.4862 & 343 & Failed SPOC comparison \\
142527279 & - & 9.78 & 8181 & 2.00 & 2.01 & 1474.3595 & 7.8284 & 651 & Nearby eclipsing binary detected \\
49670001 & - & 8.57 & 8635 & 2.16 & 2.45 & 1469.0989 & 1.1645 & 265 & Failed SPOC comparison \\
456683872 & - & 9.88 & 7760 & 1.83 & 1.96 & 2230.3949 & 0.8376 & 297 & Failed SPOC comparison \\
238058855 & - & 9.58 & 7665 & 1.79 & 1.62 & 1354.5122 & 2.9282 & 138 & Failed SPOC comparison \\
238073953 & - & 8.48 & 7632 & 1.78 & 1.98 & 1468.8892 & 0.7129 & 488 & Secondary eclipse detected \\
49245668 & - & 9.94 & 9022 & 2.28 & 1.84 & 1470.1341 & 1.7161 & 921 & Failed SPOC comparison \\
50587552 & - & 8.82 & 8025 & 1.94 & 1.53 & 1471.6227 & 3.8053 & 1241 & Failed SPOC comparison \\
260505542 & - & 8.21 & 7596 & 1.77 & 2.35 & 1329.9899 & 6.4254 & 43 & Failed SPOC comparison \\
350563225 & - & 9.08 & 7875 & 1.88 & 1.90 & 1327.9202 & 2.0497 & 1004 & Failed SPOC comparison \\
219155952 & - & 5.49 & 7851 & 1.87 & 2.35 & 1346.2500 & 6.7302 & 54 & Failed SPOC comparison \\
374858736 & - & 8.57 & 8751 & 2.20 & 2.13 & 1343.1707 & 8.4033 & 118 & Failed SPOC comparison \\
261376796 & - & 7.78 & 8145 & 1.99 & 1.82 & 1329.3773 & 9.1802 & 101 & Failed SPOC comparison \\
238190023 & - & 7.32 & 7812 & 1.85 & 1.95 & 1340.5190 & 6.3372 & 58 & Failed SPOC comparison \\
249067445 & 957 & 8.96 & 8897 & 2.24 & 1.82 & 1439.0408 & 0.8313 & 304 & Nearby eclipsing binary detected \\
237917947 & - & 9.71 & 8075 & 1.96 & 2.09 & 1326.3869 & 6.5953 & 339 & Failed SPOC comparison \\
431450742 & - & 9.72 & 8917 & 2.25 & 2.46 & 1768.1464 & 7.2324 & 524 & Failed SPOC comparison \\
470824703 & - & 8.32 & 9036 & 2.28 & 1.75 & 1767.7054 & 1.2960 & 294 & Failed SPOC comparison \\
269325108 & - & 9.49 & 8374 & 2.07 & 2.40 & 1769.8353 & 7.1752 & 262 & Failed SPOC comparison \\
343442463 & 1387 & 8.96 & 7976 & 1.92 & 1.78 & 1739.5304 & 0.7838 & 392 & Secondary eclipse detected \\
343439239 & - & 8.59 & 8998 & 2.27 & 1.91 & 1740.3855 & 4.0651 & 154 & Failed SPOC comparison \\
468812400 & - & 8.19 & 8821 & 2.22 & 2.11 & 1741.3359 & 2.0529 & 529 & Secondary eclipse detected \\
334613827 & - & 9.60 & 8993 & 2.27 & 1.98 & 1740.6024 & 1.1292 & 512 & Failed SPOC comparison \\
420126478 & - & 8.50 & 8621 & 2.15 & 1.89 & 1740.8406 & 8.0253 & 410 & Failed SPOC comparison \\
430536269 & - & 9.94 & 8757 & 2.20 & 2.30 & 1740.6207 & 1.6858 & 657 & Failed SPOC comparison \\
430231738 & - & 9.60 & 8743 & 2.19 & 1.91 & 1744.8064 & 5.7583 & 1270 & Secondary eclipse detected \\
427994565 & - & 6.12 & 8639 & 2.16 & 2.24 & 1739.7719 & 1.5429 & 132 & Secondary eclipse detected \\
331278465 & - & 9.76 & 7802 & 1.85 & 1.71 & 1739.7875 & 0.7539 & 165 & Failed SPOC comparison \\
421236698 & - & 9.71 & 7506 & 1.73 & 1.97 & 1712.1290 & 1.1305 & 1293 & Nearby eclipsing binary detected \\
422515775 & - & 9.85 & 8849 & 2.23 & 2.25 & 1748.9265 & 9.6392 & 246 & Failed SPOC comparison \\
239592686 & - & 9.67 & 8047 & 1.95 & 1.86 & 1715.4985 & 7.0249 & 1260 & Failed SPOC comparison \\
383850996 & - & 9.19 & 7861 & 1.87 & 1.62 & 1712.2741 & 2.3012 & 323 & Failed SPOC comparison \\
63700569 & - & 9.32 & 7836 & 1.86 & 2.21 & 1715.6477 & 1.5826 & 306 & Secondary eclipse detected \\
390974108 & - & 7.51 & 8541 & 2.13 & 1.74 & 1754.3147 & 7.1821 & 98 & Failed SPOC comparison \\
367102581 & 1522 & 9.40 & 9024 & 2.28 & 1.89 & 1780.2689 & 2.9799 & 494 & False Positive on ExoFOP \\
290222306 & - & 9.85 & 8761 & 2.20 & 1.55 & 1713.5492 & 1.9739 & 996 & Nearby eclipsing binary detected \\
330084158 & - & 9.93 & 8467 & 2.10 & 1.51 & 1711.8541 & 0.8722 & 845 & Secondary eclipse detected \\
210170166 & - & 9.00 & 8206 & 2.01 & 2.11 & 2797.4909 & 0.9807 & 653 & Failed SPOC comparison \\
15462531 & - & 5.60 & 8851 & 2.23 & 2.46 & 1684.9151 & 9.8984 & 164 & Failed SPOC comparison \\
88091070 & - & 7.68 & 8381 & 2.07 & 2.00 & 1684.2166 & 0.9239 & 1220 & Nearby eclipsing binary detected \\
364074479 & - & 8.89 & 8278 & 2.04 & 1.66 & 1767.1186 & 0.8314 & 700 & Secondary eclipse detected \\
187942715 & - & 9.99 & 7746 & 1.83 & 1.55 & 1685.0546 & 1.8514 & 1238 & Failed SPOC comparison \\
219599875 & - & 8.88 & 8597 & 2.15 & 2.06 & 2772.3107 & 6.6985 & 603 & Failed SPOC comparison \\
296012680 & - & 8.52 & 7881 & 1.88 & 1.72 & 1698.1243 & 1.0870 & 389 & Secondary eclipse detected \\
13116674 & - & 9.92 & 8086 & 1.96 & 1.75 & 1685.0552 & 1.8000 & 1039 & Failed SPOC comparison \\
202678853 & - & 9.58 & 8332 & 2.06 & 1.70 & 1683.9902 & 0.9404 & 1639 & Failed SPOC comparison \\
117496264 & - & 8.50 & 7705 & 1.81 & 1.76 & 1793.0813 & 1.2119 & 225 & Failed SPOC comparison \\
172060088 & - & 9.20 & 8394 & 2.08 & 1.78 & 1688.3406 & 7.4759 & 301 & Failed SPOC comparison \\
185340863 & - & 8.82 & 8771 & 2.20 & 2.25 & 1684.8948 & 1.9538 & 323 & Nearby eclipsing binary detected \\
278262895 & - & 9.91 & 8198 & 2.01 & 1.88 & 1684.3115 & 0.9688 & 406 & Secondary eclipse detected \\
185453858 & - & 9.65 & 7522 & 1.74 & 1.59 & 1685.2755 & 4.2118 & 1568 & Nearby eclipsing binary detected \\
326849316 & - & 9.60 & 8938 & 2.25 & 1.74 & 2771.0924 & 0.9371 & 1129 & Failed SPOC comparison \\
267489265 & 1132 & 9.23 & 7880 & 1.88 & 1.86 & 1683.6600 & 1.5503 & 387 & False Positive on ExoFOP \\
390977166 & - & 7.90 & 8833 & 2.22 & 1.85 & 1684.9453 & 1.0289 & 110 & Failed SPOC comparison \\
158211785 & - & 9.44 & 8259 & 2.03 & 1.95 & 1684.8716 & 1.6026 & 207 & Nearby eclipsing binary detected \\
243281362 & - & 9.41 & 7778 & 1.84 & 1.85 & 1697.6621 & 2.1780 & 702 & Failed SPOC comparison \\
120689840 & - & 9.17 & 7792 & 1.85 & 1.62 & 1685.5115 & 7.4044 & 1043 & Failed SPOC comparison \\
350067271 & - & 9.85 & 8616 & 2.15 & 1.66 & 2393.5661 & 5.2109 & 626 & Failed SPOC comparison \\
235679672 & - & 7.41 & 7527 & 1.74 & 1.56 & 1684.2171 & 1.3567 & 75 & Failed SPOC comparison \\
115472755 & - & 9.31 & 7873 & 1.88 & 1.90 & 1774.4964 & 7.9155 & 1007 & Failed SPOC comparison \\
353542443 & - & 8.68 & 7880 & 1.88 & 1.85 & 2744.6443 & 6.4633 & 1411 & Failed SPOC comparison \\
329933455 & - & 9.71 & 7640 & 1.78 & 2.43 & 2019.6076 & 1.4990 & 345 & Failed SPOC comparison \\
154067210 & - & 8.34 & 8669 & 2.17 & 1.91 & 1687.6100 & 8.3673 & 108 & Failed SPOC comparison \\
257046629 & - & 7.48 & 8485 & 2.11 & 1.71 & 1685.4794 & 7.3743 & 72 & Failed SPOC comparison \\
353971167 & - & 5.89 & 7922 & 1.90 & 2.31 & 1908.8308 & 8.6237 & 391 & Failed SPOC comparison \\
67391773 & - & 8.12 & 8619 & 2.15 & 1.83 & 1876.2741 & 7.4568 & 605 & Failed SPOC comparison \\
160393513 & - & 9.16 & 7867 & 1.88 & 1.83 & 1684.7794 & 0.9507 & 263 & Nearby eclipsing binary detected \\
417829082 & - & 7.28 & 8608 & 2.15 & 2.27 & 1848.6663 & 8.3566 & 143 & Failed SPOC comparison \\
417829085 & - & 7.62 & 8194 & 2.01 & 2.07 & 1848.6463 & 7.5499 & 217 & Failed SPOC comparison \\
235344374 & - & 8.53 & 8804 & 2.21 & 2.40 & 1470.0971 & 2.8624 & 191 & Failed SPOC comparison \\
468983464 & - & 8.79 & 7712 & 1.81 & 2.21 & 1492.2579 & 0.6478 & 900 & Secondary eclipse detected \\
117160089 & - & 8.56 & 7520 & 1.74 & 1.82 & 1872.1306 & 0.8302 & 147 & Failed SPOC comparison \\
453194892 & - & 9.99 & 8817 & 2.22 & 2.42 & 1493.1831 & 1.9978 & 564 & Failed SPOC comparison \\
16600483 & - & 8.91 & 8716 & 2.18 & 2.00 & 1851.5746 & 6.8672 & 255 & Failed SPOC comparison \\
80883443 & - & 7.43 & 8398 & 2.08 & 2.01 & 1848.5721 & 8.2882 & 134 & Failed SPOC comparison \\
73299751 & - & 9.33 & 8099 & 1.97 & 2.23 & 1497.7270 & 9.7373 & 432 & Failed SPOC comparison \\
318693043 & - & 9.57 & 8082 & 1.96 & 1.59 & 1493.6688 & 2.1942 & 735 & Failed SPOC comparison \\
291499489 & - & 9.90 & 8164 & 1.99 & 1.62 & 2202.7575 & 0.9652 & 822 & Nearby eclipsing binary detected \\
386588115 & - & 8.96 & 7739 & 1.83 & 1.66 & 2204.0876 & 2.2795 & 187 & Failed SPOC comparison \\
386527071 & - & 9.30 & 7600 & 1.77 & 1.73 & 2203.4276 & 1.3993 & 528 & Nearby eclipsing binary detected \\
191812422 & - & 7.68 & 7538 & 1.74 & 2.05 & 1848.6579 & 8.4208 & 242 & Failed SPOC comparison \\
457094212 & - & 7.22 & 8764 & 2.20 & 2.11 & 2202.4327 & 0.9554 & 88 & Failed SPOC comparison \\
159498300 & - & 9.11 & 8384 & 2.07 & 1.84 & 1848.5975 & 9.9565 & 683 & Failed SPOC comparison \\
67294909 & - & 9.82 & 8772 & 2.20 & 2.49 & 1819.3703 & 1.7908 & 261 & Nearby eclipsing binary detected \\
138167103 & - & 9.71 & 7841 & 1.87 & 1.94 & 1817.7251 & 1.9169 & 834 & Secondary eclipse detected \\
189957931 & - & 9.88 & 7626 & 1.78 & 1.94 & 1844.1128 & 1.0768 & 639 & Failed SPOC comparison \\
310785538 & - & 9.83 & 8361 & 2.07 & 1.65 & 1818.3782 & 0.8317 & 345 & Failed SPOC comparison \\
238459076 & - & 9.87 & 8125 & 1.98 & 2.25 & 2475.2044 & 0.8344 & 612 & Failed SPOC comparison \\
461697646 & - & 9.20 & 9041 & 2.29 & 1.79 & 1468.7623 & 0.5234 & 749 & Nearby eclipsing binary detected \\
81817111 & - & 9.53 & 7859 & 1.87 & 2.42 & 2475.6866 & 3.7706 & 567 & Failed SPOC comparison \\
269748569 & - & 9.11 & 8892 & 2.24 & 1.99 & 2475.6144 & 1.5867 & 234 & Failed SPOC comparison \\
414897731 & - & 9.11 & 8528 & 2.12 & 2.07 & 1818.1053 & 0.6836 & 624 & Secondary eclipse detected \\
464804110 & - & 7.37 & 7985 & 1.92 & 2.48 & 1441.3511 & 6.9626 & 144 & Failed SPOC comparison \\
385041086 & - & 8.27 & 8729 & 2.19 & 1.88 & 1792.6606 & 0.8959 & 930 & Secondary eclipse detected \\
177998439 & - & 9.84 & 7770 & 1.84 & 1.92 & 1794.8861 & 7.1518 & 613 & Failed SPOC comparison \\
252850925 & - & 8.07 & 8833 & 2.22 & 1.74 & 1798.9255 & 6.9885 & 231 & Failed SPOC comparison \\
280567859 & - & 9.85 & 7510 & 1.73 & 1.54 & 1801.3152 & 9.0695 & 505 & Failed SPOC comparison \\
85523680 & - & 9.53 & 8337 & 2.06 & 1.92 & 1794.7812 & 7.1528 & 616 & Failed SPOC comparison \\
470933866 & - & 8.49 & 8080 & 1.96 & 1.59 & 1794.8661 & 7.1988 & 355 & Failed SPOC comparison \\
83613310 & - & 9.52 & 8375 & 2.07 & 1.54 & 1766.9374 & 1.6423 & 2166 & Secondary eclipse detected \\
428121854 & - & 8.64 & 8737 & 2.19 & 2.18 & 1794.6664 & 7.2474 & 436 & Failed SPOC comparison \\
264805320 & - & 9.73 & 8501 & 2.11 & 1.79 & 1793.7160 & 1.5120 & 532 & Failed SPOC comparison \\
251171941 & - & 8.00 & 7810 & 1.85 & 2.26 & 1794.8510 & 7.1518 & 232 & Failed SPOC comparison \\
440044679 & - & 7.80 & 8048 & 1.95 & 2.07 & 1774.4860 & 7.9155 & 677 & Failed SPOC comparison \\
52712633 & - & 8.42 & 8517 & 2.12 & 2.46 & 1794.0854 & 7.5521 & 189 & Failed SPOC comparison \\
243183474 & - & 9.12 & 7524 & 1.74 & 2.05 & 1774.6297 & 7.9143 & 768 & Failed SPOC comparison \\
54464870 & 2115 & 8.29 & 8153 & 1.99 & 2.11 & 1792.6954 & 3.6944 & 249 & Nearby eclipsing binary detected \\
202864596 & - & 9.50 & 7782 & 1.84 & 1.83 & 1799.1116 & 8.1953 & 371 & Failed SPOC comparison \\
251466508 & - & 8.13 & 7583 & 1.76 & 2.13 & 1799.0910 & 7.2654 & 214 & Failed SPOC comparison \\
196748157 & - & 8.94 & 7808 & 1.85 & 2.23 & 1768.0602 & 7.1860 & 524 & Failed SPOC comparison \\
196748154 & - & 9.13 & 8269 & 2.03 & 1.88 & 1768.0852 & 7.1860 & 679 & Failed SPOC comparison \\
348527821 & - & 9.82 & 8898 & 2.24 & 1.94 & 1794.9110 & 7.1055 & 553 & Failed SPOC comparison \\
44558568 & - & 8.96 & 7585 & 1.76 & 1.97 & 1768.0144 & 7.2334 & 403 & Failed SPOC comparison \\
184684986 & - & 9.16 & 8024 & 1.94 & 1.63 & 1795.2861 & 6.9232 & 342 & Failed SPOC comparison \\
307040154 & - & 9.59 & 7772 & 1.84 & 1.84 & 1793.2768 & 3.3662 & 1133 & Secondary eclipse detected \\
348140245 & - & 7.28 & 7714 & 1.82 & 1.77 & 1794.8709 & 7.1518 & 305 & Failed SPOC comparison \\
191078049 & - & 9.30 & 8600 & 2.15 & 1.76 & 1774.6253 & 7.9143 & 770 & Failed SPOC comparison \\
67421369 & - & 9.16 & 8117 & 1.98 & 1.64 & 1792.6868 & 8.3065 & 267 & Failed SPOC comparison \\
445494173 & - & 8.42 & 8881 & 2.24 & 2.20 & 1798.5810 & 7.2215 & 398 & Failed SPOC comparison \\
347490756 & - & 9.86 & 8737 & 2.19 & 2.07 & 1792.9358 & 0.9496 & 693 & Secondary eclipse detected \\
292318065 & - & 9.02 & 8457 & 2.10 & 1.85 & 1795.2659 & 7.0118 & 319 & Failed SPOC comparison \\
327337584 & - & 9.32 & 8667 & 2.17 & 2.00 & 1792.6574 & 2.1588 & 308 & Failed SPOC comparison \\
331740746 & - & 6.77 & 8224 & 2.02 & 1.80 & 1768.0574 & 7.1860 & 299 & Failed SPOC comparison \\
437754891 & - & 8.53 & 7684 & 1.80 & 2.04 & 1774.6353 & 7.9143 & 550 & Failed SPOC comparison \\
273268991 & - & 9.63 & 9239 & 2.34 & 2.25 & 2036.9101 & 0.5709 & 460 & Secondary eclipse detected \\
273322273 & - & 6.87 & 9264 & 2.35 & 2.23 & 2040.7701 & 6.9488 & 95 & Failed SPOC comparison \\
1503583 & - & 8.07 & 9428 & 2.40 & 2.42 & 1520.6869 & 6.3258 & 181 & Failed SPOC comparison \\
381776783 & - & 8.29 & 9281 & 2.36 & 1.89 & 1798.4961 & 7.2225 & 310 & Failed SPOC comparison \\
364214419 & - & 8.82 & 9605 & 2.45 & 2.15 & 1629.5781 & 7.0249 & 259 & Failed SPOC comparison \\
455370344 & - & 9.47 & 9517 & 2.42 & 2.05 & 1627.1084 & 2.6728 & 509 & Secondary eclipse detected \\
242238486 & - & 7.24 & 9922 & 2.54 & 1.89 & 2336.3214 & 0.8752 & 329 & Secondary eclipse detected \\
359308549 & - & 7.56 & 9709 & 2.48 & 1.65 & 1599.2986 & 1.2217 & 221 & Secondary eclipse detected \\
260851474 & - & 6.70 & 9876 & 2.52 & 2.03 & 1603.5489 & 5.8033 & 108 & Failed SPOC comparison \\
438695853 & - & 7.38 & 9861 & 2.52 & 1.70 & 1604.6336 & 7.1064 & 129 & Failed SPOC comparison \\
135095841 & - & 8.17 & 9127 & 2.31 & 1.85 & 1578.2291 & 7.2008 & 215 & Failed SPOC comparison \\
9259773 & - & 8.92 & 9429 & 2.40 & 2.23 & 1572.1446 & 7.6011 & 214 & Failed SPOC comparison \\
461125245 & - & 7.56 & 9984 & 2.56 & 1.71 & 1573.3259 & 6.4960 & 119 & Failed SPOC comparison \\
146609580 & - & 7.98 & 9238 & 2.34 & 1.85 & 1547.4312 & 8.5425 & 159 & Failed SPOC comparison \\
390440616 & - & 7.58 & 9930 & 2.54 & 1.66 & 1575.1359 & 6.4880 & 133 & Failed SPOC comparison \\
437238231 & - & 8.45 & 9110 & 2.31 & 1.88 & 1550.4617 & 9.7590 & 299 & Failed SPOC comparison \\
73077367 & - & 9.50 & 9158 & 2.32 & 1.95 & 1546.1866 & 6.5534 & 249 & Failed SPOC comparison \\
72741753 & - & 9.21 & 9458 & 2.41 & 1.86 & 2286.7413 & 6.3856 & 243 & Failed SPOC comparison \\
35757914 & - & 8.87 & 9240 & 2.34 & 2.22 & 1546.2411 & 6.6110 & 193 & Failed SPOC comparison \\
33159940 & - & 9.53 & 10000 & 2.56 & 1.75 & 1545.1611 & 2.8458 & 259 & Nearby eclipsing binary detected \\
323814524 & - & 7.81 & 9914 & 2.54 & 2.02 & 1603.8700 & 6.9433 & 91 & Failed SPOC comparison \\
308083544 & - & 9.32 & 9843 & 2.52 & 2.02 & 1356.4642 & 2.1365 & 228 & Nearby eclipsing binary detected \\
308307401 & - & 9.82 & 9617 & 2.45 & 1.77 & 1332.2200 & 7.8588 & 281 & Failed SPOC comparison \\
238597883 & 1004 & 9.14 & 9219 & 2.34 & 2.15 & 1493.5211 & 3.5730 & 483 & Nearby eclipsing binary detected \\
238645122 & - & 9.24 & 9934 & 2.54 & 2.02 & 1492.9860 & 0.9113 & 357 & Secondary eclipse detected \\
89617003 & - & 7.66 & 9836 & 2.51 & 1.55 & 1539.8921 & 7.4326 & 161 & Failed SPOC comparison \\
268914142 & - & 9.63 & 9865 & 2.52 & 1.90 & 1492.4861 & 8.9188 & 363 & Failed SPOC comparison \\
144085672 & - & 8.67 & 9513 & 2.42 & 2.36 & 1523.4521 & 5.8805 & 207 & Failed SPOC comparison \\
364398081 & - & 9.54 & 9856 & 2.52 & 2.12 & 1342.9556 & 2.4091 & 376 & Failed SPOC comparison \\
112321756 & - & 9.41 & 9428 & 2.40 & 2.13 & 1495.0018 & 8.0833 & 324 & Failed SPOC comparison \\
130338349 & - & 9.91 & 9268 & 2.35 & 2.18 & 1471.2944 & 8.1270 & 409 & Failed SPOC comparison \\
372908153 & - & 7.88 & 9137 & 2.31 & 1.87 & 1341.7665 & 8.3249 & 89 & Failed SPOC comparison \\
443372514 & - & 9.44 & 9547 & 2.43 & 2.27 & 1469.3849 & 0.5972 & 1026 & Failed SPOC comparison \\
443385985 & - & 9.56 & 9736 & 2.48 & 2.26 & 1475.2333 & 8.0585 & 466 & Failed SPOC comparison \\
55401489 & - & 8.38 & 9260 & 2.35 & 1.91 & 1342.0865 & 6.7199 & 125 & Failed SPOC comparison \\
326242263 & - & 7.10 & 9380 & 2.38 & 1.60 & 1412.6958 & 8.8528 & 343 & Failed SPOC comparison \\
352027833 & - & 7.48 & 9319 & 2.37 & 1.92 & 1774.5174 & 7.9155 & 463 & Failed SPOC comparison \\
426111489 & - & 8.72 & 9378 & 2.38 & 2.25 & 1767.1842 & 0.8666 & 628 & Secondary eclipse detected \\
366870097 & - & 9.98 & 9193 & 2.33 & 2.06 & 1779.9995 & 2.9946 & 412 & Failed SPOC comparison \\
323299601 & - & 9.53 & 9238 & 2.34 & 1.53 & 1740.1967 & 0.9575 & 198 & Failed SPOC comparison \\
264405661 & - & 9.26 & 9242 & 2.34 & 1.84 & 1769.6588 & 8.8588 & 132 & Failed SPOC comparison \\
430108490 & - & 8.63 & 9269 & 2.35 & 2.12 & 1714.5987 & 9.3681 & 1127 & Failed SPOC comparison \\
126449150 & - & 8.15 & 9352 & 2.38 & 1.69 & 1711.6936 & 1.5136 & 1738 & Secondary eclipse detected \\
365683032 & 1354 & 8.79 & 9224 & 2.34 & 1.65 & 1716.3229 & 1.4288 & 1546 & Failed SPOC comparison \\
98577715 & - & 9.27 & 9445 & 2.40 & 2.20 & 1684.8965 & 1.2768 & 150 & Failed SPOC comparison \\
168906586 & - & 9.73 & 9102 & 2.30 & 2.04 & 1691.9907 & 9.9208 & 334 & Failed SPOC comparison \\
69259883 & - & 9.65 & 9248 & 2.35 & 2.20 & 1684.6424 & 0.8210 & 317 & Nearby eclipsing binary detected \\
258558086 & - & 8.49 & 9430 & 2.40 & 1.71 & 2021.3325 & 7.6598 & 484 & Failed SPOC comparison \\
137778067 & - & 9.97 & 9263 & 2.35 & 2.10 & 1686.5360 & 8.9690 & 347 & Failed SPOC comparison \\
202442982 & - & 7.89 & 9277 & 2.35 & 1.94 & 1698.5231 & 2.7389 & 810 & Secondary eclipse detected \\
285002773 & - & 7.45 & 9484 & 2.41 & 1.74 & 1876.5056 & 8.0095 & 289 & Failed SPOC comparison \\
452983279 & - & 9.93 & 9575 & 2.44 & 2.49 & 1492.6269 & 1.3628 & 385 & Failed SPOC comparison \\
247904876 & - & 9.59 & 9880 & 2.53 & 2.46 & 1469.4225 & 0.8446 & 211 & Failed SPOC comparison \\
450050496 & - & 6.79 & 9651 & 2.46 & 1.90 & 1444.8312 & 9.5935 & 194 & Failed SPOC comparison \\
82393678 & - & 8.64 & 9613 & 2.45 & 2.45 & 1792.3467 & 0.6726 & 489 & Failed SPOC comparison \\
240741578 & - & 9.03 & 9522 & 2.42 & 2.41 & 1767.9151 & 1.4391 & 182 & Nearby eclipsing binary detected \\
242604239 & - & 8.74 & 9238 & 2.34 & 2.40 & 1774.5055 & 7.9155 & 860 & Failed SPOC comparison \\
250657037 & - & 8.13 & 9240 & 2.34 & 1.76 & 1794.8862 & 7.1518 & 311 & Failed SPOC comparison \\
354253270 & - & 9.81 & 9268 & 2.35 & 2.19 & 1792.8012 & 2.6646 & 720 & Failed SPOC comparison \\
249829768 & - & 9.80 & 9463 & 2.41 & 2.28 & 1792.3612 & 1.5019 & 781 & Failed SPOC comparison \\
241112201 & - & 9.43 & 9309 & 2.36 & 1.86 & 1798.1662 & 5.9429 & 585 & Failed SPOC comparison \\
77258978 & - & 8.23 & 9245 & 2.34 & 1.97 & 1799.0463 & 7.2185 & 341 & Failed SPOC comparison \\
\enddata
\tablecomments{$T$ mag, $T_{\rm eff}$, $R_\star$, and $M_\star$ are the properties of the host star according to the TIC \citep{stassun2018tess, stassun2019}. $T_0$, $P_{\rm orb}$, and $\delta$ are the transit midpoint times, orbital periods, and transit depths estimated by the BLS periodogram. The column titled ``Vetting Outcome'' describes the reason why the TCE was classified as a false alarm or false positive. This table is published in its entirety in machine-readable format.}
\end{deluxetable*}


\bibliography{sample631}{}
\bibliographystyle{aasjournal}



\end{document}